\pdfoutput=1

\documentclass[12pt]{article}
\usepackage[latin9]{inputenc}
\usepackage{color}
\usepackage{float}
\usepackage{amsmath}
\usepackage{amssymb}
\usepackage{graphicx}
\usepackage{setspace}
\usepackage[authoryear]{natbib}
\PassOptionsToPackage{normalem}{ulem}
\usepackage{ulem}
\makeatletter

\providecommand{\tabularnewline}{\\}
\floatstyle{ruled}
\newfloat{algorithm}{tbp}{loa}
\providecommand{\algorithmname}{Algorithm}
\floatname{algorithm}{\protect\algorithmname}

\newenvironment{lyxlist}[1]
	{\begin{list}{}
		{\settowidth{\labelwidth}{#1}
		 \setlength{\leftmargin}{\labelwidth}
		 \addtolength{\leftmargin}{\labelsep}
		 }}
	{\end{list}}

\usepackage{fullpage}

\usepackage{bbm}
\usepackage[bottom]{footmisc}

\providecommand{\keywords}[1]{\textbf{\textit{Keywords: }} #1}

\newenvironment{acknowledgements}{
\section*{Acknowledgements}\begin{itshape}
}{\end{itshape}}

\title{Inverse modeling of hydrologic parameters in CLM4 via generalized polynomial chaos in the Bayesian framework}
\author{Georgios Karagiannis 
\thanks{Georgios Karagiannis: Department of Mathematical Sciences, Durham University, UK; 
georgios.karagiannis@durham.ac.uk},  
Zhangshuan Hou
\thanks{Zhangshuan Hou: Computational Mathematics, Pacific Northwest National Laboratory, Richland, WA 99352, USA; 
Zhangshuan.Hou@pnnl.gov}, 
Maoyi Huang, 
\thanks{Maoyi Huang: Computational Mathematics, Pacific Northwest National Laboratory, Richland, WA 99352, USA; 
Maoyi.Huang@pnnl.gov}, 
and Guang Lin
\thanks{Guang Lin: Department of Mathematics, School of Mechanical Engineering, 
Department of Statistics (Courtesy), 
Department of Earth, Atmospheric, and 
Planetary Sciences (Courtesy), 
Purdue University, West Lafayette, IN 47907, USA; 
guanglin@purdue.edu
}
}

\@ifundefined{showcaptionsetup}{}{%
 \PassOptionsToPackage{caption=false}{subfig}}
\usepackage{subfig}
\makeatother

\begin{document}
\begin{singlespace}

\maketitle
\begin{abstract}
In this study, the applicability of generalized polynomial chaos (gPC)
expansion for land surface model parameter estimation is evaluated.
We compute the (posterior) distribution of the critical hydrological
parameters that are subject to great uncertainty in the community
land model (CLM). The unknown parameters include those that have been
identified as the most influential factors on the simulations of surface
and subsurface runoff, latent and sensible heat fluxes, and soil moisture
in CLM4.0. We setup the inversion problem this problem in the Bayesian
framework in two steps: (i) build a surrogate model expressing the
input-output mapping, and (ii) compute the posterior distributions
of the input parameters. Development of the surrogate model is done
with a Bayesian procedure, based on the variable selection methods
that use gPC expansions. Our approach accounts for bases selection
uncertainty and quantifies the importance of the gPC terms, and hence
all the input parameters, via the associated posterior probabilities. 
\end{abstract}
\keywords{Uncertainty quantification, generalized polynomial chaos,
Bayesian inversion, US-ARM, inverse modeling}

\end{singlespace}

\section{Introduction\label{sec:Introduction}}

\textcolor{black}{Scientists use land surface models (LSMs) to quantitatively
simulate the exchange of water and energy fluxes at the Earth surface-atmosphere
interface. During the past decades, LSMs have evolved from oversimplified
schemes describing only the surface boundary conditions for general
circulation models (GCMs) to complex models that integrate modules
representing biogeochemical, hydrological, and energy cycles at the
surface-atmosphere {[}\citet{pitman2003evolution}{]}. Built upon
mathematical formulations of the laws of physics, the model parameters
are usually associated with certain physical meaning and have influences
on the major model outputs such as water and energy fluxes. It is
a common hypothesis that the parameters are measurable and transferable
to locations sharing the same physical properties or site conditions,
as assumed in the Project for Intercomparison of Land Surface Parameterization
Schemes (PILPS) {[}\citet{bastidas1999sensitivity,henderson1995project,henderson1996predicting}{]}.
However, default assignment of parameter values are actually inappropriate
according to {[}\citet{bastidas1999sensitivity,rosero2010quantifying}{]}.
Meanwhile, given the high dimensionality of the parameter space and
complexity of the land surface system, more studies are needed to
understand what parameters are more uncertain and what the potential
is for using observations to constrain or calibrate the uncertain
parameters to better capture uncertainty in the resulting land surface
states {[}\citet{hou2012sensitivity,huang2013uncertainty}{]}. This
study aims at quantifying the uncertainties related to a subset of
parameters in a community LSM named the Community Land Model (CLM),
which is the land component within the Community Earth System Model
(CESM) (formerly known as the Community Climate System Model (CCSM)
{[}\citet{collins2006formulation,gent2010improvements,lawrence2011parameterization}{]}.}

\textcolor{black}{There are different sources of uncertainties associated
with LSMs, and they include model structural uncertainty due to simplified
assumptions or representations of the actual processes or phenomena.
Many such assumptions are only valid under specific conditions. In
addition, LSMs are also subject to uncertainty related to input parameter
values, particularly because a large number of input parameters such
as those associated with land cover and land use and soil properties
are not directly measurable at the scales of its applications. A common
practice in land surface modeling has been to define a set of default
parameter values that are globally applicable. Efforts has been made
by the land surface modeling community in the last two decades to
deal with uncertainty in model parameters, data, and model structure.
The focus of this study is to reduce uncertainty in model parameters
via generalized polynomial expansion and Bayesian inversion.}

Stochastic inversion for a high-dimensional parameter space is computationally
demanding. In order to address this problem, surrogate models can
be used as alternatives to the numerical simulators. Ensemble simulations,
which are required to develop surrogate models, can be performed efficiently
in a task-parallel manner on supercomputing facilities. But surrogate
development itself is a non-trivial effort. The use of surrogates
in the calibration of climate models or LSMs is particularly uncommon.
In {[}\citet{ray2015bayesian,huang2016surrogate}{]}, the authors
used various surrogates (e.g., polynomials and/or universal kriging)
to calibrate hydrological parameters of CLM 4.0 using measurements
of latent heat fluxes. Two competing models were used for the model-data
mismatch to estimate a composite of measurement error and (a crude
estimate of) the structural error of CLM. In {[}\citet{gong2015multi}{]},
the authors used adaptive surrogate-based optimization to perform
parameter estimation of the Common Land Model using six observables
jointly; 12 independent parameters were (deterministically) calibrated.
{[}\citet{sargsyan2014dimensionality}{]} attempted to construct surrogates
for five variables of interests from CLM4 with prognostic carbon and
nitrogen modules turned on (i.e., CLM4-CN) using Bayesian compressive
sensing (BCS) in combination with polynomial chaos expansions (PCEs).
They found that the input-output relationship in CLM4-CN could be
composed of qualitatively different regimes (i.e., live or dead vegetation
regimes associated with different regions in the parameter space),
so that clustering-based and classification - piecewise PCE construction
is needed.

This study evaluates the applicability of using gPC for CLM4 hydrological
model calibration. We present a fully Bayesian procedure, based on
\citep{KaragiannisLin2014}, which couples fully Bayesian statistics,
variable selection, and generalised polynomial chaos surrogate models
to address the uncertainty quantification and model inversion problem
in CLM4. The procedure produces a cheap mathematical/statistical approximation
of the model output (latent heat flux) as a function of a set of model
parameters. Bayesian inversion of the model parameters given observations
is performed by using the produced cheap gPC surrogate model instead
of the expensive computer model output in the likelihood function,
and then performing Bayesian parametric inference facilitated by Markov
chain Monte Carlo methods. The method allows dimension reduction and
selection of the important model parameters which significantly influence
the output parameter by computing inclusion posterior probabilities.

The layout of the paper is as follows: in Section \ref{sec:Data-set},
we describe the study site, input data, and the conducted numerical
simulations, and present the parameters of interest that we calibrate;
In Section \ref{sec:Bayesian-methodology}, we present the inversion
methodology using gPC in the Bayesian framework; In Section \ref{sec:Application},
we evaluate the inversion results; In Section \ref{sec:Conclusions},
we draw our conclusions.

\section{Dataset and Parameterization \label{sec:Data-set}}

{The FLUXNET database (www.fluxdata.org) contains half-hourly observations
of ecosystem CO2, heat fluxes and meteorological data of more than
250 sites worldwide and for a total of 960 site-years. The study site
in this study is US-ARM (ARM Southern Great Plains site, Lamont, Oklahoma)
{[}\citet{fischer2007spatiotemporal,torn2016ameriflux}{]}. It has
a vegetation type of croplands, covered with temporary crops followed
by harvest and a bare soil period (e.g., single and multiple cropping
systems), a humid subtropical climate, and clay-type soil texture.

Observational data used in parameter estimation are observed latent
heat fluxes and runoff measurements, which are processed and gap-filled
to obtain daily and monthly averaged data. We consider 10 critical
hydrological parameters that are likely to have dominant impacts on
the simulation of surface and subsurface runoff, latent and sensible
fluxes, and soil moisture as suggested in existing literature {[}\citet{hou2012sensitivity,NiuYangDickinsonGulden2005,NiuYangDickinsonGuldenSu2007,oleson2008improvements,oleson2010technical}{]}.
The selected parameters are fmax, Cs, fover, fdrai, qdrai,max (denoted
as Qdm hereinafter), Sy, b, Ys, Ks, and \ensuremath{\theta}s. Explanations
of the 10 parameters and their prior information are shown in Table
1 in \citet{hou2012sensitivity}. Prior distributions of the parameters
were derived based on entropy theory and 256 samples were generated
using quasi Monte Carlo sampling. Numerical simulations corresponding
to sampled parameter sets were conducted, which yield the data matrix
of inputs (i.e., realizations of the 10 parameters) and outputs (i.e.,
latent heat fluxes), which enables development of response surfaces
or surrogates that can be used for sensitivity analysis, parameter
ranking, and model calibration. }

\section{Bayesian methodology\label{sec:Bayesian-methodology}}

We describe a synergy of Bayesian methods aiming at quantifying the
importance of input CLM4 model parameters, calibrating these parameters
against real measurements, as well as building a surrogate model describing
the input-output relation in CLM4 model, in the Bayesian framework.

\subsection{Bayesian inverse problem setup \label{sec:Methodology_Inverse}}

Bayesian inverse methods allow the uncertainty quantification of input
parameters of a computer model from observations in a probabilistic
manner \citep{MarzoukXiu2009}. Bayesian inference is performed through
the posterior distribution which is derived according the Bayes theorem
that requires the specification of two components: the likelihood
function representing the information from the measurements, and the
prior distribution representing the researcher's prior information
about the uncertain parameters.

We consider that the output value observed $u^{\text{f}}\in\mathbb{R}^{d_{u}}$
is associated to some unknown input $\xi$ via a forward model (e.g.
CLM4) $u(\cdot):\mathbb{R}^{d_{\xi}}\rightarrow\mathbb{\mathbb{R}}$,
and possibly contaminated by some additive observational noise $\varepsilon^{\text{\text{f}}}$
(residuals); namely $u^{\text{f}}=u(\xi)+\varepsilon^{\text{\text{f}}}$.
A reasonable assumption is to model $\varepsilon^{\text{\text{f}}}$
to be Normally distributed with mean zero and variance $\sigma^{2}$.
This is justified by central limit theorem arguments, and the act
that in the noise term there are accumulated several insignificant
random measurement errors from observations and modeling errors due
to model parameterization. A reasonable assumption is to model $\varepsilon^{\text{\text{f}}}$
to be Normally distributed with mean zero and variance $\sigma^{2}$.
Thus, we consider a likelihood: 
\[
\mathcal{L}(u^{\text{f}}|\xi)=\text{N}_{d_{u}}(u^{\text{f}}|u(\xi),\text{diag}(\sigma_{1}^{2},...,\sigma_{d_{u}}^{2})).
\]

To account the uncertainty about the input parameters we assign prior
distribution $\xi\sim\pi_{\xi}(\cdot)=\prod_{i=1}^{d_{\xi}}\pi_{\xi_{i}}(\cdot)$,
where $\pi_{\xi_{i}}(\xi_{i})$ are considered to be shifted Beta
distributions $\text{Beta}(a_{\xi_{i}},b_{\xi_{i}},\xi_{i,\min},\xi_{i,\max})$,
with known $\xi_{i}\in[\xi_{i,\min},\xi_{i,\max}]$ and $a_{\xi_{i}}>0$,
$b_{\xi_{i}}>0$. Here, $\xi_{i,\min}$, and $\xi_{i,\max}$ can be
specified because a reasonable range of the model parameter is often
a priori known. Also $a_{\xi_{i}}$ and $b_{\xi_{i}}$ can be specified
by using the method of moments or the maximum entropy method \citep{berger2013statistical}
because often reasonable values for the mean and variance of the model
parameter are a priori known. To account for uncertainty about the
unknown parameter $\{\sigma_{j}^{2}\}$, we assign an prior distribution
$\sigma_{j}^{2}\sim\pi_{\sigma_{j}^{2}}(\cdot)$, for $j=1,...,d_{u}$.
A computationally convenient choice for the priors of the variance
$\{\pi_{\sigma_{j}^{2}}(\cdot)\}$ is the Inverse Gamma prior distribution
$\text{IG}(a_{\sigma^{2}},b_{\sigma^{2}})$ with parameters $a_{\sigma^{2}}>0$,
and $b_{\sigma^{2}}>0$, as it is semi-conjugate prior to the likelihood.
By considering the likelihood variance $\sigma^{2}$ as random unknown
parameters and treating the problem in the Bayesian framework, we
let the data decide a proper value for $\sigma^{2}$ through the posterior
distribution. The prior hyper-parameters are considered to be fixed
values and pre-defined by the researcher.

The posterior distribution of $\xi,\sigma^{2}$ given the observations
$u^{\text{f}}$ is 
\begin{align}
\pi_{\xi,\sigma^{2}}(\xi,\sigma^{2}|u^{\text{f}}) & =\frac{\mathcal{L}(u^{\text{f}}|\xi,\sigma^{2})\pi_{\xi}(\xi)\pi_{\sigma^{2}}(\sigma^{2})}{\int\mathcal{L}(u^{\text{f}}|\xi,\sigma^{2})\pi_{\xi}(\xi)\pi_{\sigma^{2}}(\sigma^{2})\text{d}(\xi,\sigma^{2})},\label{eq:post_xi}
\end{align}
by using the Bayes' theorem, and marginal distribution $\pi_{\xi}(\xi|u^{\text{f}})$
quantifies the uncertainty about the model parameters $\xi$.

The posterior distribution density (\ref{eq:post_xi}) is usually
intractable. If $u(\cdot)$ was known or cheap to be computed, inference
on $\xi$ could be obtained using standard Markov chain Monte Carlo
(MCMC) methodology \citep{ChristianCasella2004}. In Algorithm \ref{alg:MCMCalg_inv},
we present a simple MCMC sampler that updates iteratively in two blocks
$\xi$ and $\sigma^{2}$. The output sample $\{(\xi^{(t)},\sigma^{2,(t)})\}_{t=1}^{T}$
of Algorithm \ref{alg:MCMCalg_inv} can be used to perform inference
on $\xi$ and $\sigma^{2}$ e.g. expectation of any function $h(\cdot)$
of $\xi$ can be approximated as 
\[
\text{E}(h(\xi))\approx\frac{1}{T}\sum_{t=1}^{T}h(\xi^{(t)})
\]
for large $T$.

In the present application of the CLM4 model, $u(\cdot)$ is prohibitively
expensive to be computed iteratively because the forward model (CLM4)
is too expensive to run. This prevents us from using directly this
method, and particularly Algorithm \ref{alg:MCMCalg_inv}. To overcome
this issue, we build a cheap but accurate proxy (called surrogate
model) for $u(\cdot)$, and plug it in (\ref{eq:post_xi}). The construction
of such a surrogate model, in the Bayesian framework, is discussed
in Section \ref{sec:Methodology_surrogate}.

\begin{algorithm}
\protect\caption{Blocks of the MCMC sweep \label{alg:MCMCalg_inv}}

\begin{lyxlist}{00.00.0000}
\item [{Block I   }] \uline{Update }%
\mbox{%
$\xi$%
}\uline{ :} 
\end{lyxlist}
\begin{itemize}
\item Simulate $\xi$ from a Metropolis-Hastings transition probability
that targets $\pi_{\xi|\sigma^{2}}(\xi|u^{\text{f}},\sigma^{2})$,
where $\pi_{\xi,\sigma^{2}}(\xi|u^{\text{f}},\sigma^{2})\propto\text{N}(u^{\text{f}}|u(\xi),\sigma^{2})\text{N}(\xi|\mu_{\xi},\sigma_{\xi}^{2})$. 
\end{itemize}
\begin{lyxlist}{00.00.0000}
\item [{Block II  }] \uline{Update }%
\mbox{%
$\sigma^{2}$%
}\uline{:} 
\end{lyxlist}
\begin{itemize}
\item Draw $\sigma^{2}$ from $\text{IG}\left(\frac{1}{2}+a_{\sigma^{2}},\frac{1}{2}\left|u^{\text{f}}-u(\xi)\right|_{2}^{2}+b_{\sigma^{2}}\right)$. 
\end{itemize}
\end{algorithm}

\subsection{Surrogate model specification \label{sec:Methodology_surrogate}}

We describe a fully Bayesian procedure for building a surrogate model,
to be used as a cheap but accurate proxy of $u(\cdot)$ in \eqref{eq:post_xi},
based on gPC expansions and MCMC methods. The highlight is that apart
from evaluating a surrogate model, the procedure is able to quantify
the importance of each PC basis, via inclusion posterior probabilities,
which allows the selection of the important input parameters: \texttt{Fmax,
Cs, Fover, Fdrai, Qdm, Sy, B, Psis, Ks, thetas}.

\subsubsection{Generalized polynomial chaos expansion\label{sub:Generalized-polynomial-chaos}}

We consider the output parameter $u(\xi)$ as a function of the $d_{\xi}$-dimensional
vector of random input variables $\xi\in\Xi$ that admits distribution
$f(\cdot)$.

The output parameter $u(\xi)$ can be represented by an infinite series
of PC bases $\left\{ \psi_{\alpha}(\cdot)\right\} $ and PC coefficients
$\left\{ c_{\alpha}\right\} $ in the tensor form: 
\begin{equation}
u(\xi)=\sum_{\alpha\in\mathbb{N}_{0}^{d_{\xi}}}c_{\alpha}\psi_{\alpha}(\xi),\label{eq:gPC}
\end{equation}
for $\xi\sim f(\cdot)$ \citep{Xiu2010}. We denote multi-indices
$\alpha:=(\alpha_{1},...,\alpha_{d_{\xi}})$ of size $d_{\xi}$ that
are defined on a set of non-negative integers $\mathbb{N}_{0}^{d_{\xi}}:=\{(\alpha_{1},...,\alpha_{d_{\xi}}):\ \alpha_{j}\in\mathbb{N}\cup\{0\}\}$.
The family of polynomial bases $\{\psi_{\alpha}(\cdot);\ \alpha\in\mathbb{N}_{0}^{d_{\xi}}\}$
contains multidimensional orthogonal polynomial bases with respect
to the probability measure $f(\cdot)$ of $\xi$. Each multidimensional
PC basis $\psi_{\alpha}(\cdot)$ results as a tensor product of univariate
orthogonal polynomial bases $\psi_{\alpha_{j}}(\cdot)$ of degree
$\alpha_{j}\in\mathbb{N}_{0}^{1}$ namely: 
\begin{equation}
\psi_{\alpha}(\xi)=\prod_{j=1}^{d_{\xi}}\psi_{\alpha_{j}}(\xi_{j}),\ \alpha_{j}\in\mathbb{N}_{0}^{1},\label{eq:tensor_full}
\end{equation}
where $\text{E}_{f}(\psi_{\alpha_{j}}(\xi)\psi_{\alpha_{j'}}(\xi))=Z_{j}\delta_{0}(j-j')$,
for $j,\ j'=1,...,d_{\xi}$ and $Z_{j}=\text{E}_{f}(\psi_{\alpha_{j}}^{2}(\xi))$.

It is common in practice, but not a panacea, for the family of PC
bases $\left\{ \psi_{\alpha}(\cdot)\right\} $ to be pre-defined so
that they are orthogonal with respect to the distribution $f(\text{d}\cdot)$.
In this way, many common distributions can be associated with a specific
family of polynomials, e.g. the Askey family \citep{XiuKarniadakis2002}.
In this work, we focus on the use of Jacobi polynomial bases \citep{XiuKarniadakis2002}
which can be defined recursively as : 
\begin{align*}
\psi_{0}(z^{(\xi)}) & =1;\\
\psi_{1}(z^{(\xi)}) & =\frac{1}{2}[a-b+(a+b-2)z_{\xi}];\\
\psi_{j}(z^{(\xi)}) & =\frac{(2j+a+b-1)[(2j+a+b)(2j+a+b-2)z^{(\xi)}+a^{2}+b^{2}]}{2j(j+a+b)(2j+a+b-2)}\psi_{j-1}(z^{(\xi)})\\
 & \quad\quad\quad-\frac{2(j+a+b-1)(j+b-1)(2j+a+b)}{2j(j+a+b)(2j+a+b-2)}\psi_{j-2}(z^{(\xi)}),\\
 & \quad\quad\quad\quad\quad\quad j=2,...,p_{\xi}-1,
\end{align*}
where $z^{(\xi)}$ is a linear transformation $z^{(\xi)}:[\xi_{\min},\xi_{\max}]\rightarrow[-1,1]$,
and $\xi_{\min}$, $\xi_{\max}$ are the minimum and maximum of $\xi$.

In practice, a truncated version of (\ref{eq:gPC}) is used by considering
a finite set of available PC bases. Traditionally, the \textit{total
truncated rule} is used, which results in the expansion form: 
\begin{equation}
u_{p_{\xi}}(\xi):=\sum_{\alpha\in A}c_{\alpha}\psi_{\alpha}(\xi),\label{eq:gPC_trunc}
\end{equation}
that accounts for only a finite set of multi-indices $A$ such that
$A=\{\alpha\in\mathbb{N}^{d_{\xi}}:\ \sum_{i=1}^{d_{\xi}}\alpha_{i}\le p_{\xi}\}$
with cardinality $m_{\xi}:=\frac{(p_{\xi}+d_{\xi})!}{p_{\xi}!d_{\xi}!}$.
Other truncation rules can be adopted \citep{BlatmanSudret2011,SargsyanSaftaNajmDebusschereRicciutoTHornton2013}.


The evaluation of the gPC expansion is challenging. Although the PC
coefficients $\{c_{\alpha};\ \alpha\in\mathbb{N}_{0}^{d_{\xi}}\}$
are equal to $c_{\alpha}=\text{E}_{f}(u(\xi)\psi_{\alpha}(\xi))/Z_{\alpha}$
where $Z_{\alpha}=\text{E}_{f}((\psi_{\alpha}(\xi))^{2})$, for $a\in\mathbb{N}_{0}^{d_{\xi}}$
\citep{Xiu2010}, they are not available in closed form due to the
intractable integration in the expectation. Moreover, in high-dimensional
scenarios if a high degree of accuracy is required, the number of
unknown PC coefficients is of order $d_{\xi}^{p_{\xi}}$ and grows
rapidly with the dimension $d_{\xi}$ and PC degree $p_{\xi}$. This
causes computational problems such as over-fitting \citep{DoostanOwhadi2011,KaragiannisLin2014}.
Reduction of the PC degree or careless omission of PC bases, in order
to reduce the number of the unknowns, may lead to a significant increase
of bias and provide inaccurate surrogate models. Hence, there is a
particular interest in keeping in the gPC only the inputs or bases
that significantly affect the output. The Bayesian procedure in Section
\ref{sub:Computations} effectively addresses these matters.

\subsubsection{Bayesian training procedure\label{sub:Computations}}

We describe a stochastic and automatic Bayesian procedure which evaluates
accurately the PC coefficients and the gPC surrogate model, while
it allows the selection of the the significant PC bases, and hence
input model parameters. This procedure is able to trade off efficiently
between the bias (caused by omitting bases) and the over-fitting are
required. Furthermore, it can select the significant PC bases and
estimate the PC coefficients simultaneously, while providing credible
intervals.

We assume there is available a training dataset $\mathcal{D}=\left\{ (u_{j},\xi_{j})\right\} {}_{j=1}^{n_{\xi}}$,
where $n_{\xi}$ is the size of the dataset, $\xi_{j}$ denotes the
random input value, and $u_{j}:=u(\xi_{j})$ denotes the output value
corresponding to the $j$-th input value $\xi_{j}$. Given the training
dataset $\mathcal{D}$ and the gPC expansion, it is 
\begin{align}
u_{j} & =u_{p_{\xi}}(\xi_{j})+\epsilon_{j}\ ,\text{ for}\ j=1,...,n_{\xi},\label{eq:gPC_system}
\end{align}
where $\epsilon_{j}\in\mathbb{R}$ is a residual term, associated
to the $j$-th datum of $\mathcal{D}$. Eq. \ref{eq:gPC_system} can
be written in matrix form: 
\begin{equation}
u=Xc+\epsilon,\label{eq:gPC_system_matrix}
\end{equation}
where $u:=(u_{j};\ j=1:n_{\xi})^{\top}$, $\epsilon:=(\epsilon_{j};\ j=1:n_{\xi})^{\top}$,
$X_{a}:=(\psi_{a}(\xi_{j});\ j=1:n_{\xi})$, and $X:=(X_{a};\ a\in A)$
is an $n_{\xi}\times m_{\xi}$ dimensional matrix of basis functions.

Following, we formulate the Bayesian model : The likelihood function
$\mathcal{L}(u|c,\sigma^{2}):=\mathcal{L}(\{u_{j}\}|\{\xi_{j}\},c,\sigma^{2})$
is: 
\begin{align}
\mathcal{L}(u|c,\sigma^{2}) & =\prod_{j=1}^{n_{\xi}}\mathsf{N}\left(u_{j}|\psi(\xi_{j})^{\top}c,\sigma^{2}\right);\label{eq:likelihood_model}\\
 & =\mathsf{N}\left(u|X^{\top}c,\mathbb{I}_{m}\sigma^{2}\right),\nonumber 
\end{align}
where $\mathsf{N}(\cdot|\mu,\sigma^{2})$ denotes the Normal density
with mean $\mu$ and variance $\sigma^{2}$. The choice of the likelihood
is merely a modelling choice. Here, the likelihood function can be
considered as a measure of goodness-of-fit of the truncated gPC expansion
to the training data-set, where the statistical discrepancy between
the real model and the gPC expansion, for a given training data-set,
is quantified by the residual term $\{\epsilon_{j}\}$. We consider
the following hierarhical prior model $\pi(c,\gamma,\sigma^{2},\lambda,\rho)$:
\begin{align}
c_{a}|\gamma_{a},\sigma^{2},\lambda & \sim\gamma_{a}\textsf{N}(c_{a}|0,\sigma^{2}/\lambda)+(1-\gamma_{a})\delta_{0}(c_{a}),\ a\in A;\nonumber \\
\gamma_{a} & \sim\text{Bernoully}(\rho),\ a\in A;\label{eq:prior_model}\\
\sigma^{2}|a_{\sigma},b_{\sigma} & \sim\text{IG}(a_{\sigma},b_{\sigma});\nonumber \\
\lambda|a_{\lambda},b_{\lambda} & \sim\mathsf{\text{G}}(a_{\lambda},b_{\lambda});\nonumber \\
\rho|a_{\rho},b_{\rho} & \sim\text{\ensuremath{\mathsf{Beta}}}(a_{\rho},b_{\rho}),\nonumber 
\end{align}
where $a_{\lambda}$, $b_{\lambda}$, $a_{\rho}$, and $b_{\rho}$
are fixed prior hyper parameters, and predetermined by the researcher.
In the Bayesian framework, inference on the unknown parameters of
the model can be performed based on the posterior distribution 
\begin{equation}
\pi(c,\gamma,\sigma^{2},\lambda,\rho|\mathcal{D})=\frac{\mathcal{L}(u|c,\sigma^{2})\pi(c,\gamma,\sigma^{2},\lambda,\rho)}{\int\mathcal{L}(u|c,\sigma^{2})\pi(c,\gamma,\sigma^{2},\lambda,\rho)\text{d}(c,\gamma,\sigma^{2},\lambda,\rho)}.\label{eq:posterior_distr}
\end{equation}
Particular interest lies on the computation of the inclusion probabilities
$\pi(\gamma_{a}|\mathcal{D})$ that refer to the marginal posterior
probability that the $a$-th basis is important, and the $\pi(c_{a}|\mathcal{D},\gamma_{a})$
that refers to the posterior density of the $a$-th PC coefficient.

To fit the Bayesian model, we resort to Markov chain Monte Carlo (MCMC)
methods because the above posterior distribution (\ref{eq:posterior_distr})
is intractable and cannot be sampled directly. The conjugate prior
model (\ref{eq:prior_model}) allows the design of a Gibbs sampler
\citep{Hans2010,GemanGeman1984} whose blocks involve sampling form
the full conditional distributions of the parameter updated. In Algorithm
\ref{alg:MCMCalg}, we represent a pseudo-code of one sweep of the
Gibbs samples, along with the associated full conditional distributions.
We highlight that the procedure is fully automatic because there is
no need to tune the algorithmic parameters involved in Algorithm \ref{alg:MCMCalg}.
The notation $c_{-a}$ ( and $X_{-a}$) refers to the vector $c$
(and matrix $X$) excluding the $a$-th element (and column). Moreover,
we denote $m_{\xi,\gamma}=\sum_{a\in A}\gamma_{a}$.

\begin{algorithm}
\protect\caption{Blocks of the Gibbs sweep}

\label{alg:MCMCalg} 
\begin{lyxlist}{00.00.0000}
\item [{Block I   }] \uline{Update }%
\mbox{%
$\left\{ (\gamma_{a},c_{a})\right\} $%
}\uline{ :} For $a\in A$, 
\end{lyxlist}
\begin{enumerate}
\item Compute $\mu_{a}$, and $s_{a}^{2}$ where 
\begin{align*}
\mu_{a} & =\left(X_{a}^{\top}X_{a}+\lambda\right)^{-1}X_{a}^{\top}\left(u-X_{-a}c_{-a}\right),\\
s_{a}^{2} & =\sigma^{2}(X_{a}^{\top}X_{a}+\lambda)^{-1}
\end{align*}
\item Update $\gamma_{a}$: draw $\gamma_{a}$ from $\mathsf{\text{Bernoulli}}(P_{a}^{(\gamma)})$,where
\[
P_{a}^{(\gamma)}=\left[1+\frac{1-\rho}{\rho}\sqrt{\frac{2\pi\sigma^{2}}{\lambda}}\mathsf{N}(0|\mu_{a},s_{a}^{2})\right]^{-1},
\]
\item Update $c_{a}$: draw $c_{a}$ from $\pi(c_{a}|u,X,\gamma,c_{-a},\rho,\sigma^{2},\lambda)$
where 
\[
\pi(c_{a}|u,X,\gamma,c_{-a},\rho,\sigma^{2},\lambda)=\begin{cases}
\delta_{0}(c_{a}) & ,\ \text{if}\ \gamma_{a}=0\\
\mathsf{N}(c_{a}|\mu_{a},s_{a}^{2}) & ,\ \text{if}\ \gamma_{a}=1
\end{cases}.
\]
\end{enumerate}
\begin{lyxlist}{00.00.0000}
\item [{Block II  }] \uline{Update }%
\mbox{%
$\sigma^{2}$%
}\uline{:} 
\end{lyxlist}
\begin{itemize}
\item Draw $\sigma^{2}$ from $\text{IG}\left(\frac{n_{\xi}}{2}+\frac{m_{\xi,\gamma}}{2}+a_{\sigma},b_{\sigma}+\frac{n_{\xi}}{2}\left|u-Xc\right|_{2}+\frac{1}{2}\lambda\left|c\right|_{2}^{2}\right)$. 
\end{itemize}
\begin{lyxlist}{00.00.0000}
\item [{Block III }] \uline{Update }%
\mbox{%
$\lambda$%
}\uline{:} , 
\end{lyxlist}
\begin{itemize}
\item Draw $\lambda$ from $\text{\ensuremath{\mathsf{G}}}\left(\frac{m_{\xi,\gamma}}{2}+a_{\lambda},\frac{1}{2\sigma^{2}}\left|c_{a}\right|_{2}^{2}+b_{\lambda}\right)$. 
\end{itemize}
\begin{lyxlist}{00.00.0000}
\item [{Block IV  }] \uline{Update }%
\mbox{%
$\rho$%
}\uline{:} 
\end{lyxlist}
\begin{itemize}
\item Draw $\rho$ from $\mathsf{\text{Beta}}(m_{\xi,\gamma}+a_{\rho},m_{\xi}-m_{\xi,\gamma}+b_{\rho})$. 
\end{itemize}
\end{algorithm}

In order to evaluate the surrogate model, as well as quantify the
importance of the input model parameters, we consider two fully Bayesian
approaches: the \textit{Bayesian model averaging} \citep{HoetingMadiganRafteryVolinsky1999}
most suitable in cases that the predictive ability of the surrogate
model is of main interest, and the \textit{median probability model}
\citep{BarbieriBerger2004} most suitable for cases where interest
in discovering a sparse (or parsimonious) representation of the stochastic
solution or selecting important input model parameters.

\paragraph{Bayesian model averaging:}

The evaluation of the gPC expansion (\ref{eq:gPC_trunc}) can be performed
by Bayesian model averaging (BMA) \citep{HoetingMadiganRafteryVolinsky1999}
if the predictive ability of the gPC expansion is of main interest.

We consider a Gibbs sample $\left\{ \left(\gamma^{(t)},c^{(t)},\sigma^{2,(t)},\lambda^{(t)},\rho^{(t)}\right)\right\} _{t=1}^{T}$
generated by Algorithm \ref{alg:MCMCalg}. Estimates and associated
standard errors of $\left\{ c_{a}\right\} $ can be computed by the
ergodic averages according to the standard Bayesian practice \citep{ChristianCasella2004,HoetingMadiganRafteryVolinsky1999},
e.g $\hat{c}_{a}=\frac{1}{T}\sum_{t=1}^{T}c_{a}^{(t)}$ and $\text{s.e.}(\hat{c}_{a})=s_{a}^{c}\sqrt{\frac{\varrho_{a}^{c}}{T}}$
where $s_{a}^{c}$ is the sample standard deviation and $\varrho_{a}^{c}$
is the integrated autocorrelation time of $\left\{ c_{a}^{(t)};\ t=1:T\right\} $
for $a\in A$. Estimates for $\mu$, $v$ and $u(\xi)$ can be computed
by Monte Carlo integration using the ergodic average of quantities
in Gibbs sample, for instance: 
\begin{align}
\hat{u}(\xi) & =\frac{1}{T}\sum_{t=1}^{T}(\sum_{a\in A}c_{a}^{(t)}\psi_{a}(\xi))=\sum_{a\in A}\hat{c}_{a}\psi_{a}(\xi);\label{eq:gpc_est_bma}\\
\hat{P_{a}} & =\frac{1}{T}\sum_{t=1}^{T}\mathbbm{1}(\gamma_{a});\nonumber \\
\hat{\mu} & =\frac{1}{T}\sum_{t=1}^{T}c{}_{0}^{(t)}=\hat{c}_{0}, & \hat{v} & =\frac{1}{T}\sum_{t=1}^{T}\sum_{\alpha\in A-\{0\}}(c{}_{\alpha}^{(t)})^{2}Z_{\alpha};\nonumber \\
\hat{\sigma}^{2} & =\frac{1}{T}\sum_{t=1}^{T}\sigma^{2,(t)}, & \hat{\lambda} & =\frac{1}{T}\sum_{t=1}^{T}\lambda^{(t)};\nonumber \\
\hat{c}_{a} & =\frac{1}{T}\sum_{t=1}^{T}c_{a}^{(t)}, & \hat{\rho} & =\frac{1}{T}\sum_{t=1}^{T}\rho^{(t)}.\nonumber 
\end{align}

\paragraph{Median probability model based evaluation:}

A parsimonious (or sparse) surrogate model involving only significant
basis functions and important input model parameters can be obtained
by examining the estimated inclusion probabilities $\pi(\gamma_{a}|\mathcal{D})$.
A suitable probabilistic basis selection mechanism is the median probability
model (MPM) \citep{BarbieriBerger2004}.

Given a Gibbs sample $\left\{ \left(\gamma^{(t)},c^{(t)},\sigma^{2,(t)},\lambda^{(t)},\rho^{(t)}\right)\right\} _{t=1}^{T}$
drawn from Algorithm \ref{alg:MCMCalg}, the marginal inclusion posterior
probabilities, namely the posterior distribution that the PC basis
is significant, can be estimated as $\hat{P}_{a}=\hat{\pi}(\gamma_{a}|\mathcal{D})=\frac{1}{T}\sum_{t=1}^{T}\gamma_{a}^{(t)}$.
According to the MPM rule, the inclusion parameters are estimated
as $\hat{\gamma}_{a}^{\text{(MPM)}}=\mathbbm{1}(\hat{P}_{a}\in(0.5.1))$,
for $a\in A$. Significant PC coefficients $\{c_{a}\}$, and PC bases
$\{\psi_{a}(\cdot)\}$ are those whose marginal inclusion probabilities
are such that $\hat{P}_{a}>0.5$ (and hence $\hat{\gamma}_{a}^{\text{(MPM)}}=1$).

After the selection of the significant PC bases according to the aforementioned
MPM rule, inference about the unknown quantities of interest can be
performed by re-running the Gibbs sampler Algorithm \ref{alg:MCMCalg}
for fixed $\gamma$ equal to $\hat{\gamma}^{\text{(MPM)}}$. The new
Gibbs sample $\left\{ \left(c'^{(t)},\sigma'{}^{2,(t)},...\right)\right\} _{t=1}^{T}$
can be used to perform inference and estimation. For instance, estimates
for $\mu$, $v$ and $u(\xi)$ can be computed by Monte Carlo integration
and using the equations of the estimators in (\ref{eq:gpc_est_bma}).
Note that a number of the coefficients $\{c'{}_{a}^{(t)}\}$ will
be constantly zero for all $t=1,...,T$. The reason is because, unlike
in BMA approach, here we consider only a single subset of PC bases,
and hence the inclusion parameter is a fixed parameter equal to $\gamma^{(t)}=\gamma^{(\text{MPM})}$,
for $t=1,...,T$.

The MPM approach allows the selection of the important input model
parameters that significantly affect the output. If an input parameter
$\xi_{j}$ is not represented by any significant PC basis in the gPC
expansion, it would be reasonable to consider that input parameter
$\xi_{j}$ does not significantly affect the output model and hence
be omitted from the analysis.

\section{Analysis of the US-ARM data-set\label{sec:Application}}

Here we consider the US-ARM data set. The main interest lies on computing
the posterior distributions of the $10$ (random) input parameters
of CLM4 \\
 $\xi=(\text{Fmax, Cs, Fover, Fdrai, Qdm, Sy, B, Psis, Ks, \ensuremath{\theta}s})$,
given an observed latent heat flux (LH) measurement $u^{(\text{f})}$
is in Table \ref{tab:Observed-vale}. 
\begin{table}
\center %
\begin{tabular}{|c|c|c|c|c|}
\hline 
Months  & \multicolumn{4}{c|}{Measurement}\tabularnewline
\hline 
\hline 
Jan-Apr  & $15.542$  & $22.017$  & $41.365$  & $59.095$\tabularnewline
\hline 
May-Jul  & $58.377$  & $58.813$  & $45.107$  & $41.362$\tabularnewline
\hline 
Aug-Dec  & $31.250$  & $28.645$  & $17.635$  & $12.778$\tabularnewline
\hline 
\end{tabular}

\caption{Observed value $u^{(\text{f})}$\label{tab:Observed-vale}, for the
US-ARM data-set}
\end{table}

We apply the above methodology which involves two stages: (i) build
a surrogate model to replace the accurate but expensive forward model
CLM4 according to the methodology in Section \ref{sec:Methodology_surrogate},
and (ii) conduct inversion of the $10$ inputs, according to the theory
in Section \ref{sec:Methodology_Inverse}.

\paragraph*{Surrogate model building step~~~~}

For each month, we build a surrogate model that maps the input of
CLM4 $\xi$ to the output LH $u$. For this type of dataset, \citet{hou2012sensitivity}
and \citet{SunHouHuangTianRuby2013} observed that the dependency
between the output parameters LH corresponding to different months
is weak and hence can be neglected. Therefore, here we build surrogates
models for each month independently. An advantage of assuming independence
is that it leads to a simpler parameterization for the statistical
model, which is easier to treat and interpret.

To build the gPC expansion, we follow the procedure in Section \ref{sec:Methodology_surrogate}.
For the design of the gPC expansion, we consider PC bases from the
Jacobi polynomial family. The parameters of the Jacobi PC bases are
set according to the prior information of the input of CLM4 in \citep[Table 1]{hou2012sensitivity}
by matching the moments of the corresponding shifted Beta distribution
to which they are orthogonal. We consider the prior model (\ref{eq:prior_model})
with hyper-parameters $a_{\lambda}=10^{-3}$, $b_{\lambda}=10^{-3}$,
$a_{\sigma}=10^{-3}$, $b_{\sigma}=10^{-3}$, $a_{\rho}=1$, and $b_{\rho}=1$.
This choice of hyper-parameters leads to weakly informative priors.
This is a reasonable choice because there is lack of prior information
about the parameterization of the surrogate model. For training the
gPC expansion, we used the training data set US-ARM. We run Algorithm
\ref{alg:MCMCalg} for $2\cdot10^{5}$ iterations where the first
$10^{5}$ were discarded as burn in.

\begin{figure}
\center \subfloat[January]{\includegraphics[scale=0.2]{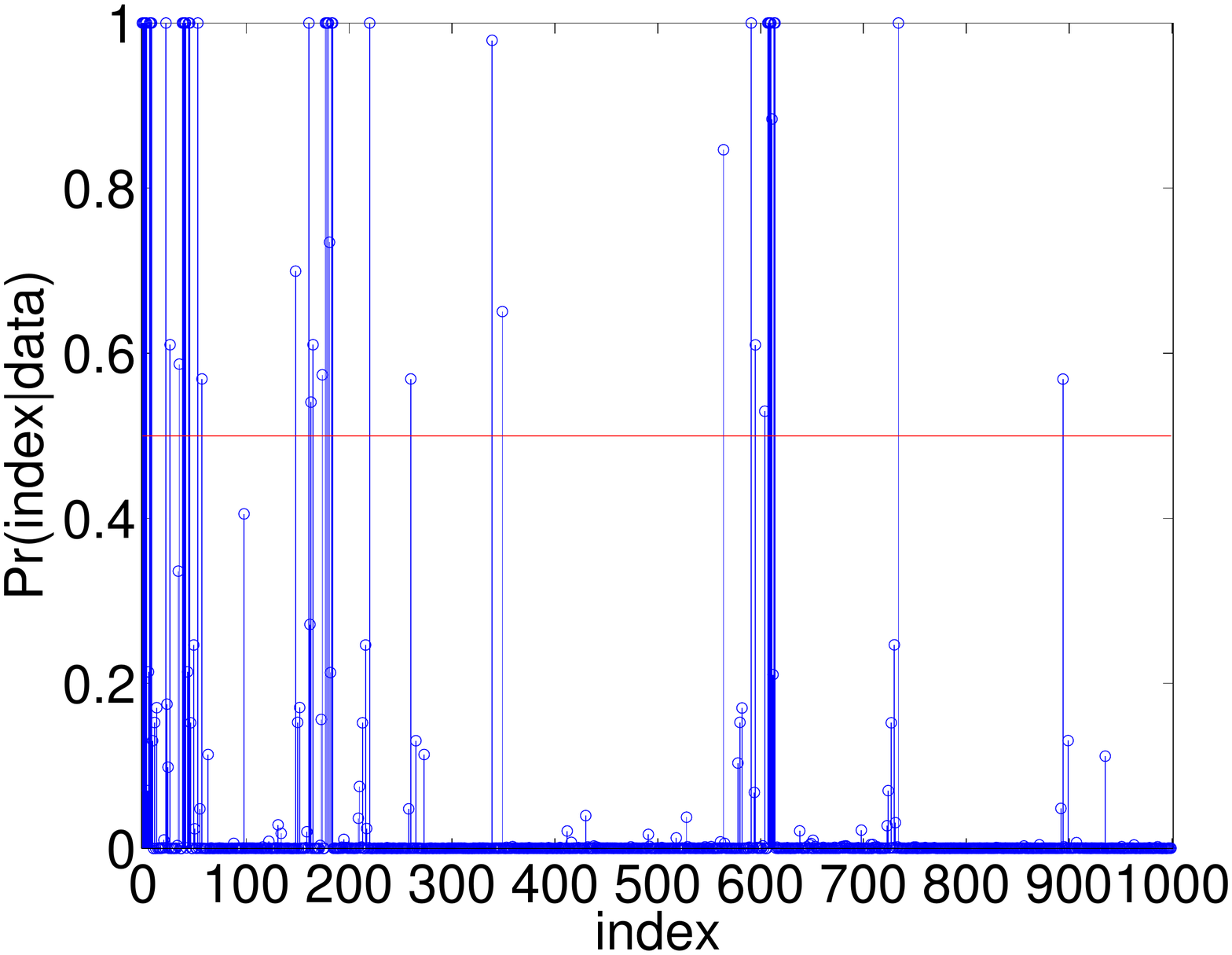}

}\subfloat[February]{\includegraphics[scale=0.2]{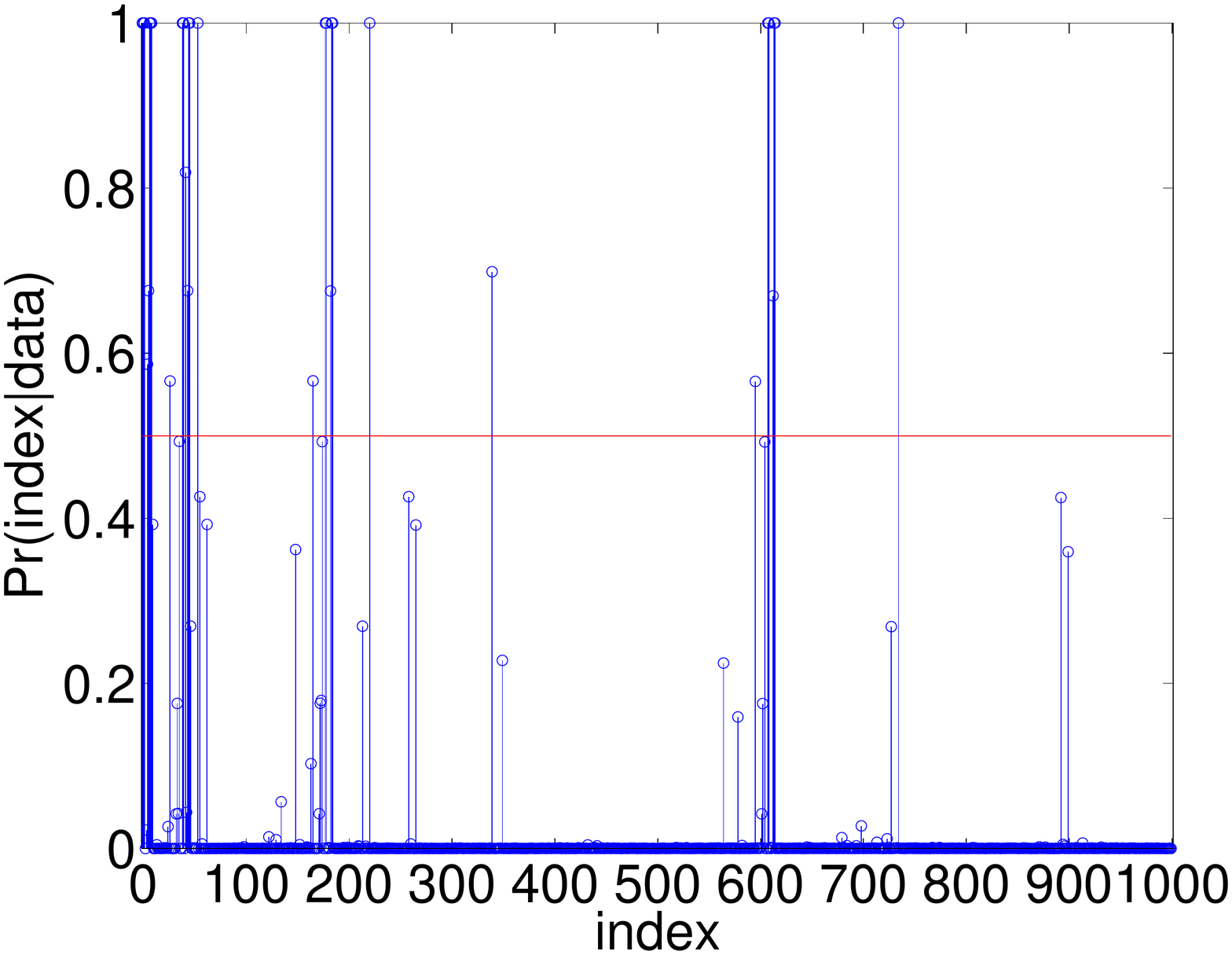}

}

\subfloat[March]{\includegraphics[scale=0.2]{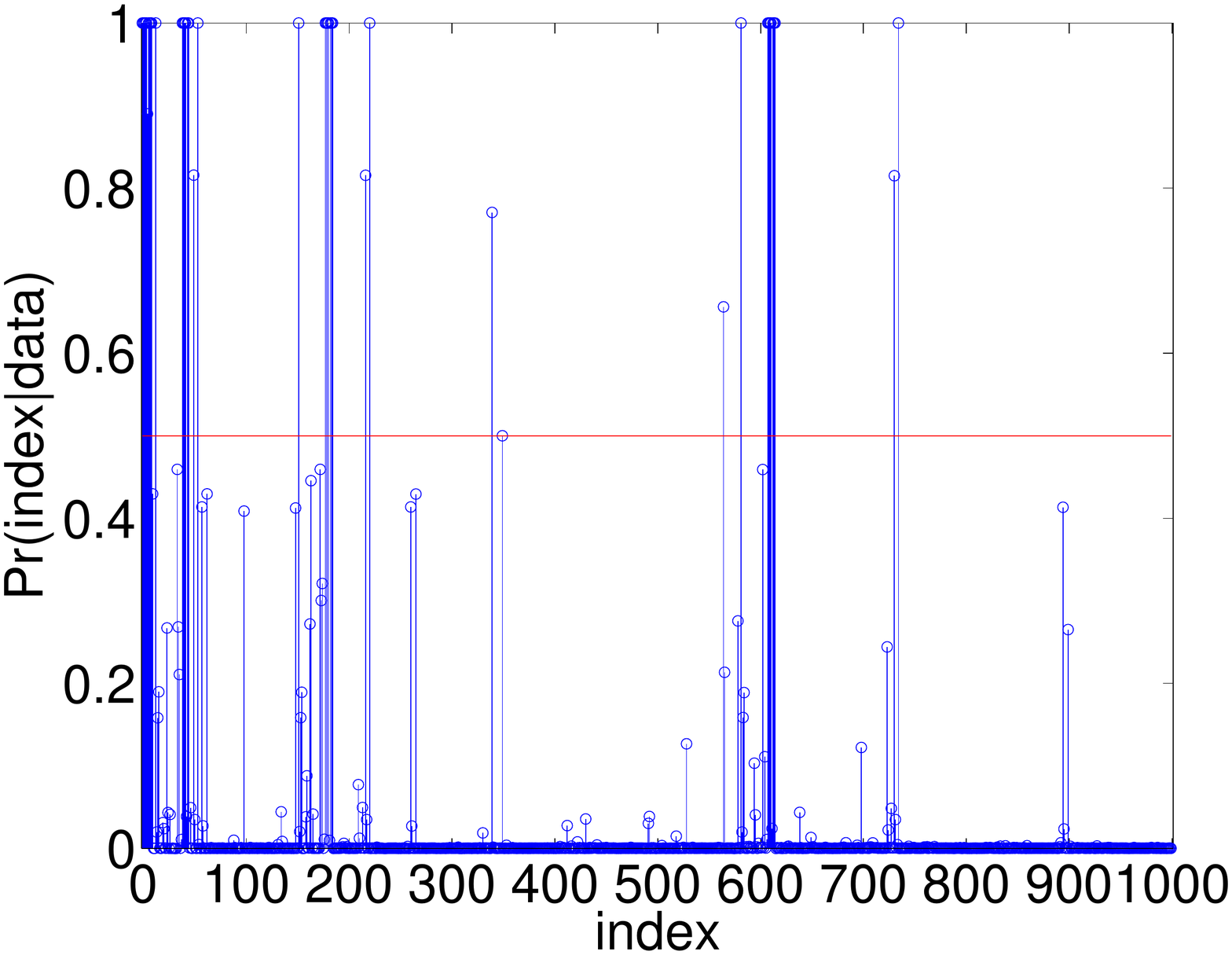}

}\subfloat[April]{\includegraphics[scale=0.2]{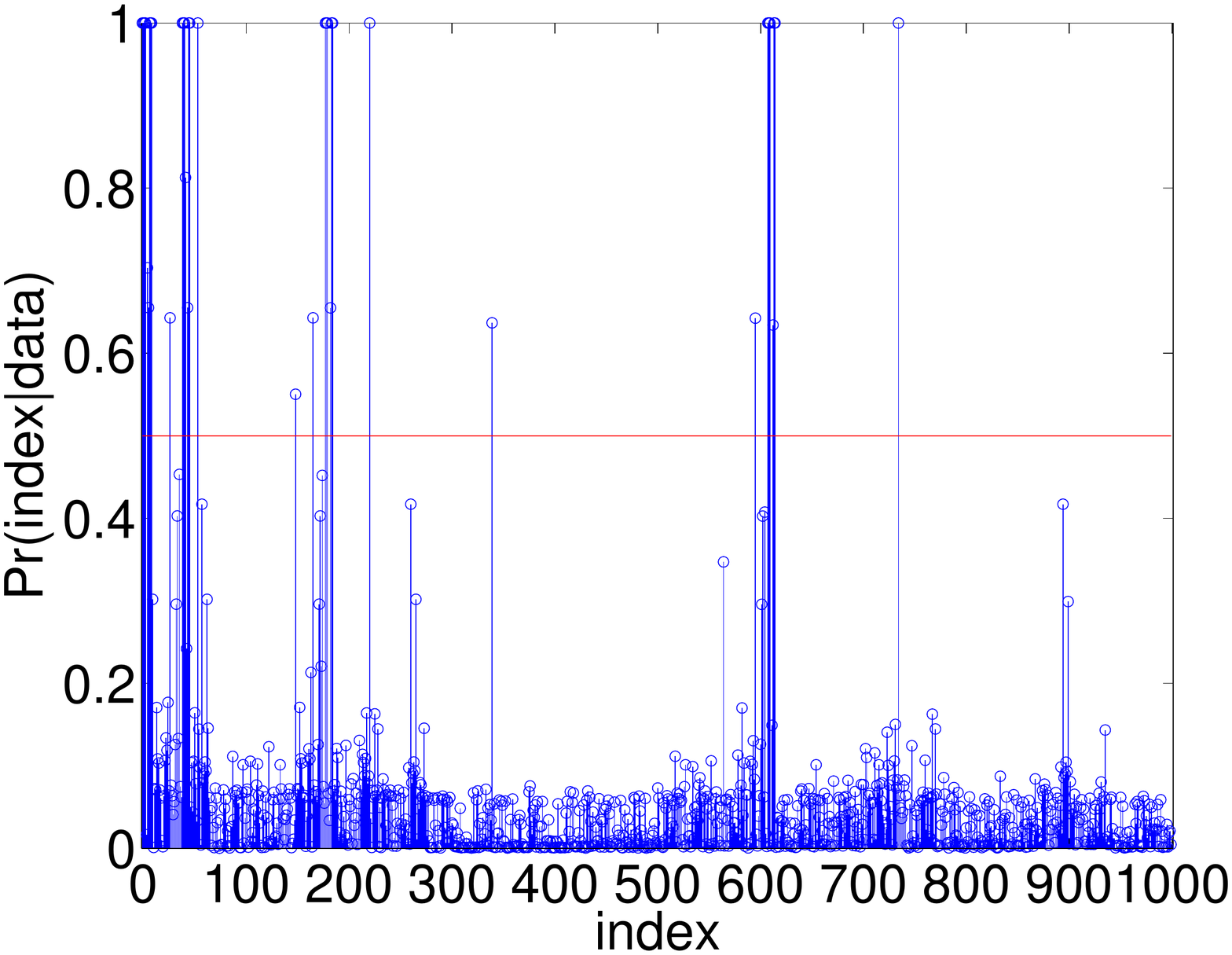}

}

\subfloat[May]{\includegraphics[scale=0.2]{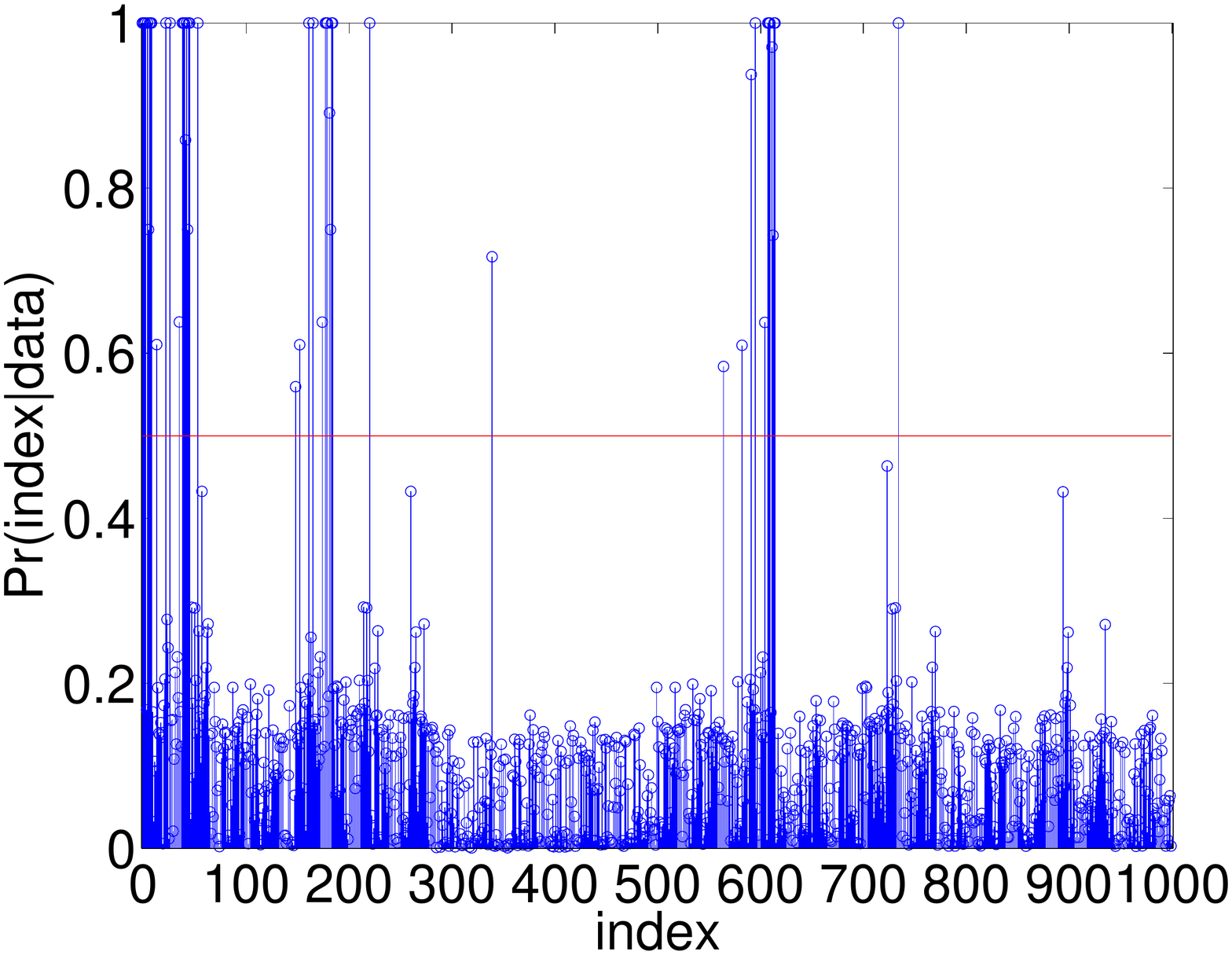}

}\subfloat[June]{\includegraphics[scale=0.2]{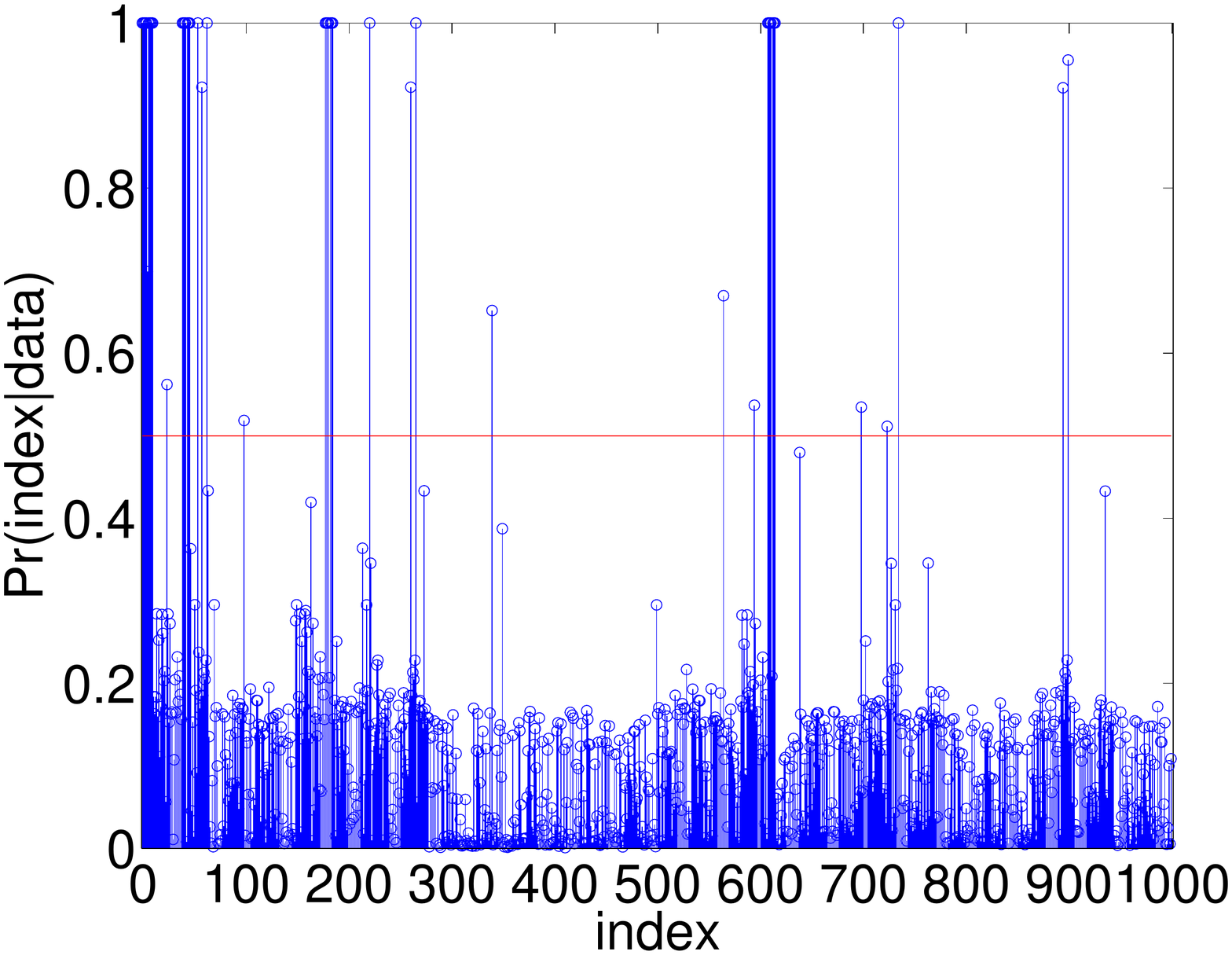}

}

\protect\caption{Plots of the posterior marginal inclusion probabilities; January -
June \label{fig:Pr_gamma_A}}
\end{figure}

\begin{figure}
\center \subfloat[July]{\includegraphics[scale=0.2]{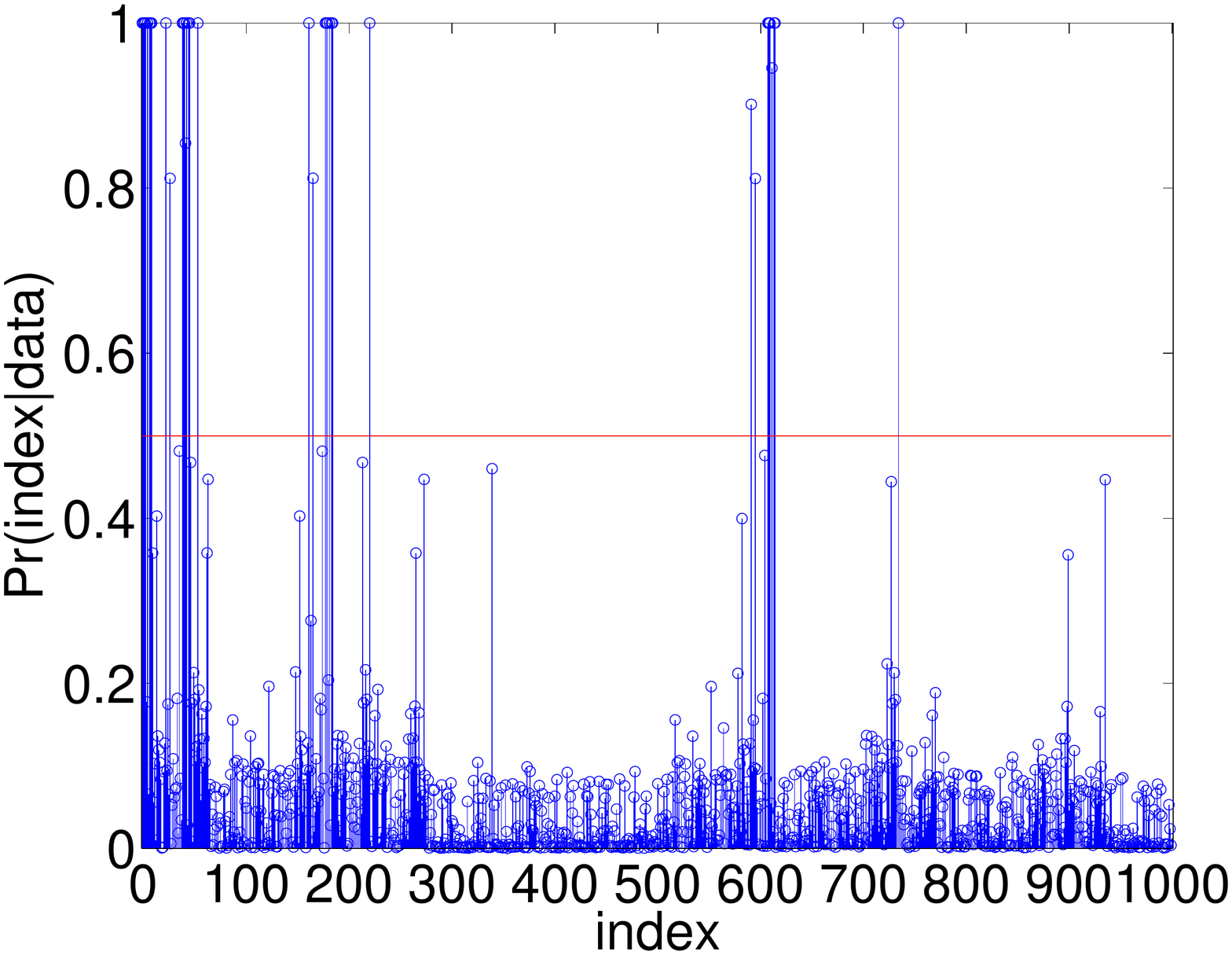}

}\subfloat[August]{\includegraphics[scale=0.2]{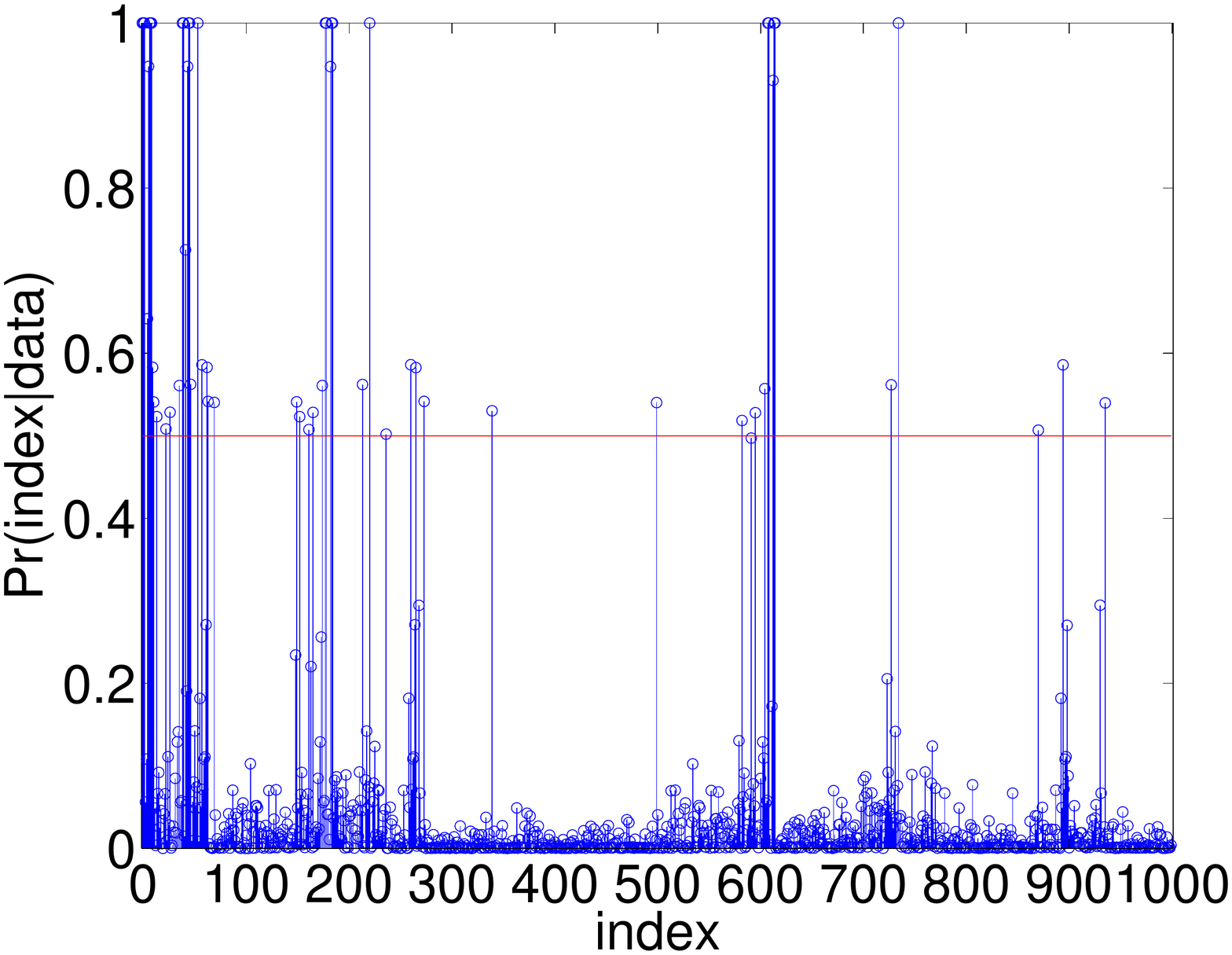}

}

\subfloat[September]{\includegraphics[scale=0.2]{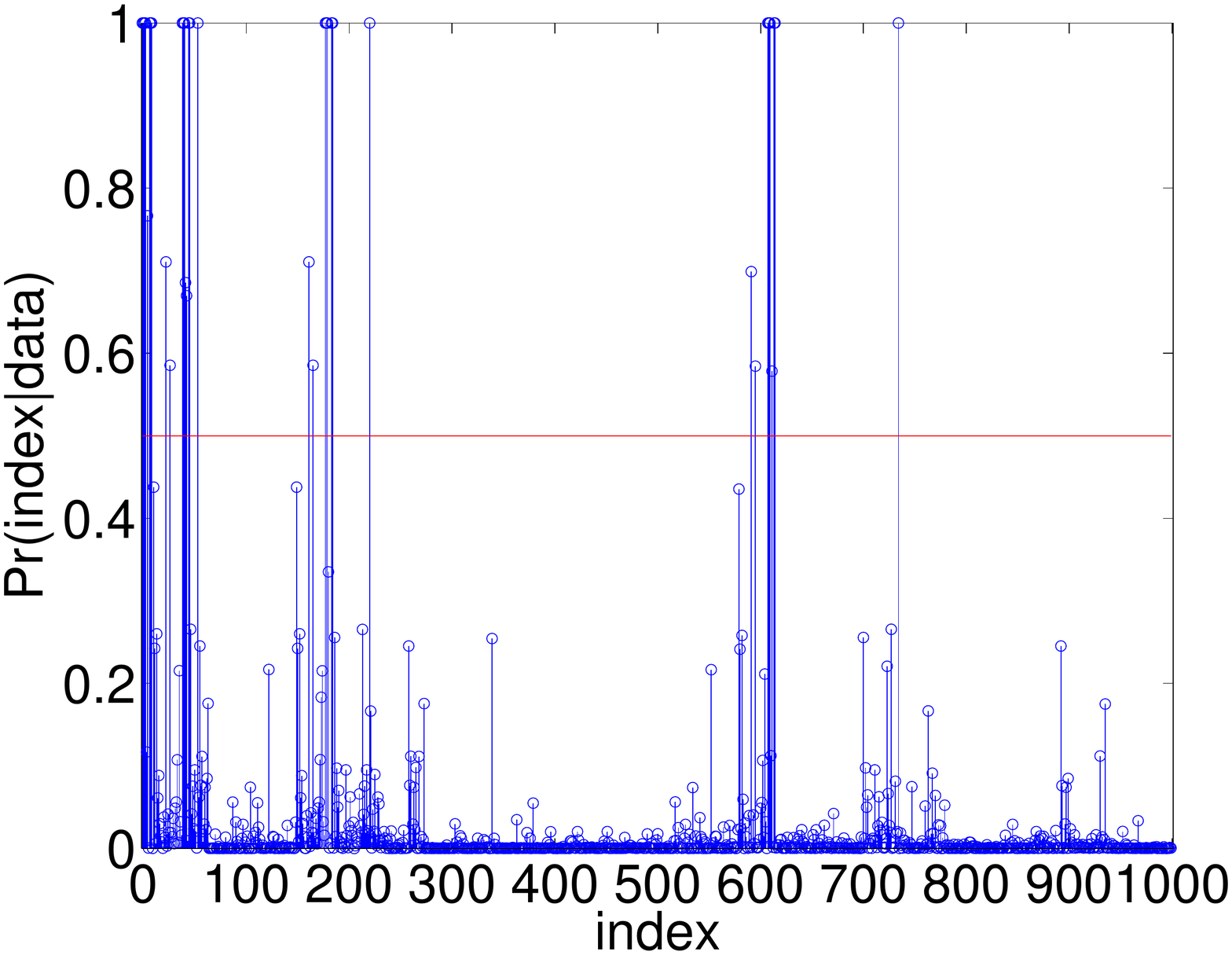}

}\subfloat[October]{\includegraphics[scale=0.2]{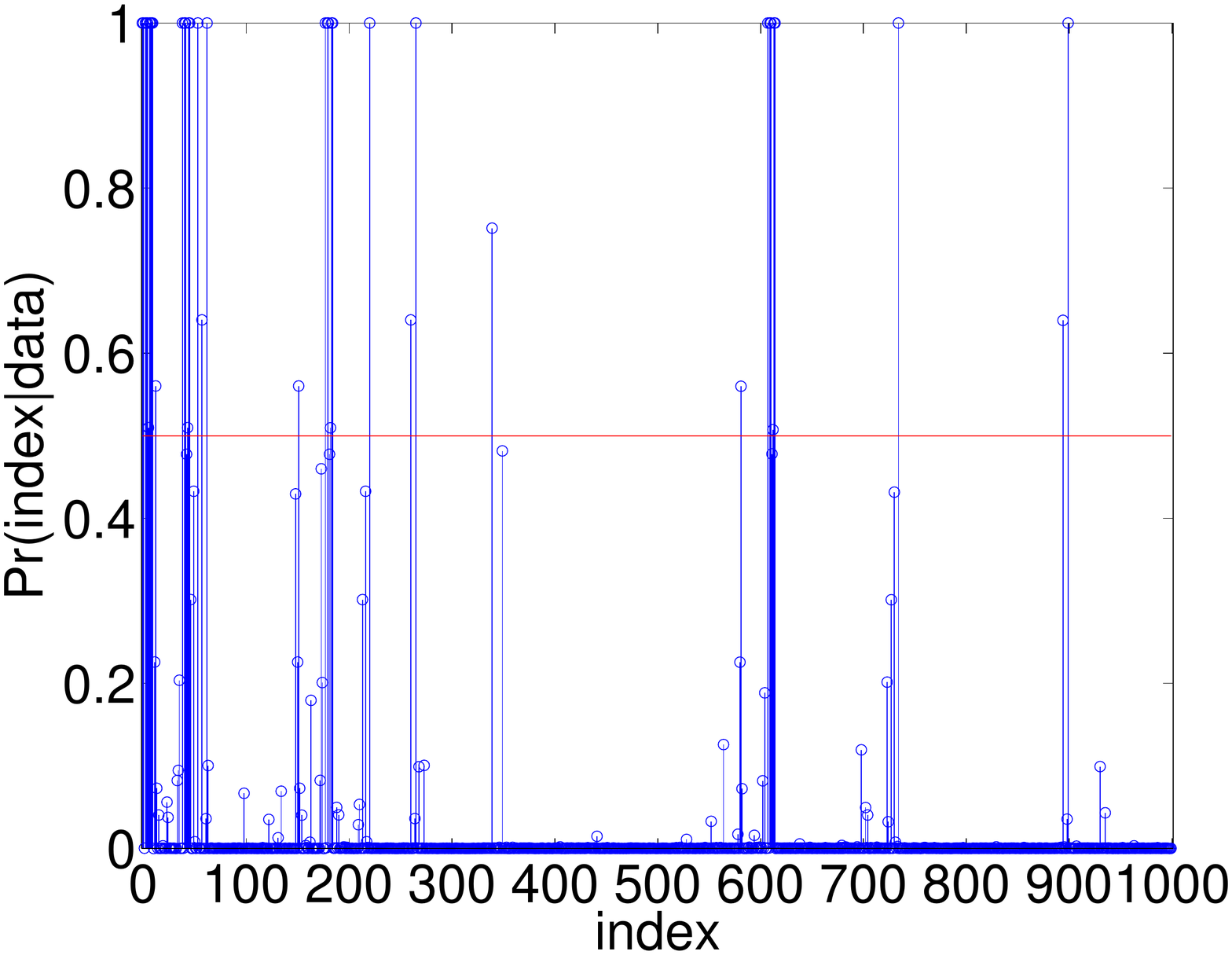}

}

\subfloat[November]{\includegraphics[scale=0.2]{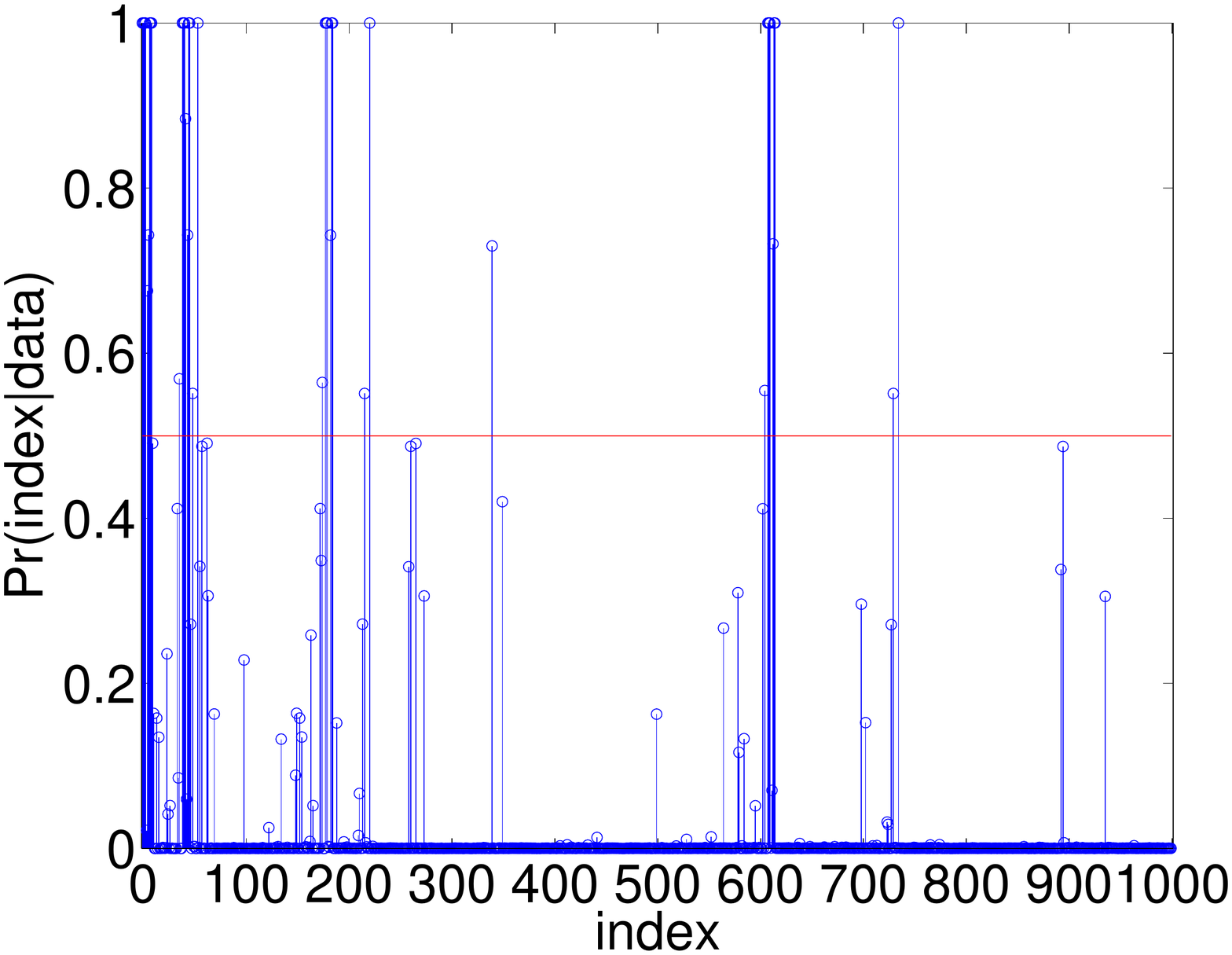}

}\subfloat[December]{\includegraphics[scale=0.2]{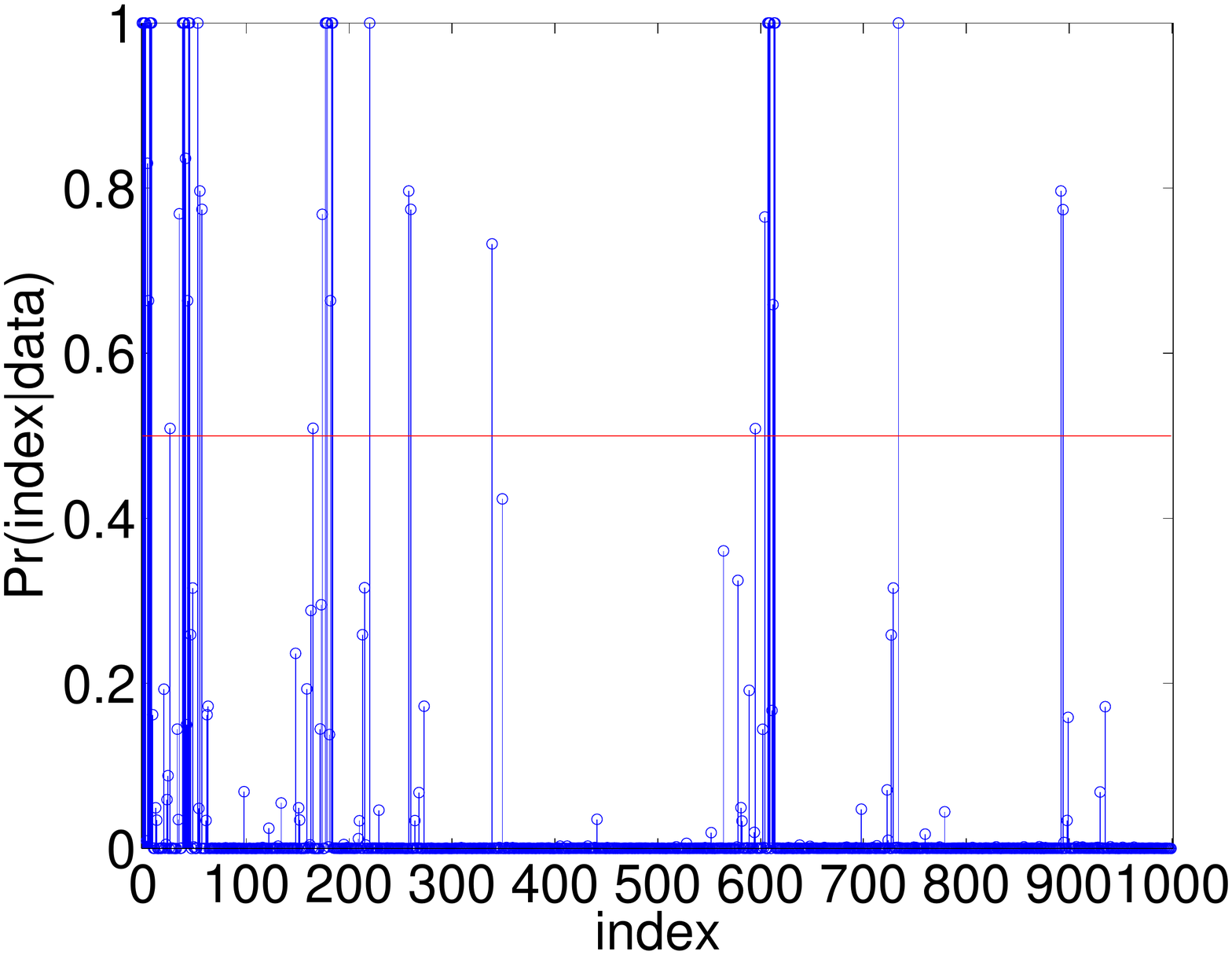}

}

\protect\caption{Plots of the posterior marginal inclusion probabilities; July - December
\label{fig:Pr_gamma_B}}
\end{figure}

In Figures \ref{fig:Pr_gamma_A}-\ref{fig:Pr_gamma_B}, we present
the marginal posterior probabilities $\{\Pr(\gamma_{j}|\mathcal{D})\}$
computed by the ergodic average of the occurrences of the corresponding
PC bases in the Gibbs sample. We observe that during the period May-August
the marginal inclusion probabilities are higher than those of the
rest months. This indicates that the input parameters of CLM4 may
have larger impact on the output LH during those months. From the
hydrology point of view, this is expected because LH is higher on
average and has larger variability during these months, and effects
of hydrological parameters are expected to be more pronounced in the
summer months.

We can infer that the input parameters Fdrai, Qdm, and B are significant
according to the MPM rule. This is because these input parameters
are represented by significant PC basis functions; namely the corresponding
marginal inclusion probabilities in Figures \ref{fig:Pr_gamma_A}-\ref{fig:Pr_gamma_B}
are greater than $0.5$. From the hydrology perspective this is reasonable,
because these parameters are major factors controlling the drainage
and runoff generation, which in turn impact heat fluxes. The results
are also consistent with the previous work in {[}\citet{,hou2012sensitivity}{]}.

\begin{figure}
\center \subfloat[January]{\includegraphics[scale=0.2]{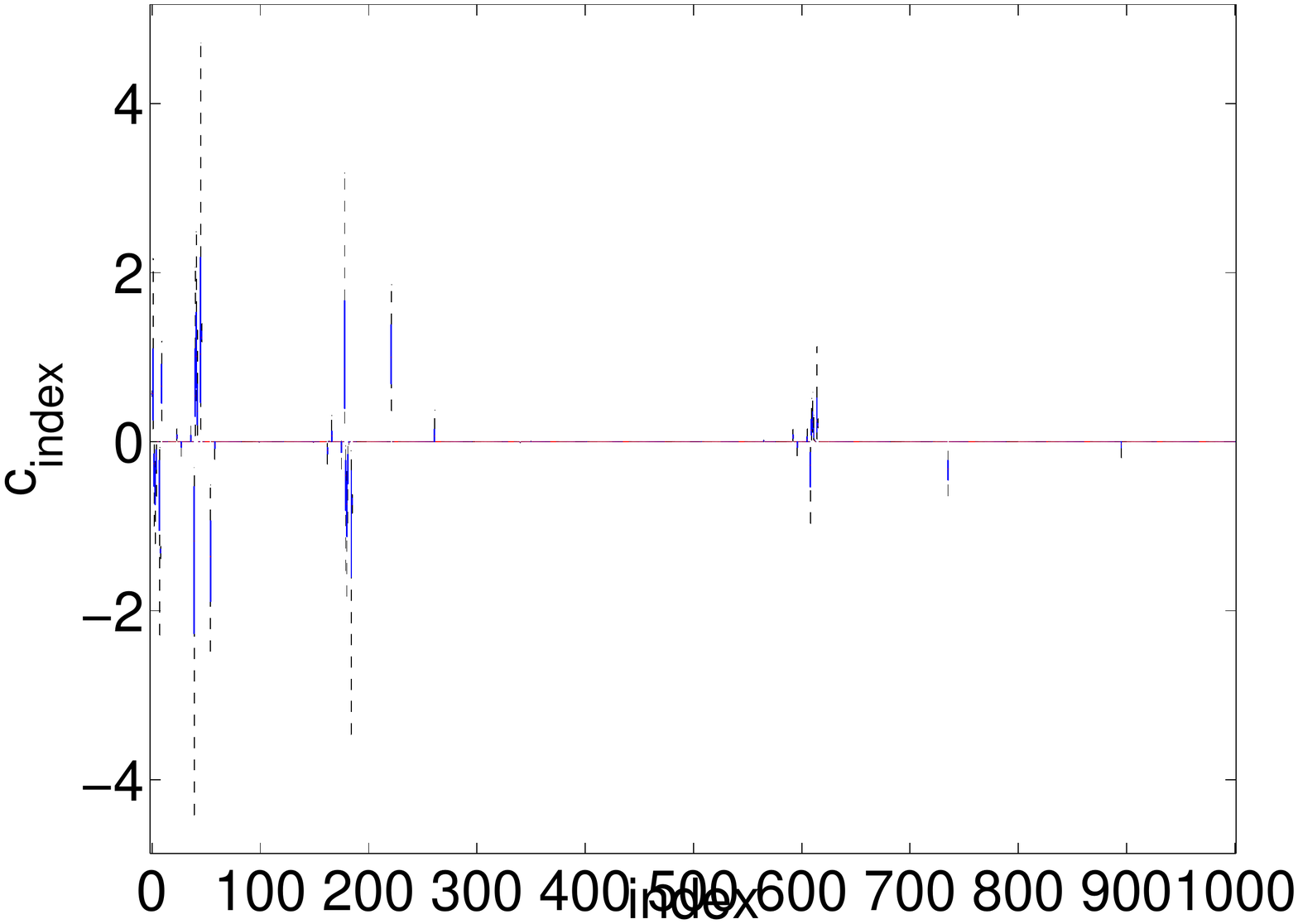}

}\subfloat[February]{\includegraphics[scale=0.2]{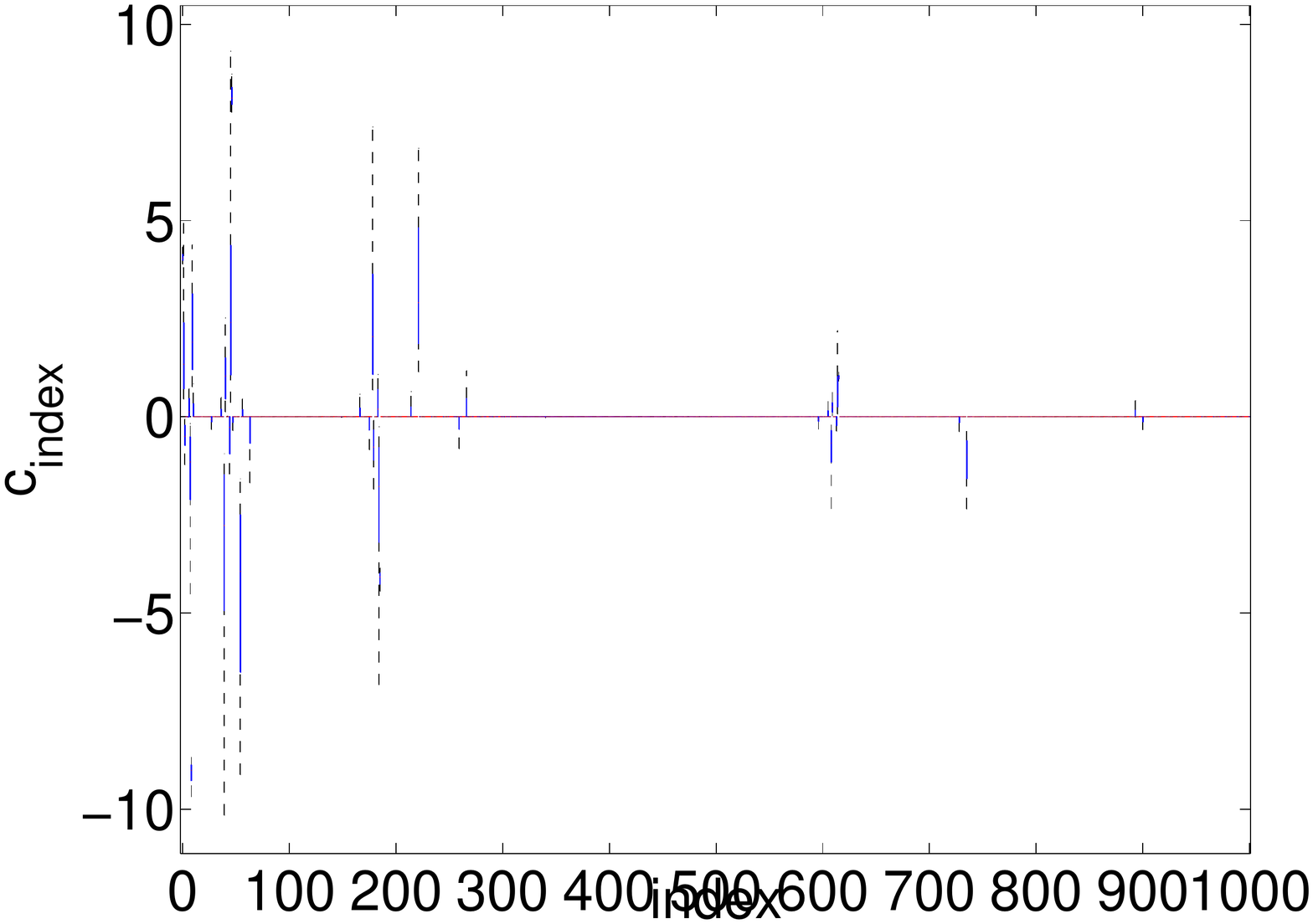}

}

\subfloat[March]{\includegraphics[scale=0.2]{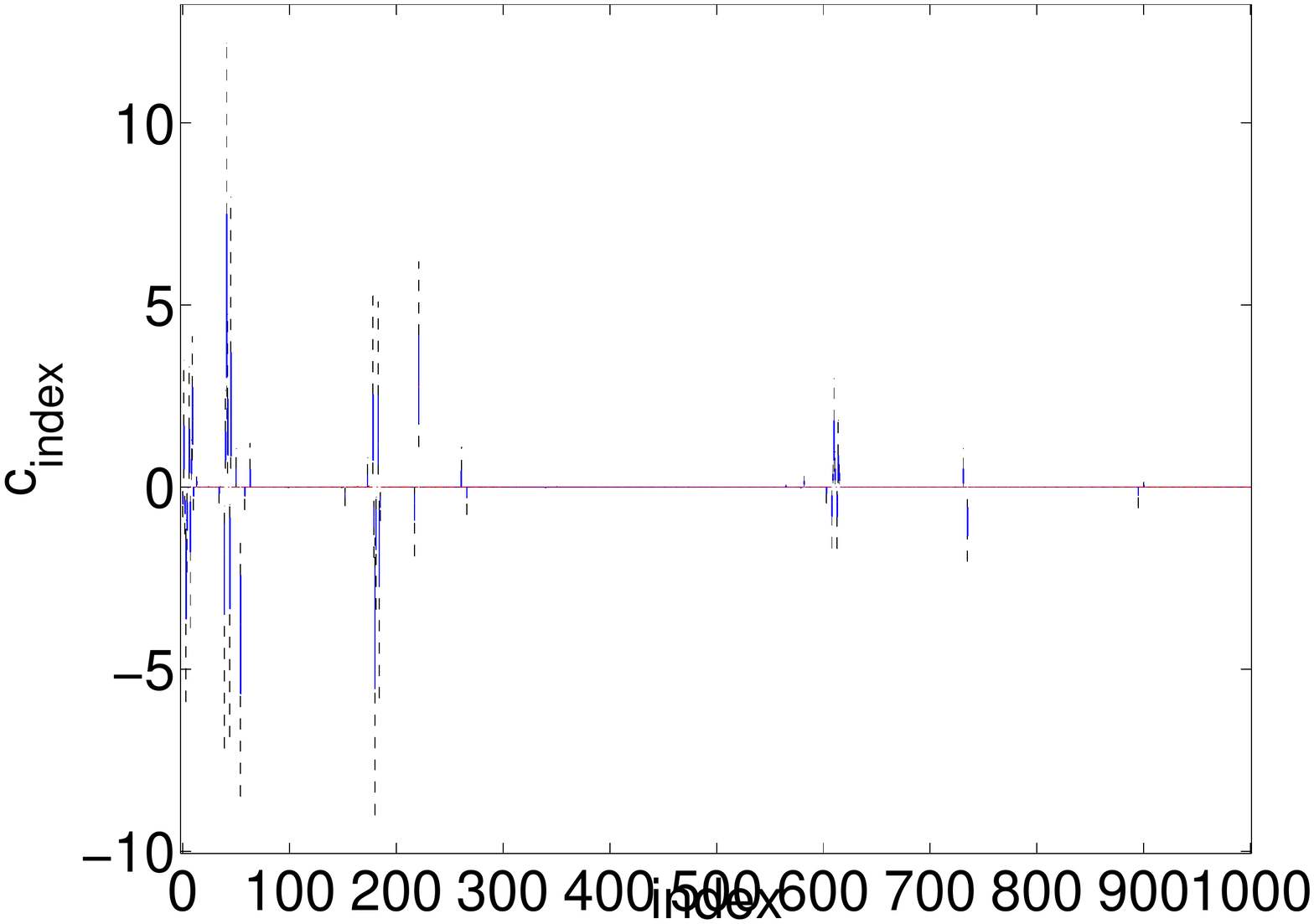}

}\subfloat[April]{\includegraphics[scale=0.2]{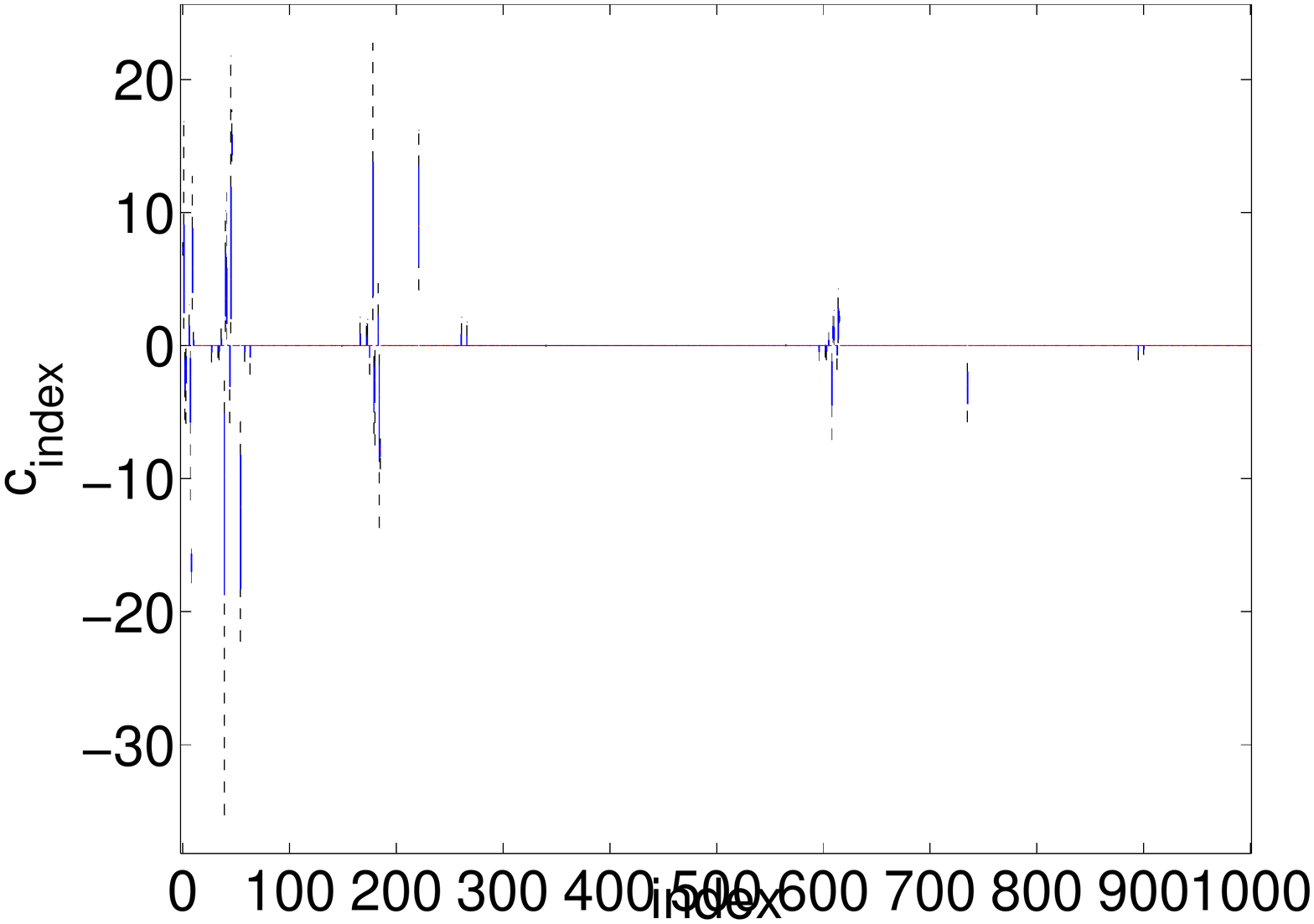}

}

\subfloat[May]{\includegraphics[scale=0.2]{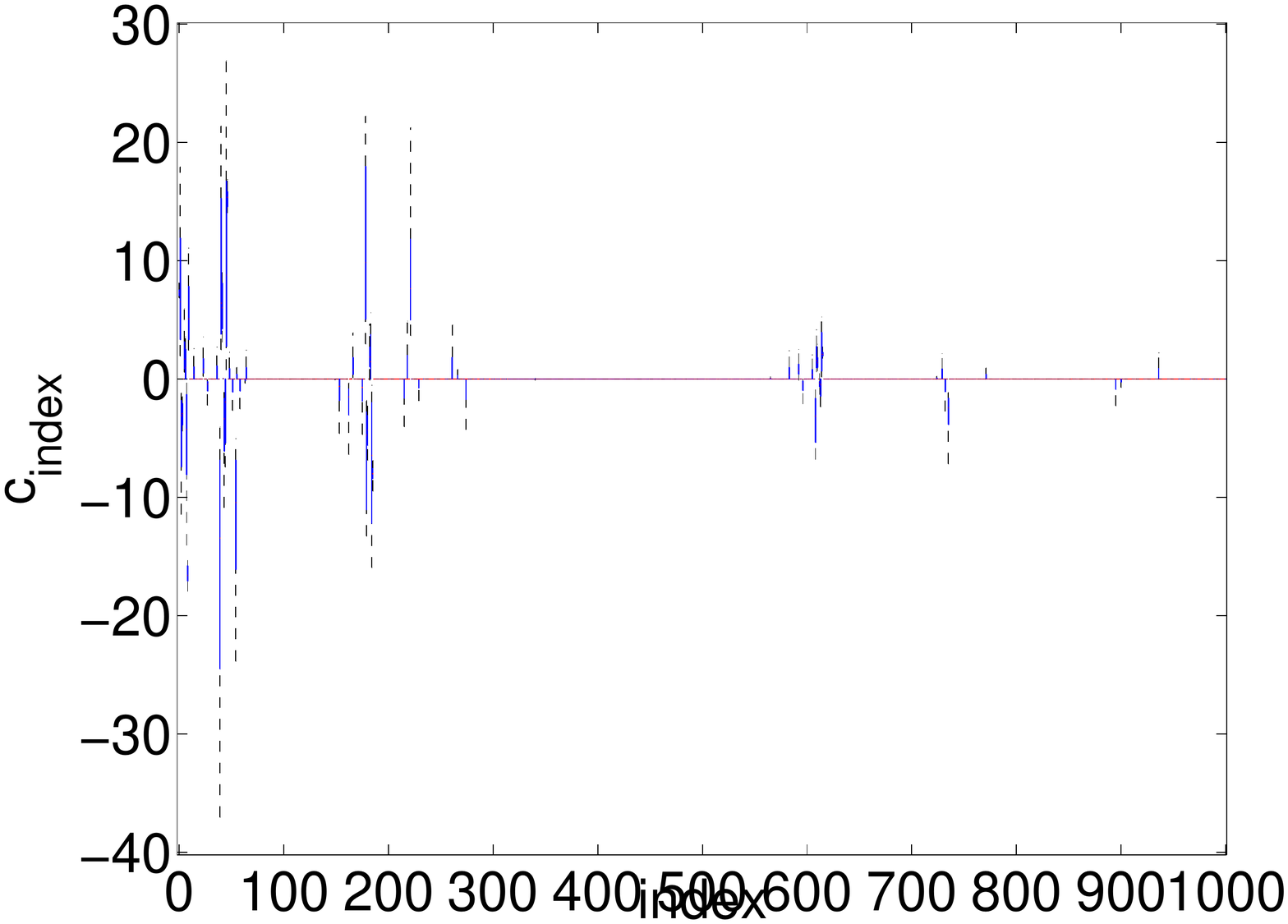}

}\subfloat[June]{\includegraphics[scale=0.2]{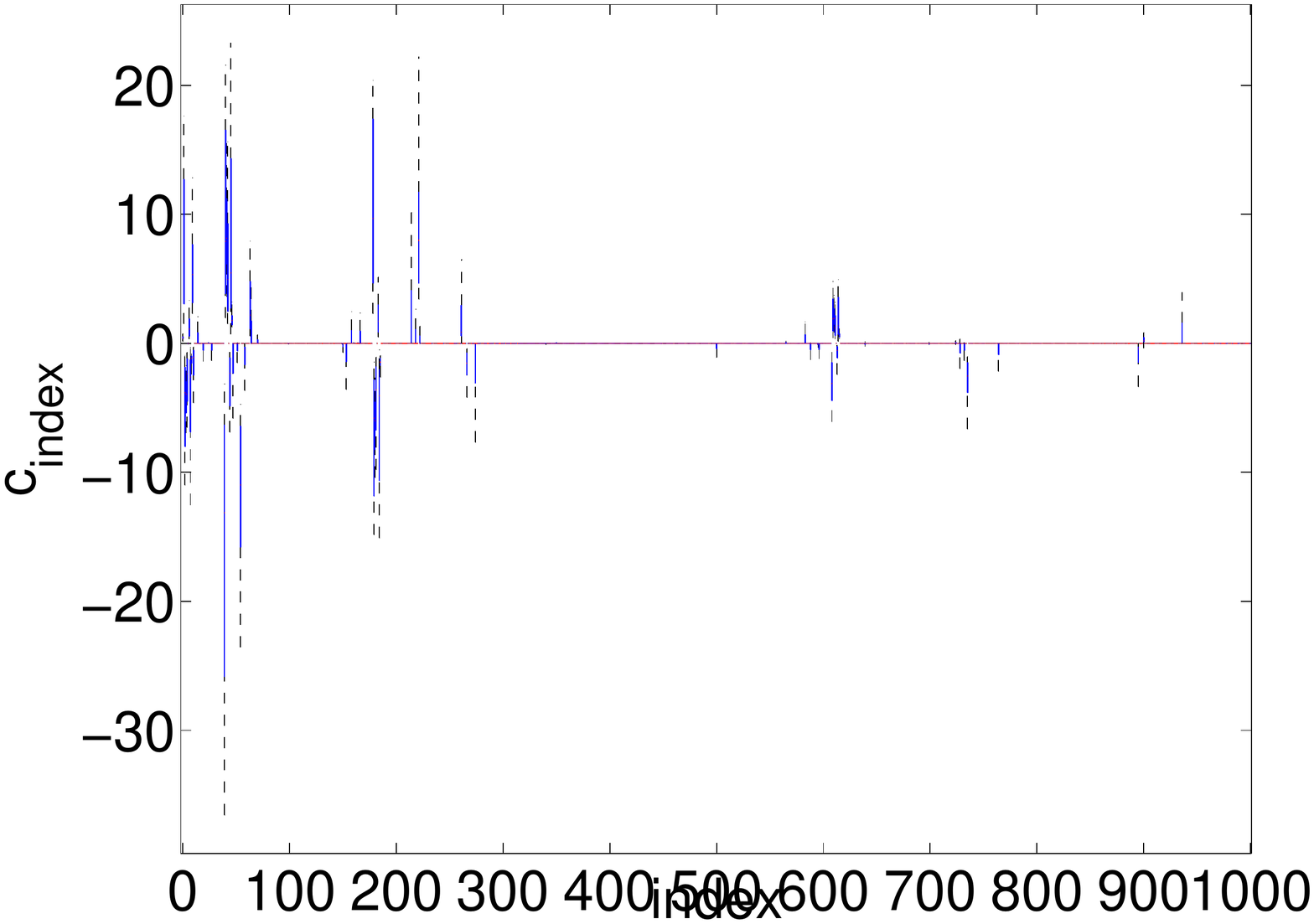}

}

\protect\caption{Boxplots of the posterior PC coefficients; January - June. \label{fig:BP_beta_A}}
\end{figure}

\begin{figure}
\center \subfloat[July]{\includegraphics[scale=0.2]{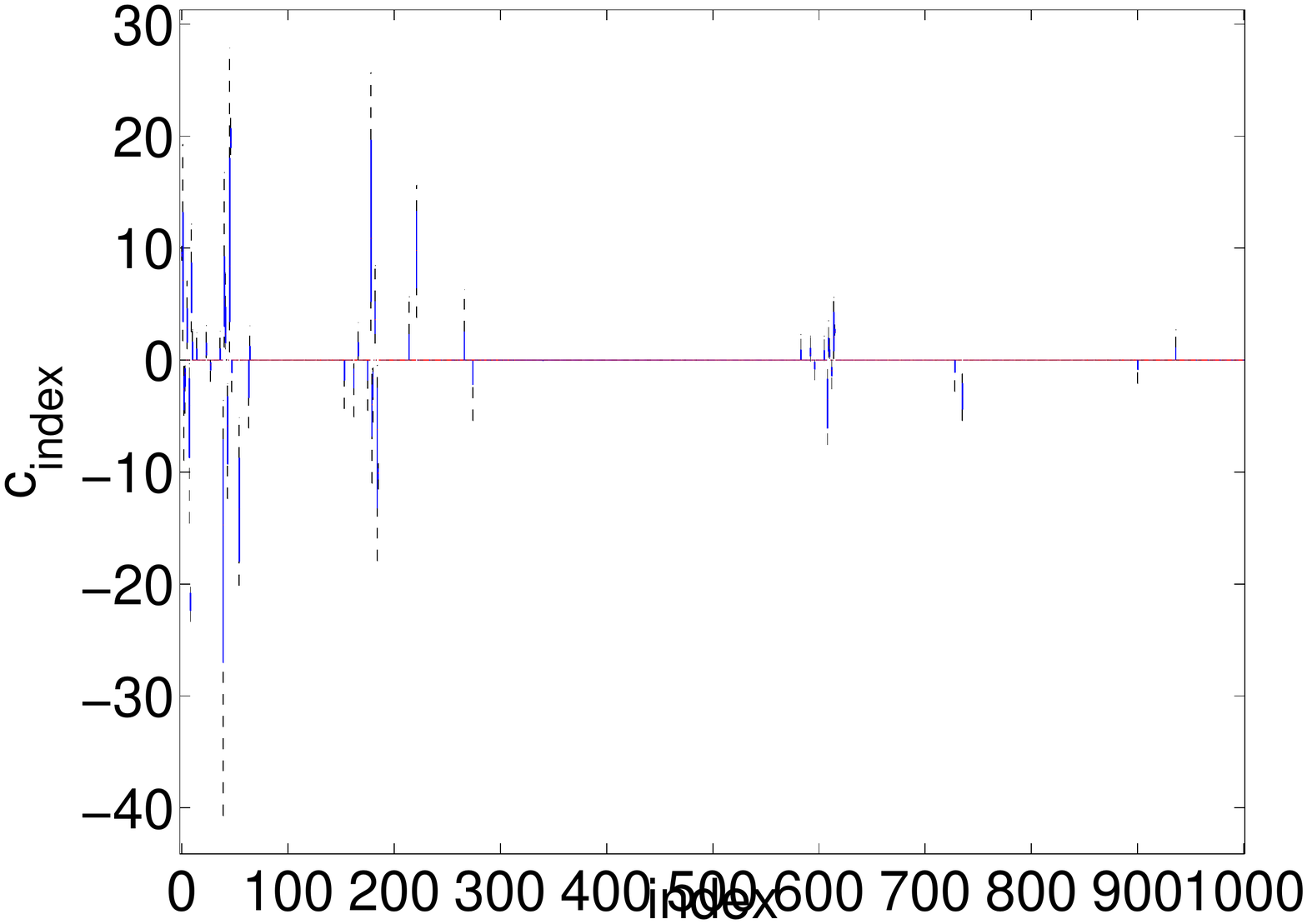}

}\subfloat[August]{\includegraphics[scale=0.2]{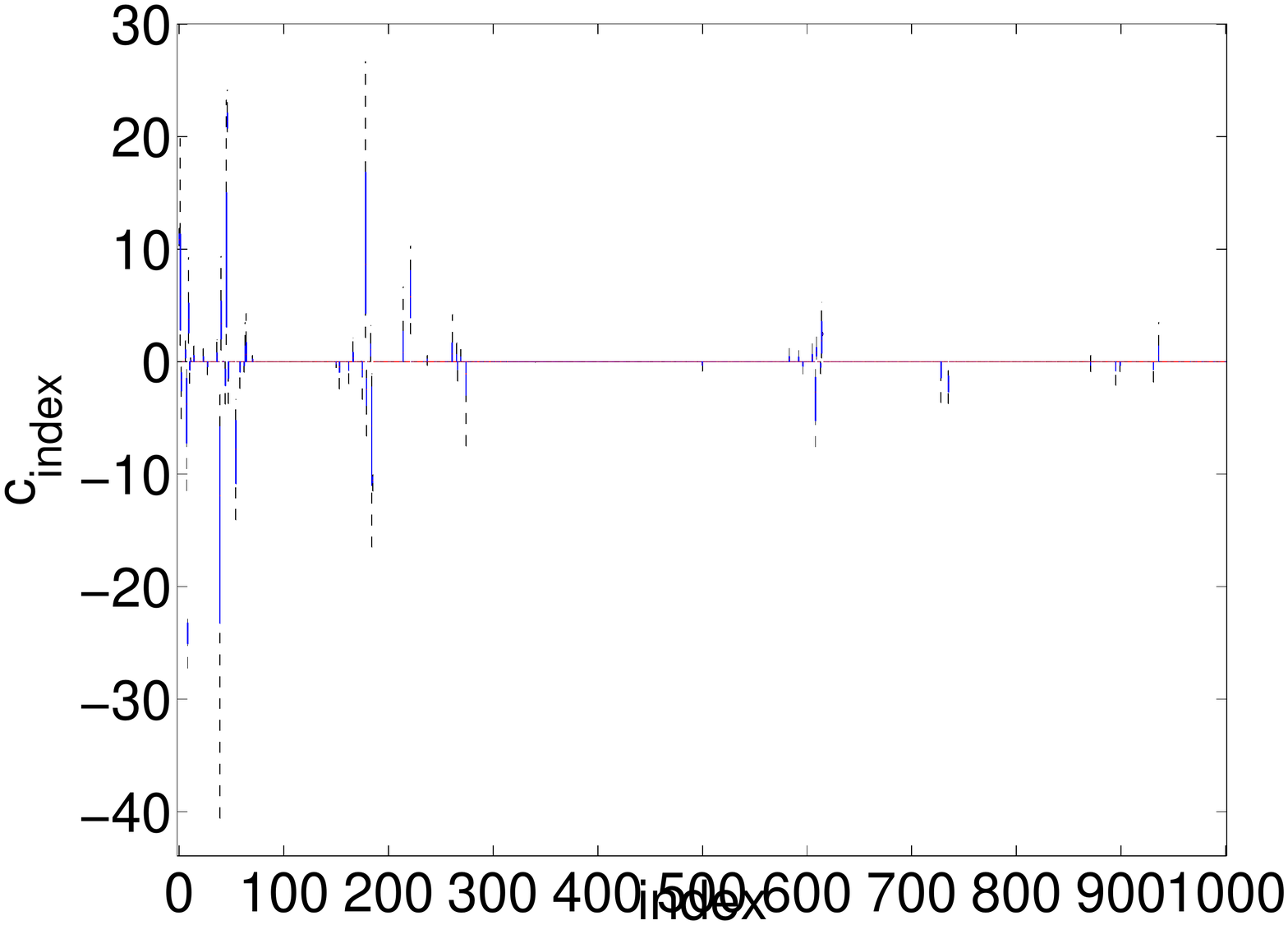}

}

\subfloat[September]{\includegraphics[scale=0.2]{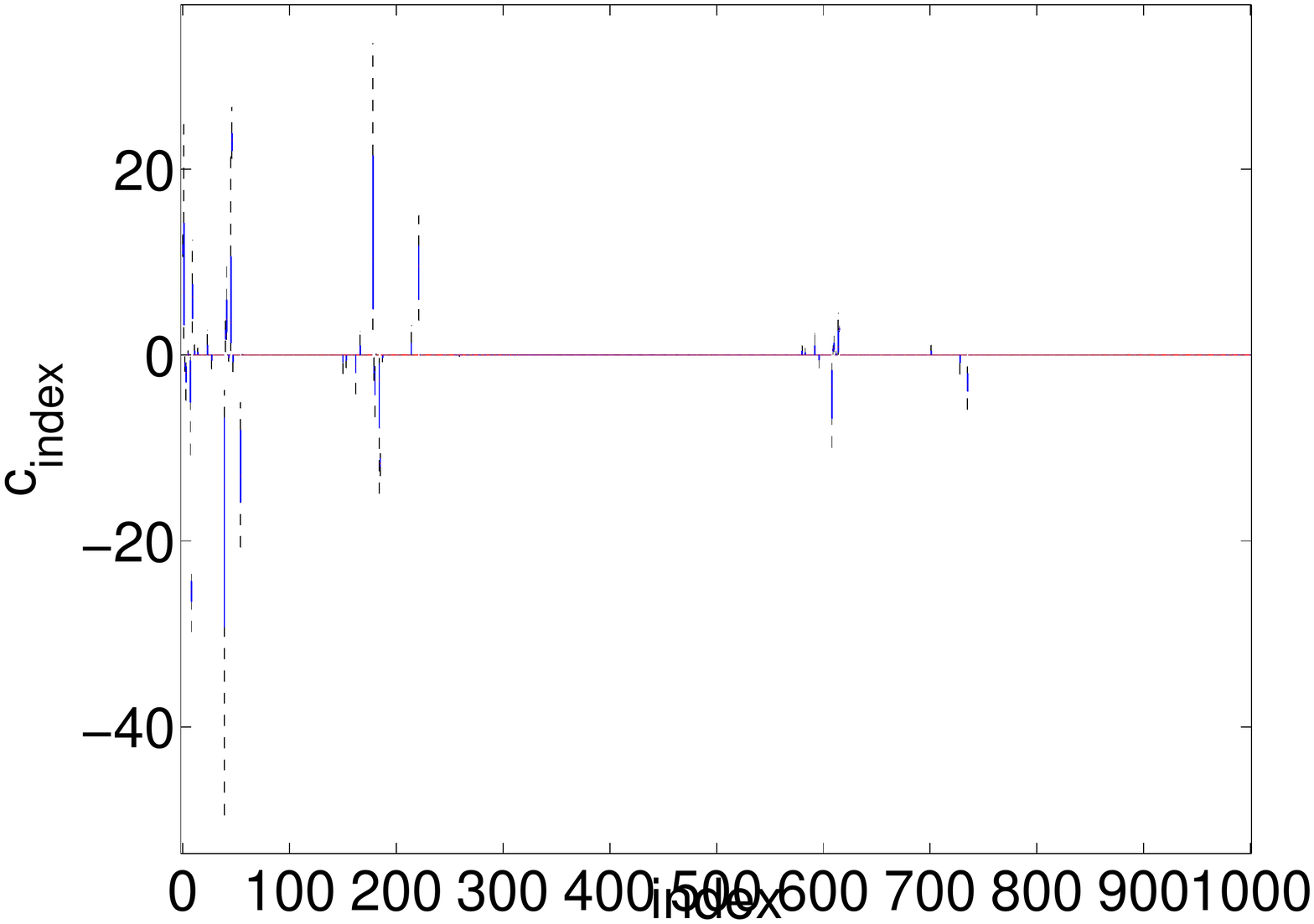}

}\subfloat[October]{\includegraphics[scale=0.2]{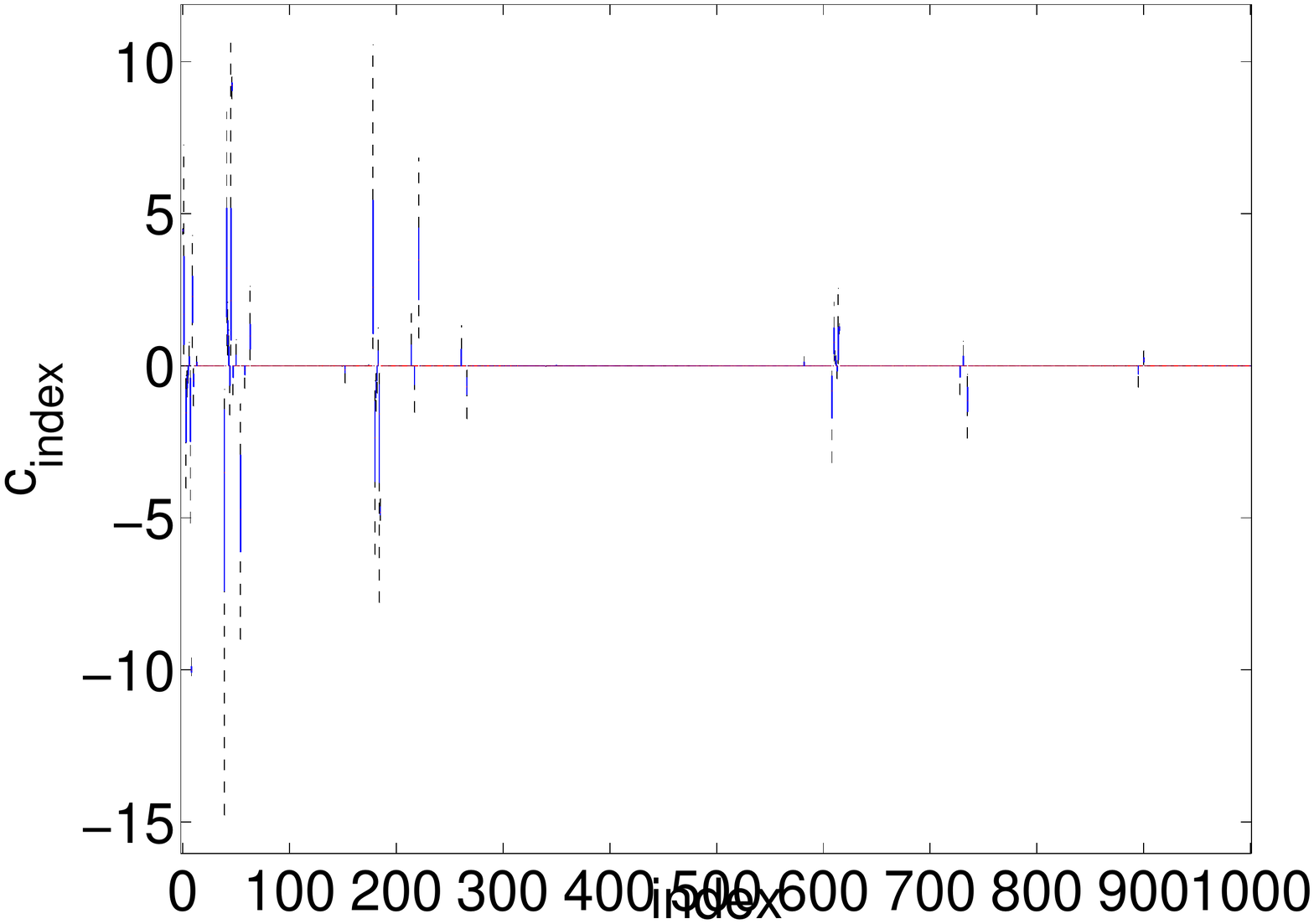}

}

\subfloat[November]{\includegraphics[scale=0.2]{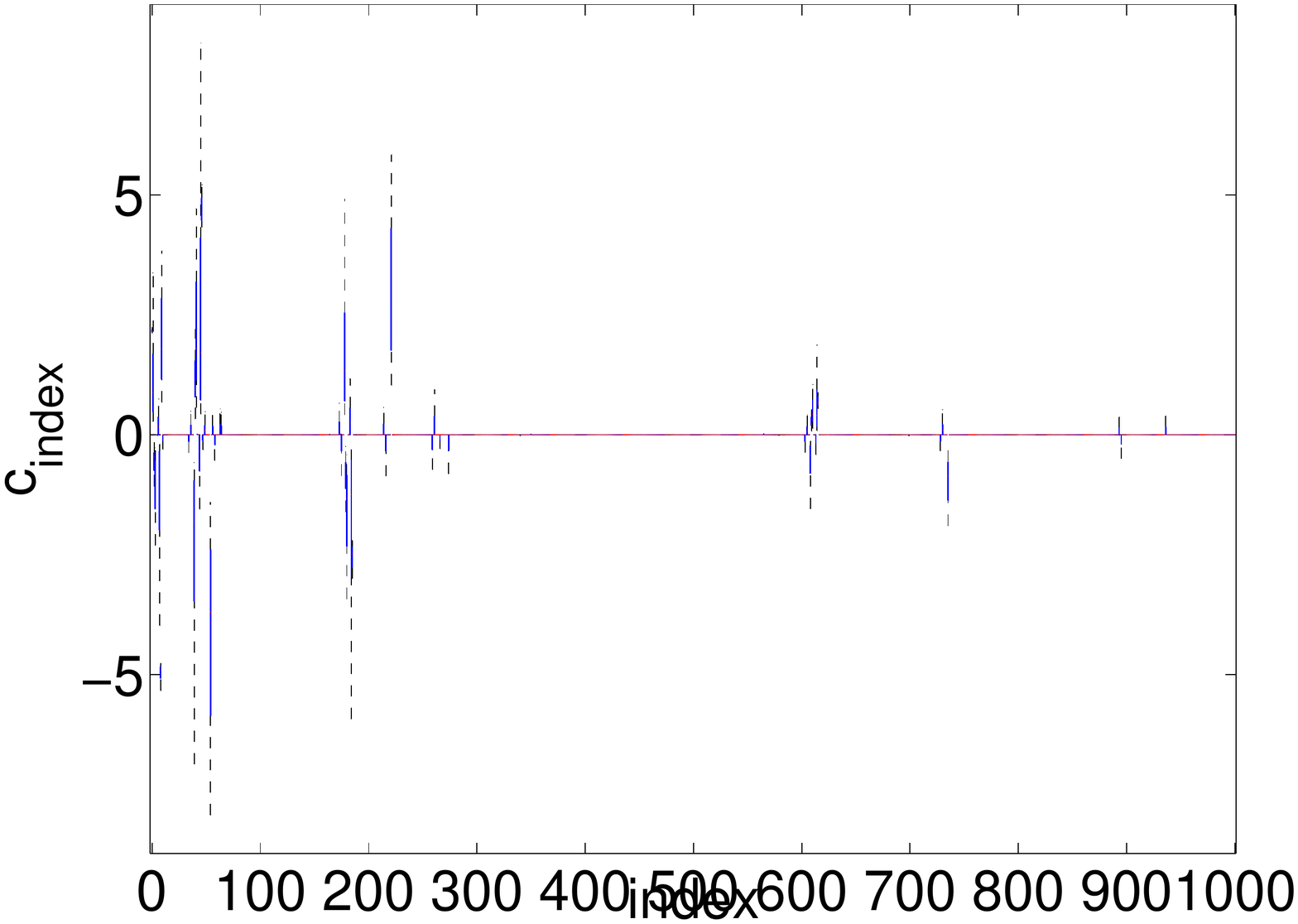}

}\subfloat[December]{\includegraphics[scale=0.2]{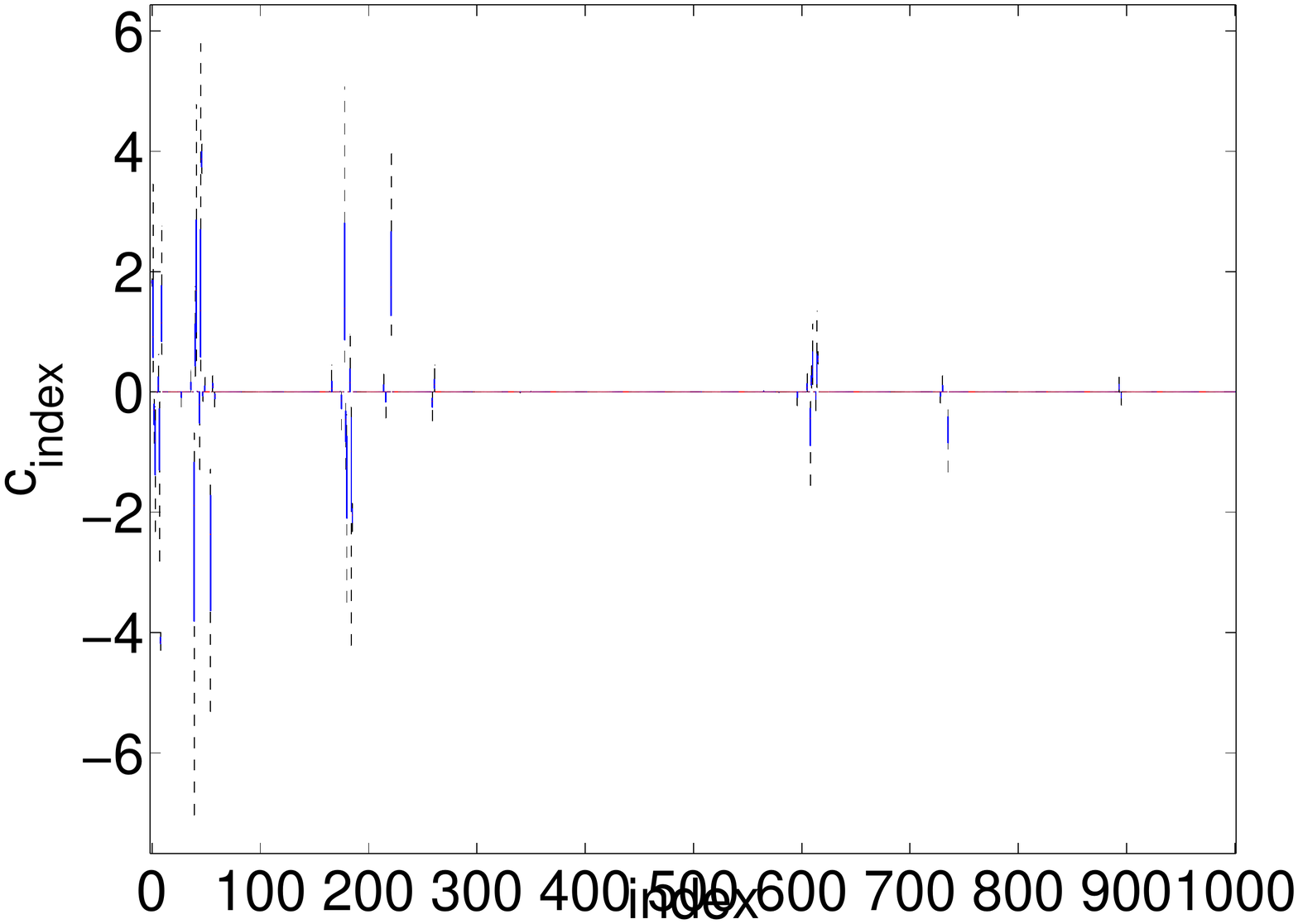}

}

\protect\caption{Boxplots of the posterior PC coefficients; July - December. \label{fig:BP_beta_B}}
\end{figure}

In Figures \ref{fig:BP_beta_A}-\ref{fig:BP_beta_B}, we use box-plots
to represent the posterior density estimates of the PC coefficients
generated by the Gibbs sampler. We observe that the coefficients with
narrow bounds around the zero value correspond to non-significant
PC bases in Figures \ref{fig:Pr_gamma_A}-\ref{fig:Pr_gamma_B}, namely
those with $\Pr(\gamma_{j}|\mathcal{D})<0.5$. That shows that the
method is consistent. Moreover, we observe that the significant PC
coefficients that correspond to the period May-August have in general
larger absolute values. This indicates that the variance of the output
LH during those months is larger compared to that of the rest months,
which is as expected.

The gPC expansion of LH as a function of the input parameters $\xi$
can be evaluated according to the estimator in (\ref{eq:gpc_est_bma})
which is the ergodic average of the Gibbs sample.

\subsection*{Inversion step:}

We calibrate the $10$ CLM parameters $\xi=(\text{Fmax, Cs, Fover, Fdrai, Qdm, Sy, B, Psis, Ks, \ensuremath{\theta}})$
against the measurement of the parameter LH $u^{(\text{f})}$ in Table
\ref{tab:Observed-vale}, as in Section \ref{sec:Methodology_Inverse}.
We consider Beta priors on the CLM parameters whose hyper-parameters
are specified by using the method of moments and based on the prior
information in \citet{hou2012sensitivity}.

Calibration is performed by running the MCMC sampler (Algorithm \ref{alg:MCMCalg_inv})
and evaluating the posterior distributions according to the procedure
in Section \ref{sec:Methodology_Inverse}. Even though, in the previous
step, we detected that the input model parameters are Fdrai, Qdm,
and B are the significant ones, we also consider them for calibration
to obtain information about them as well. In order to make the MCMC
sampler tractable, we replace the forward model CLM4 $u(\cdot)$,
in Algorithm \ref{alg:MCMCalg_inv}, with the estimated gPC expansion
that serves as a surrogate model. We run the MCMC sampler (Algorithm
\ref{alg:MCMCalg_inv}) for $2\cdot10^{4}$ iterations and discard
the first $10^{4}$ as burn in.

\begin{figure}
\center\subfloat[fmax]{\includegraphics[scale=0.2]{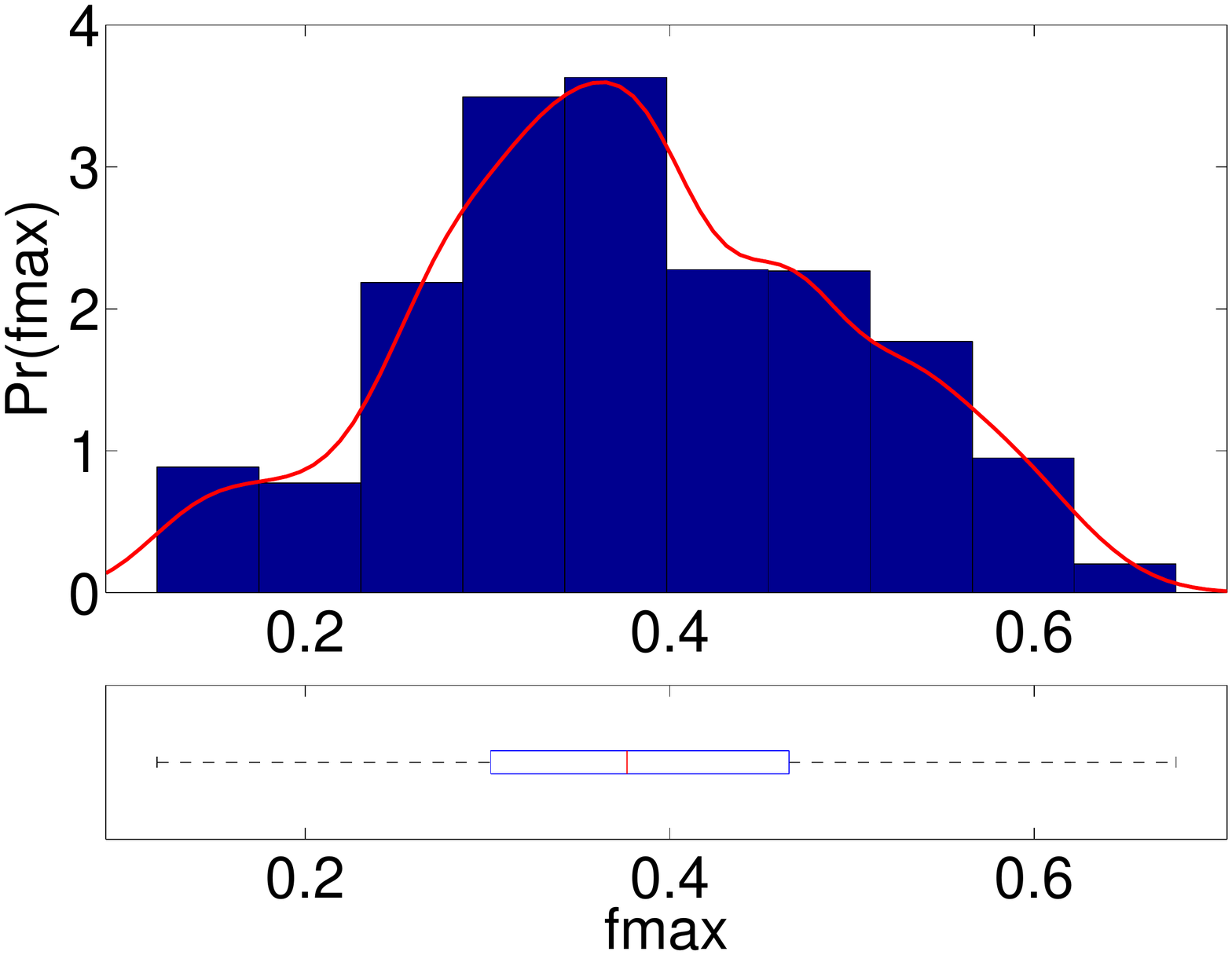}

}\subfloat[Cs]{\includegraphics[scale=0.2]{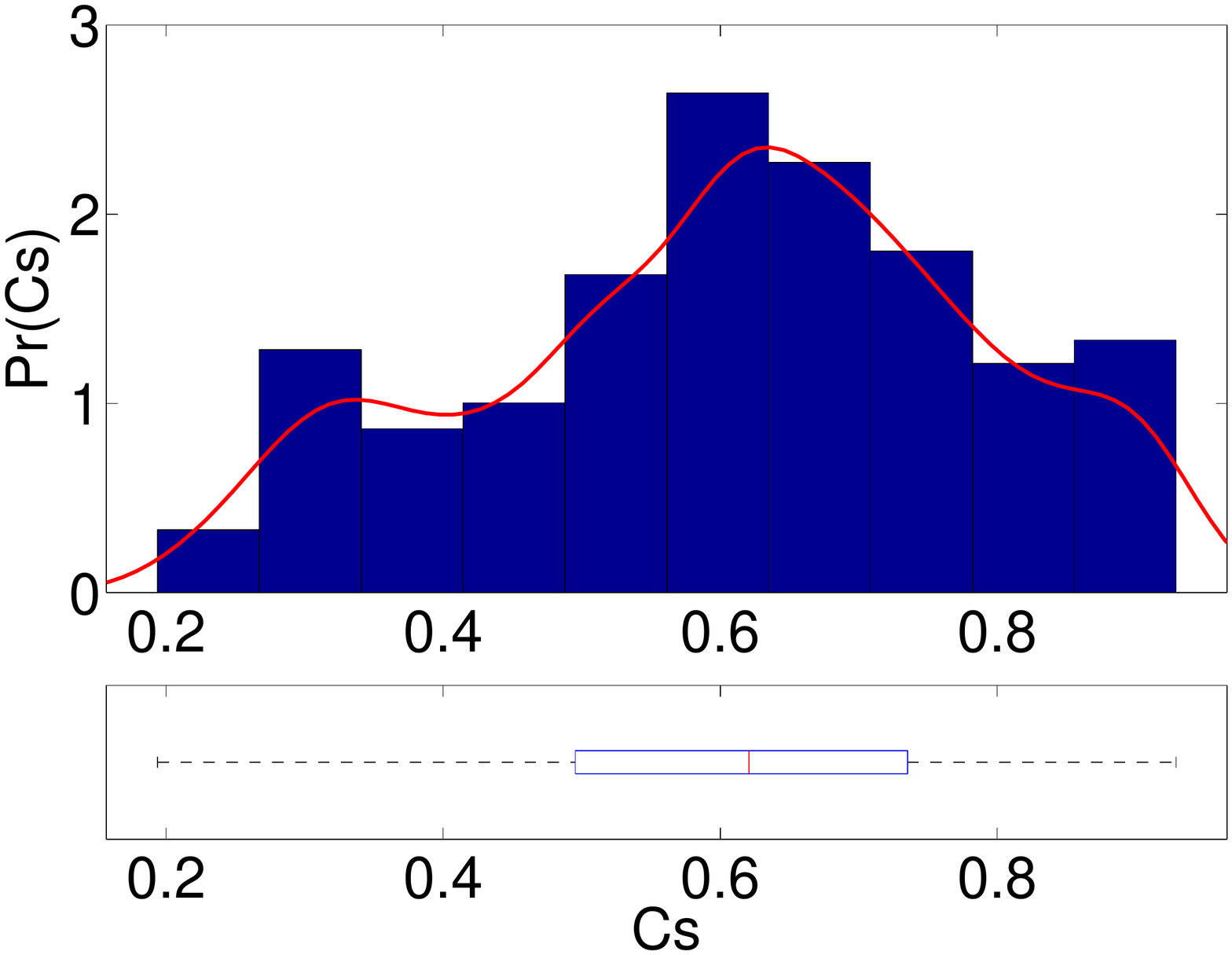}

}

\subfloat[fover]{\includegraphics[scale=0.2]{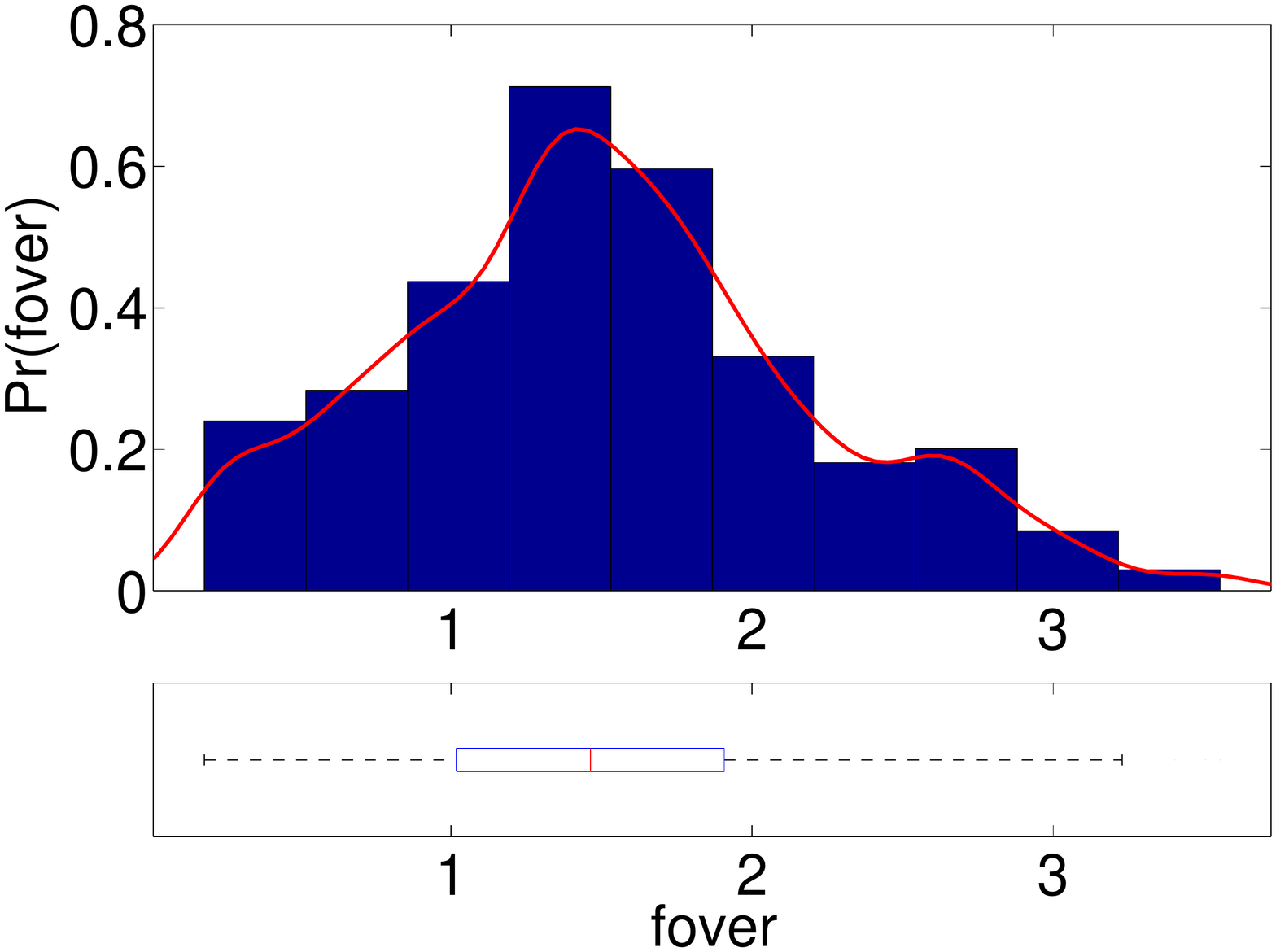}

}\subfloat[fdral]{\includegraphics[scale=0.2]{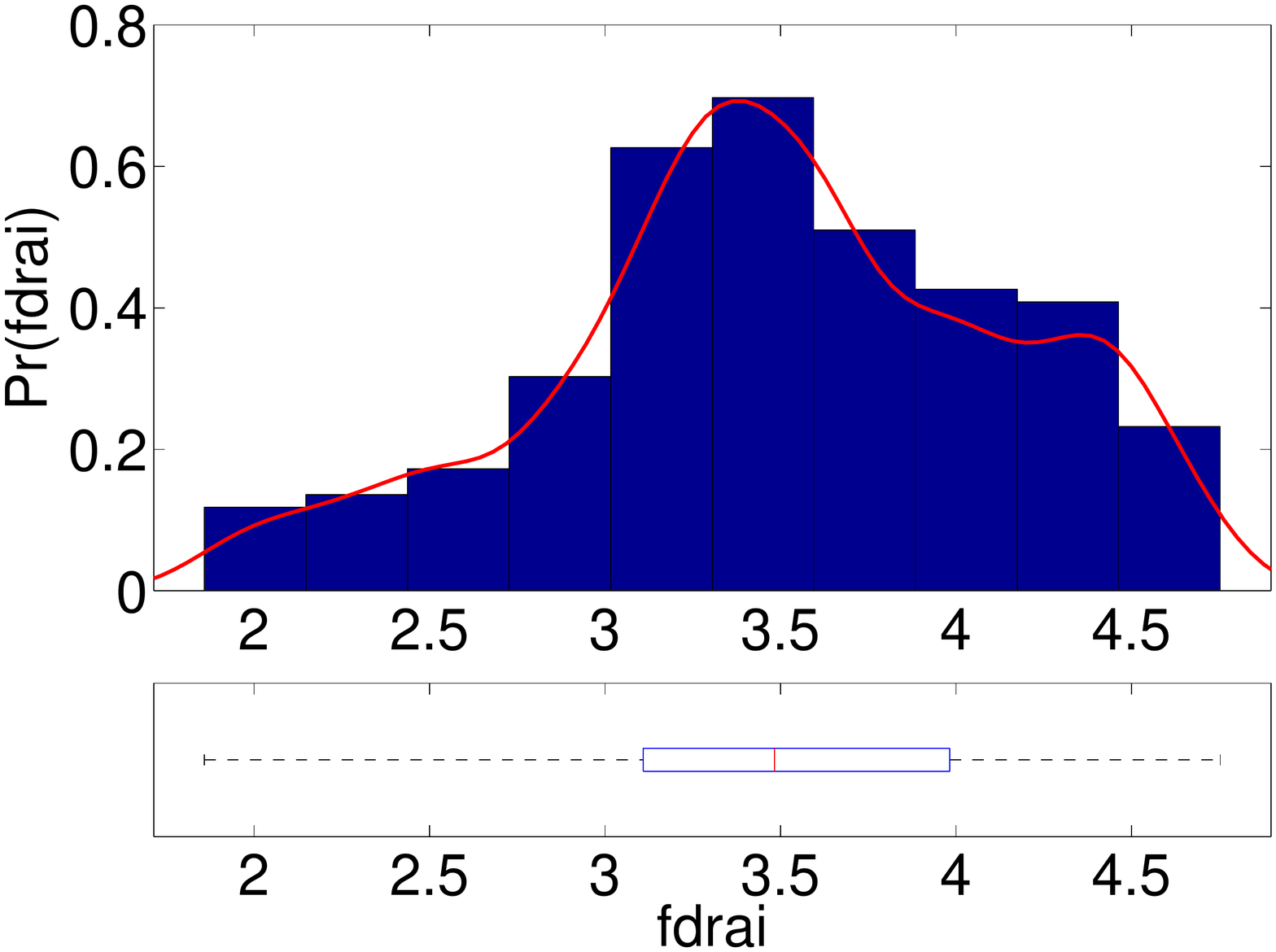}

}

\subfloat[Qdm (log)]{\includegraphics[scale=0.2]{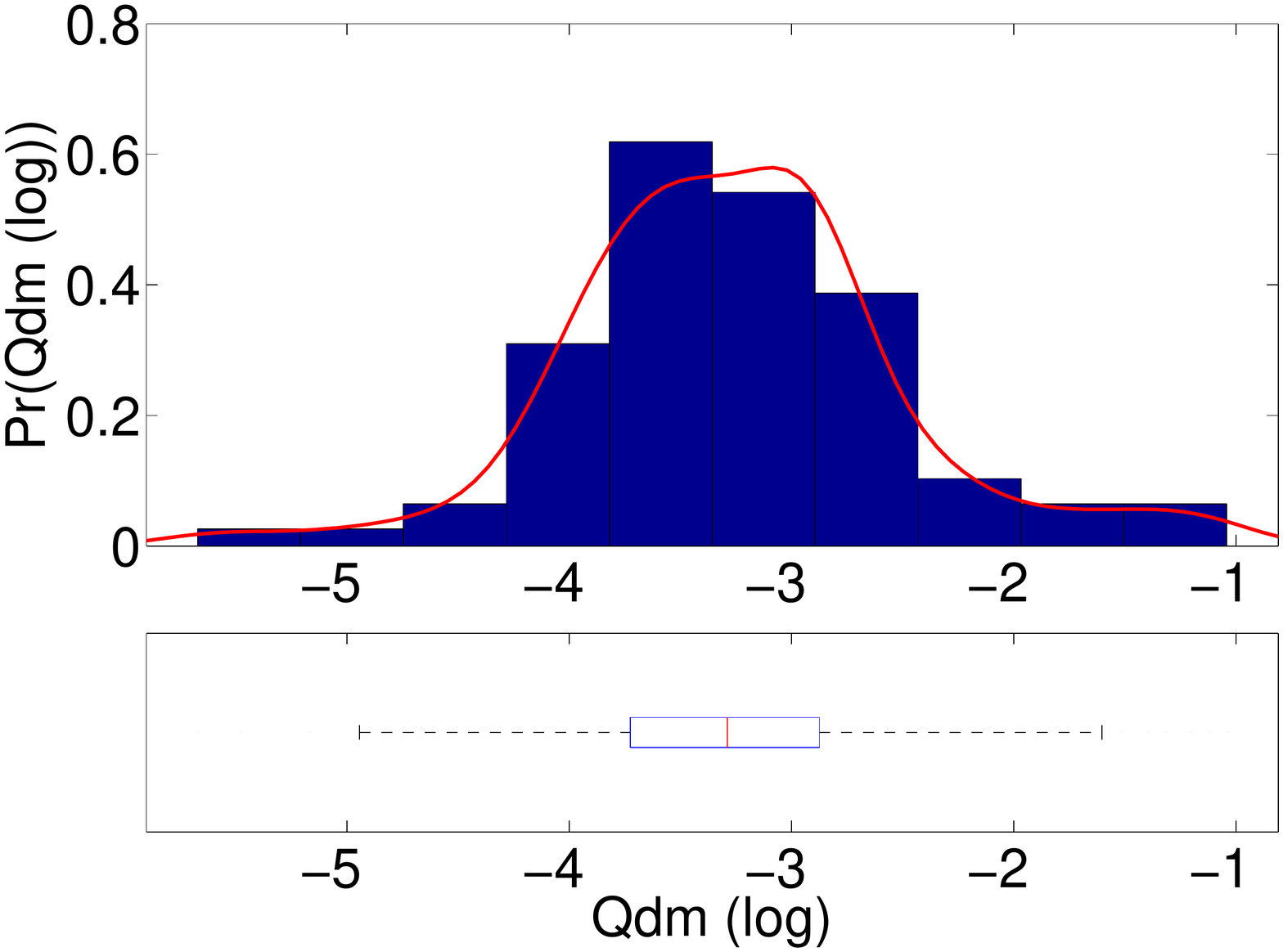}

}

\protect\caption{A posteriori distributions of the input parameters of CLM4 for a given
output $u^{\text{f}}$ (BMA evaluation) \label{fig:BP_xi_inv_x1x5}}
\end{figure}

\begin{figure}
\center\subfloat[Sy]{\includegraphics[scale=0.2]{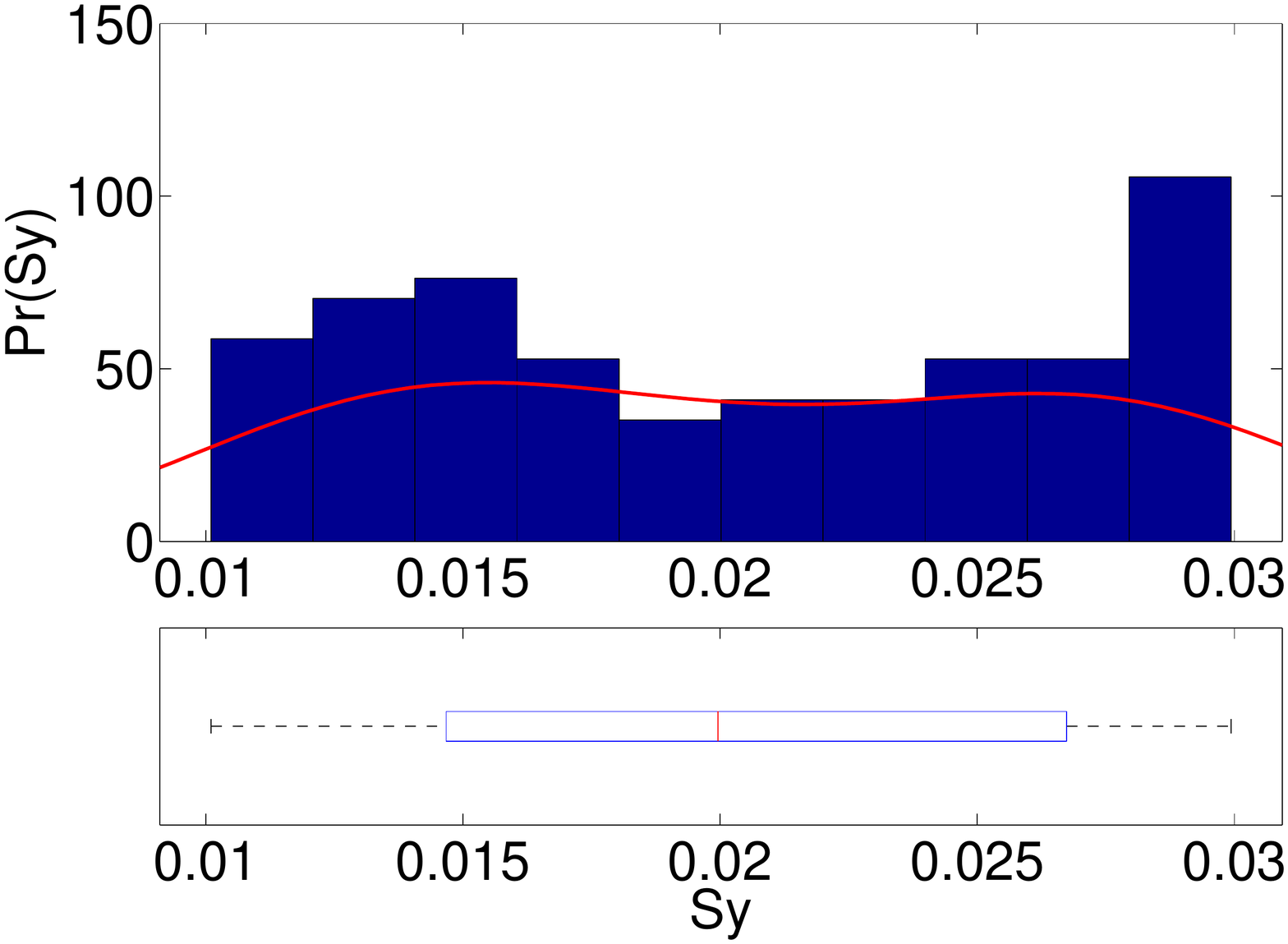}

}\subfloat[b]{\includegraphics[scale=0.2]{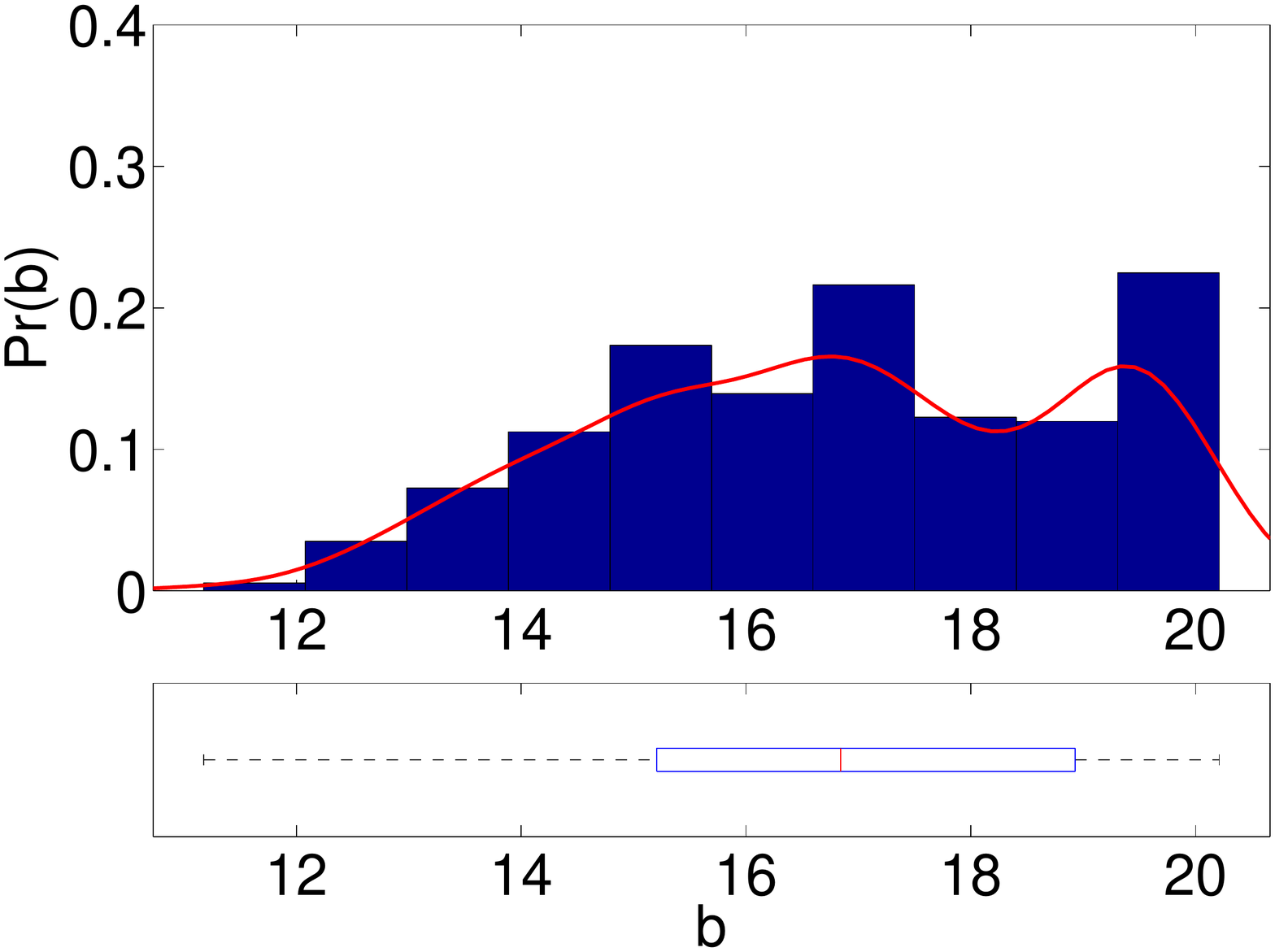}

}

\subfloat[psi (log)]{\includegraphics[scale=0.2]{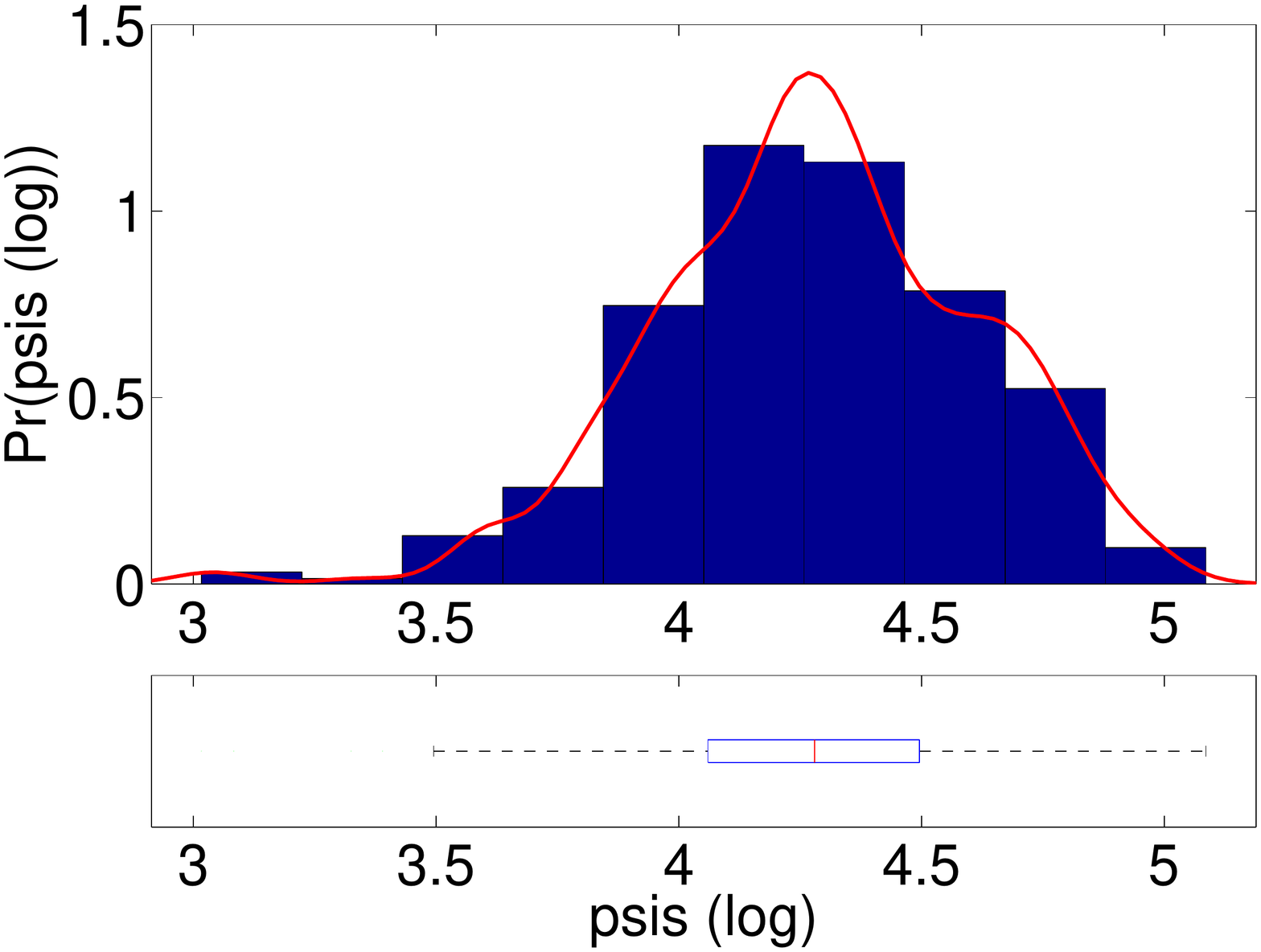}

}\subfloat[Ks (log)]{\includegraphics[scale=0.2]{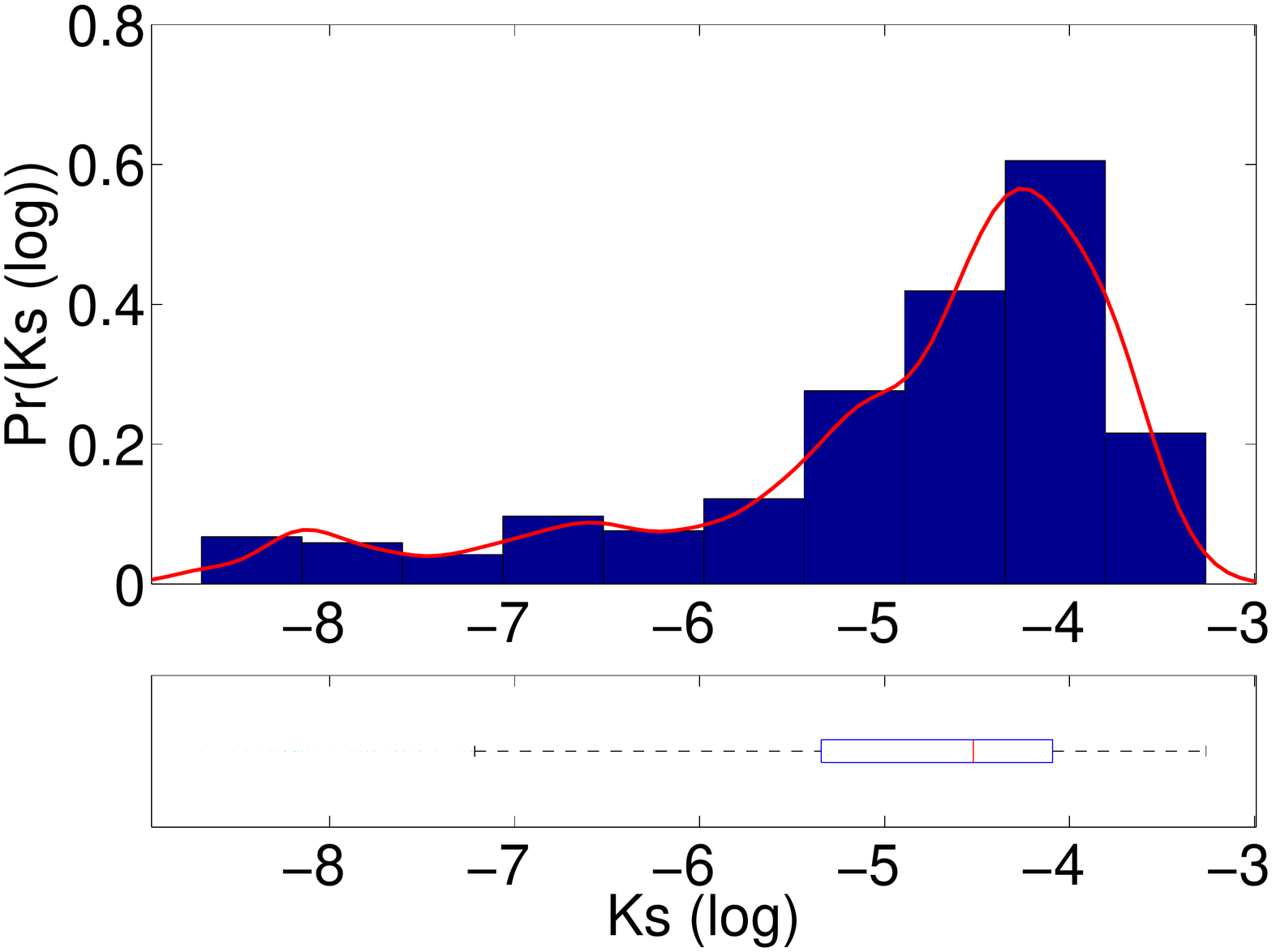}

}

\subfloat[$\theta$]{\includegraphics[scale=0.2]{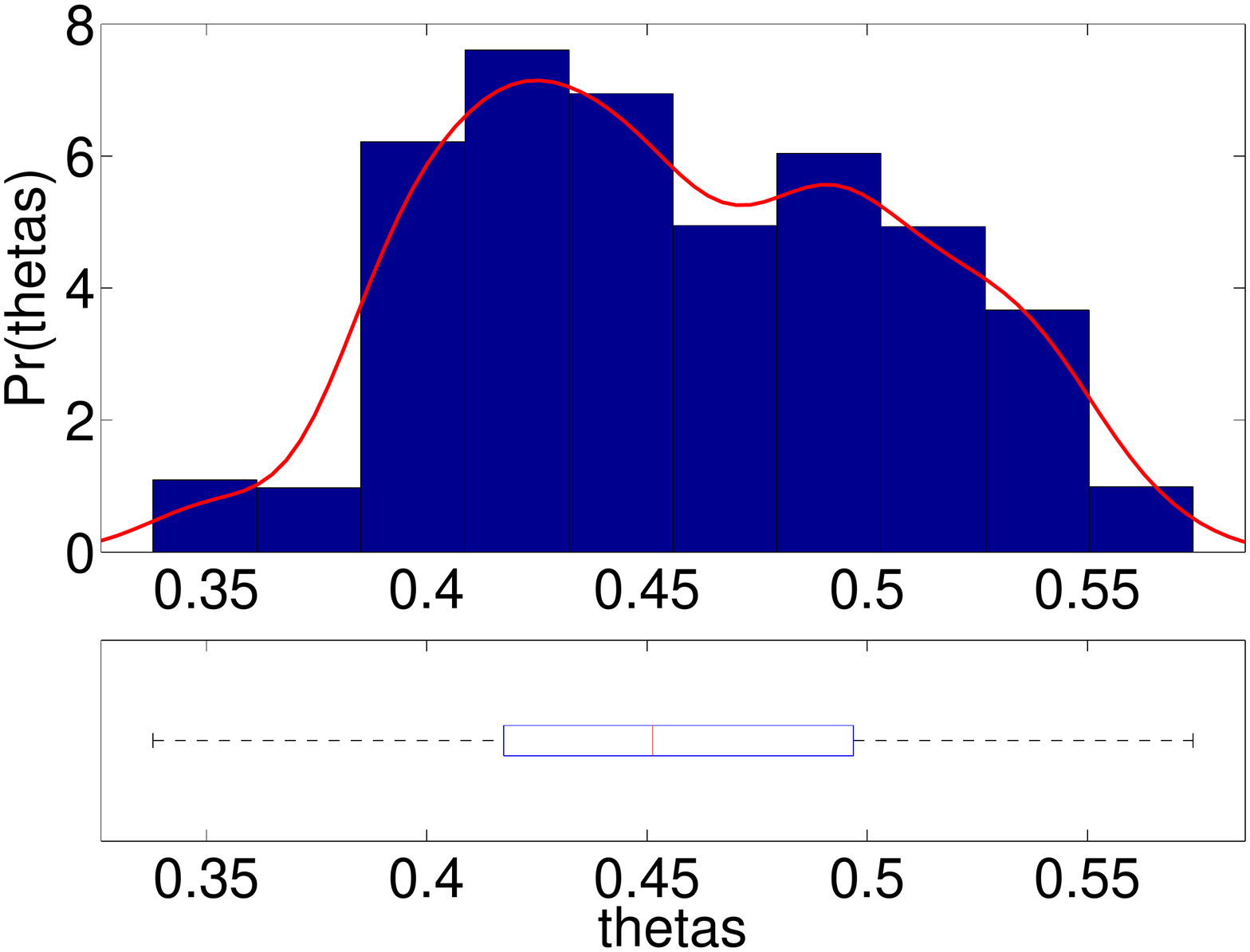}

}

\protect\caption{A posteriori distributions of the input parameters of CLM4 for a given
output $u^{\text{f}}$ (BMA evaluation) \label{fig:BP_xi_inv_x6x10}}
\end{figure}

In Figures \ref{fig:BP_xi_inv_x1x5}-\ref{fig:BP_xi_inv_x6x10}, we
present the estimated posterior densities of the input parameters
of CLM4, as generated by the MCMC sampler. The blue bars correspond
to the histogram estimate while the red line correspond to the kernel
density estimate. These posterior distributions allow us to find a
reasonable range of input values that correspond to the given value
of output $u^{\text{f}}$. The associated box-plots of the marginal
posteriors of the model parameters at each individual figure indicate
the range of the main posterior density. We can see that we have successfully
managed to shorten the ranges of the possible values for most of the
input model parameters. For instance, we observe that the main density
on the marginal posterior distribution of density of Qdm is around
the area $[-4,-2.5]$ in $\log_{10}$ scale.

\subsection*{Model validation}

\begin{figure}
\center\subfloat[January]{\includegraphics[scale=0.2]{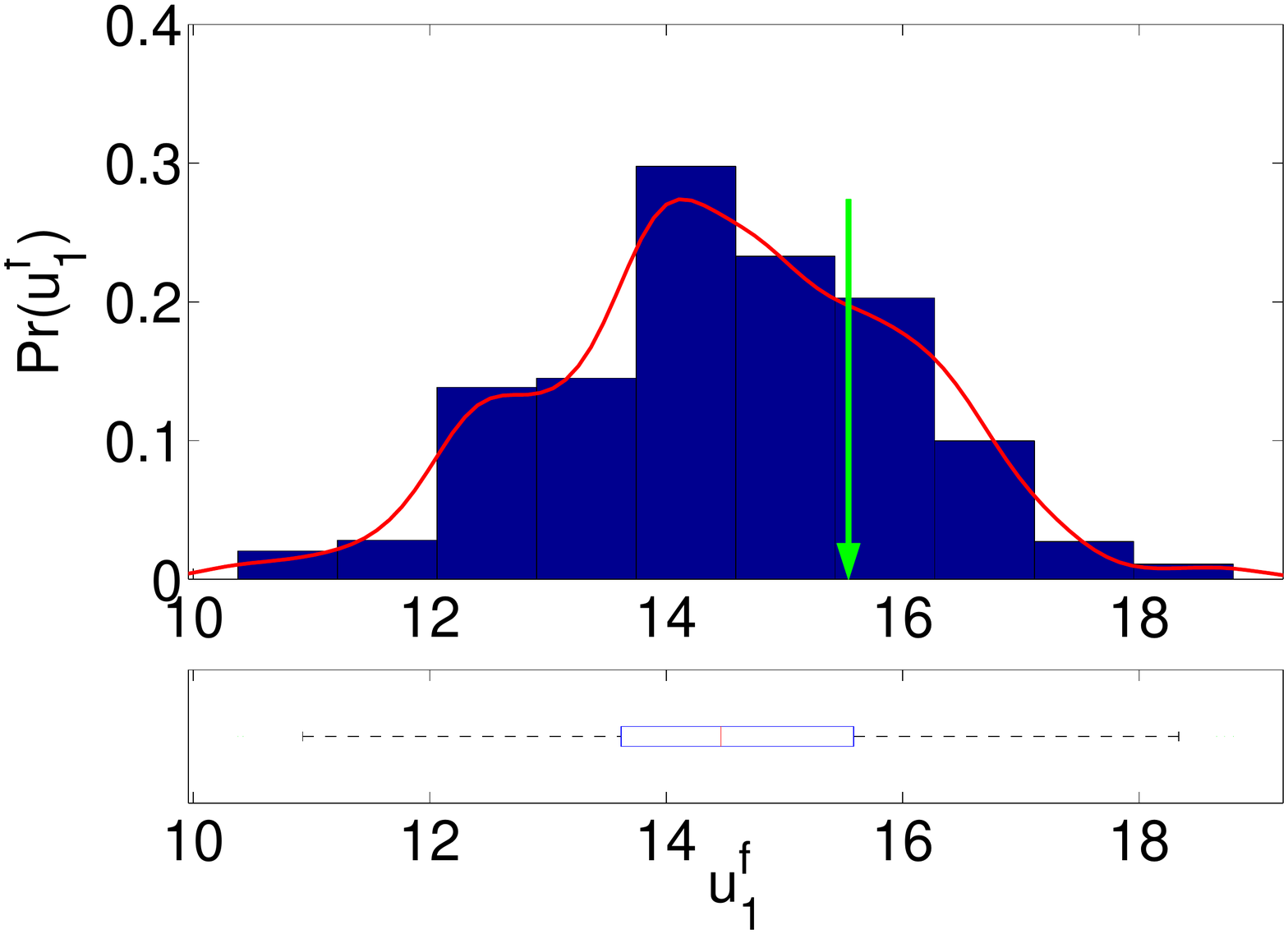}

}\subfloat[February]{\includegraphics[scale=0.2]{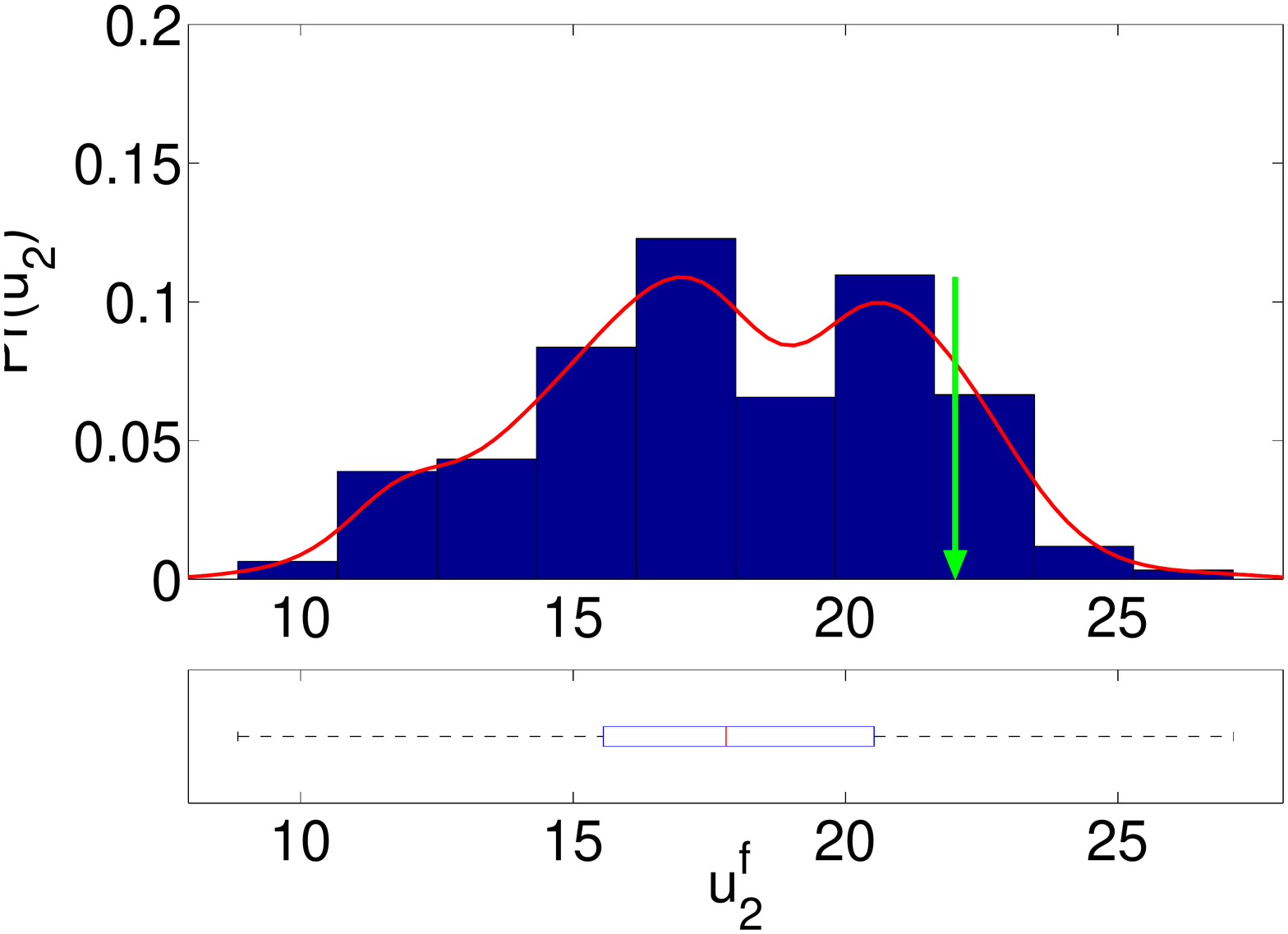}

}

\subfloat[March]{\includegraphics[scale=0.2]{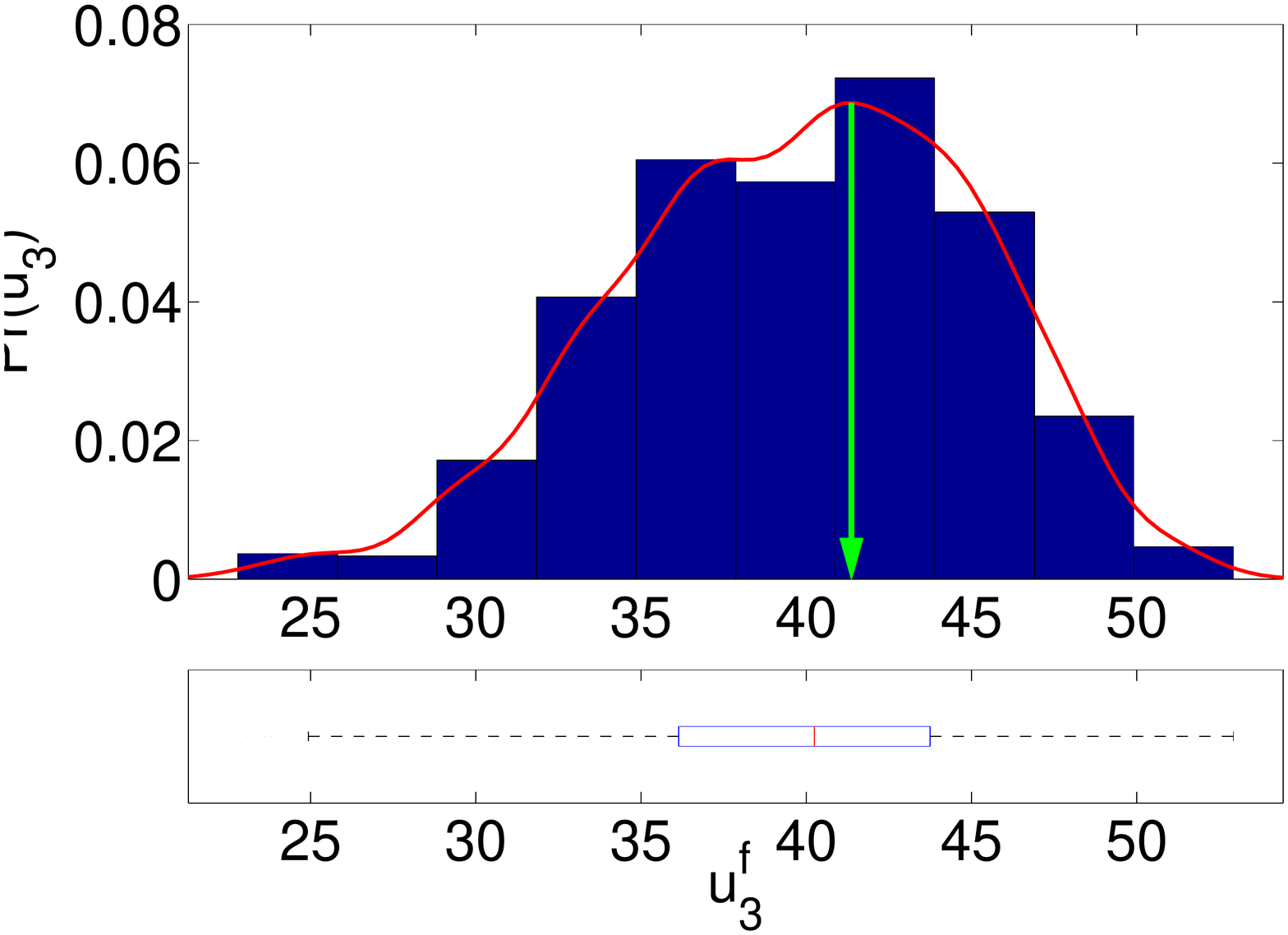}

}\subfloat[April]{\includegraphics[scale=0.2]{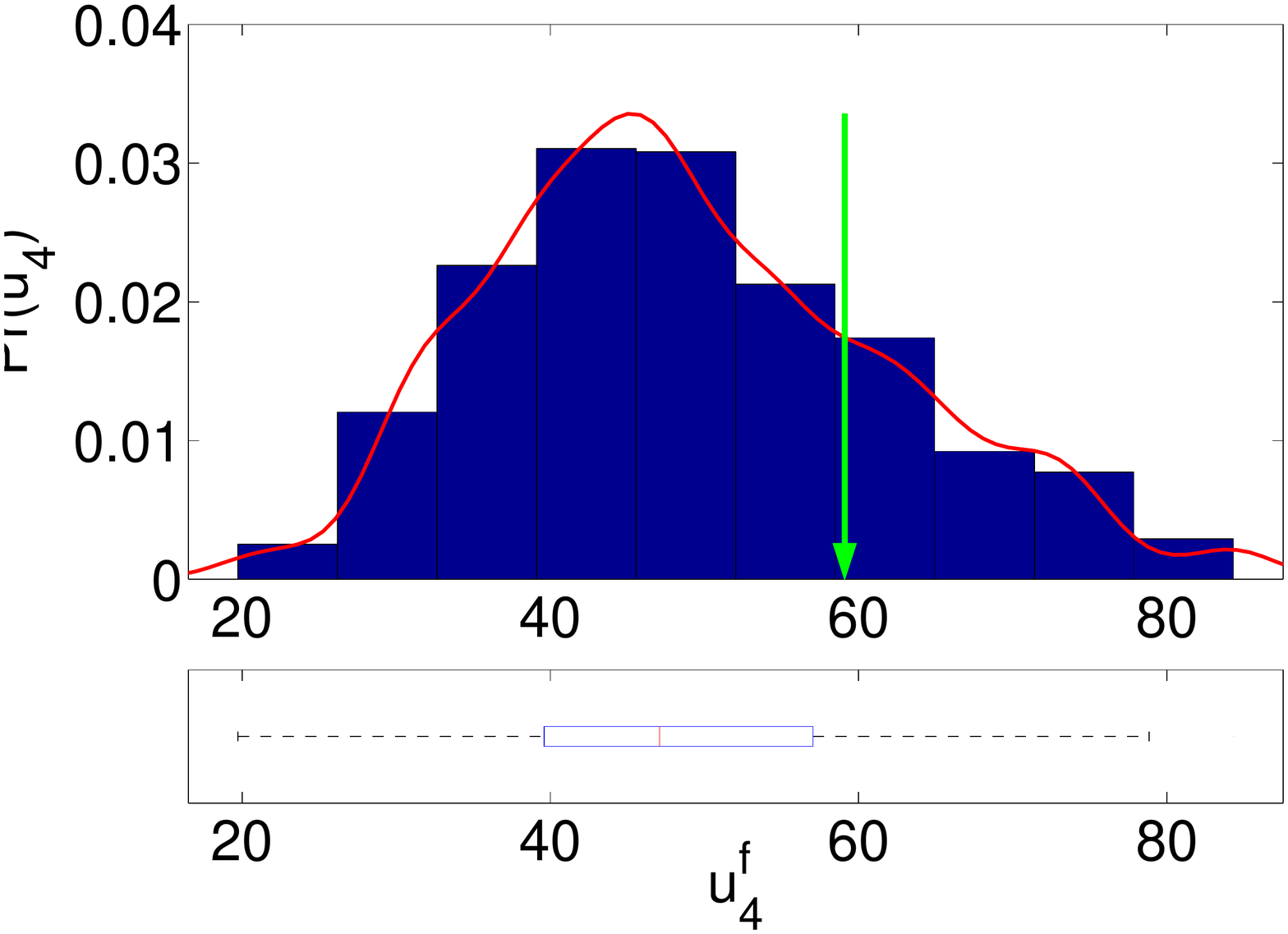}

}

\subfloat[May]{\includegraphics[scale=0.2]{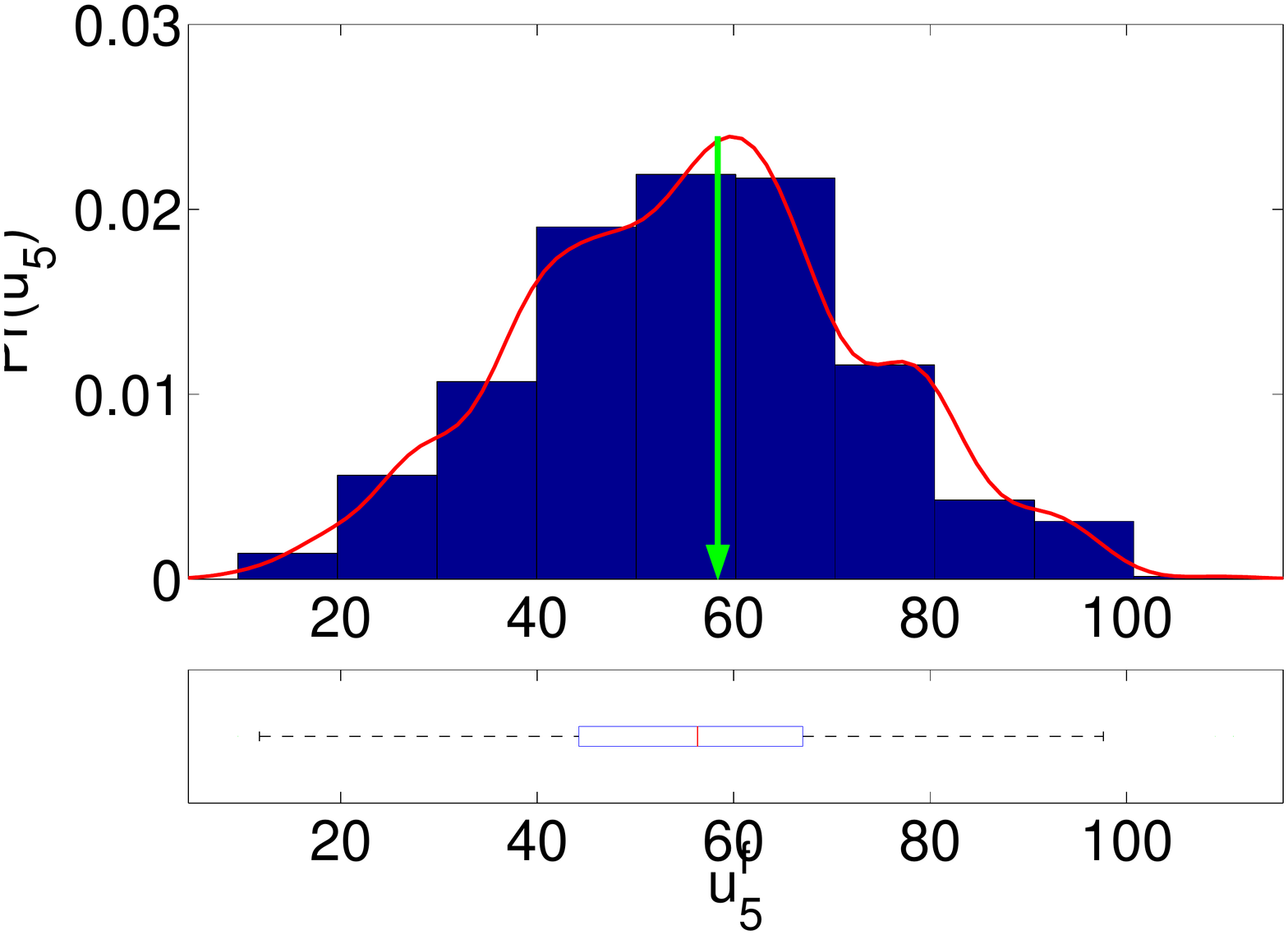}

}\subfloat[June]{\includegraphics[scale=0.2]{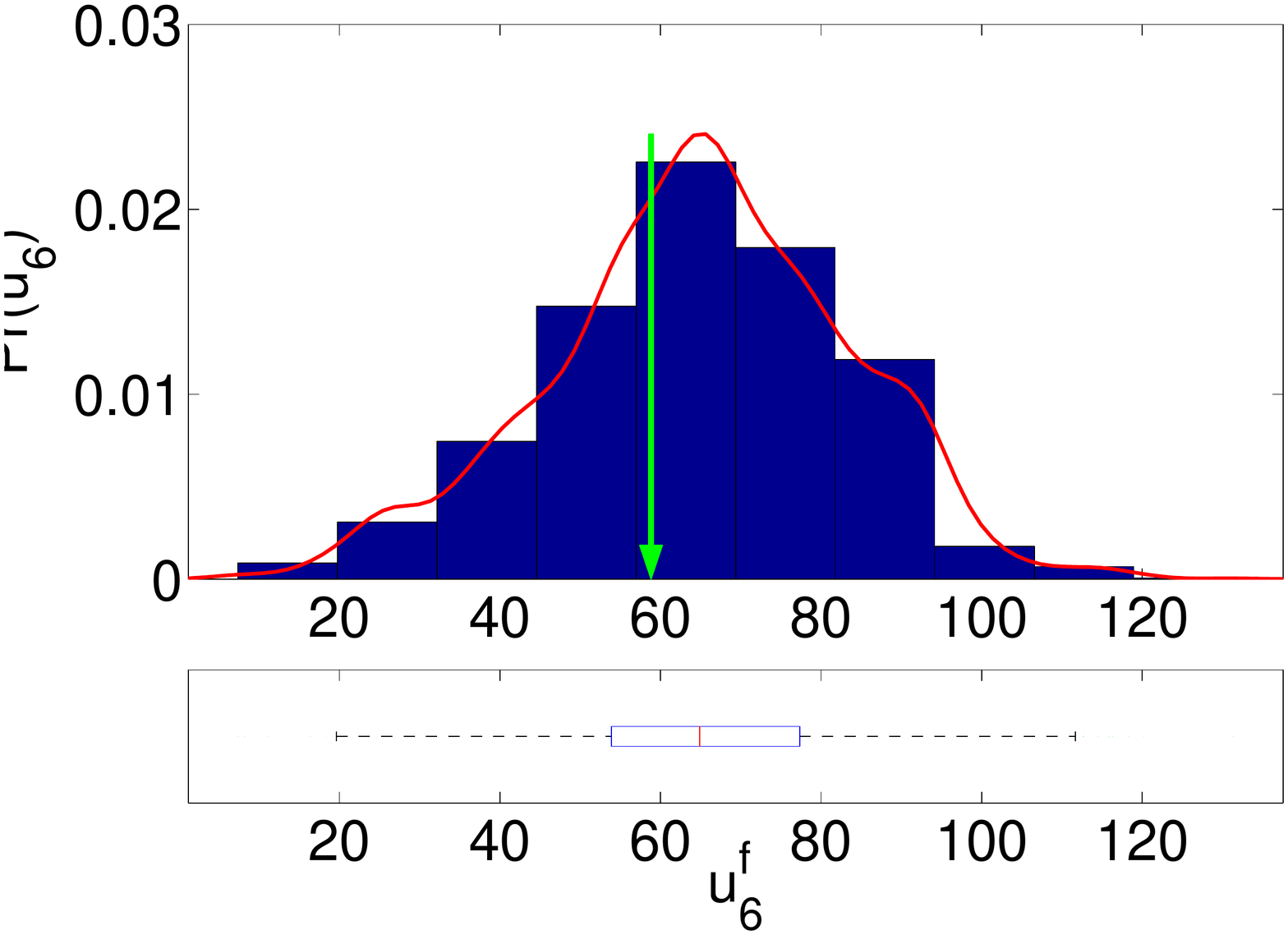}

}

\protect\caption{Distributions of the output LH associated to input parameters of CLM4
drawn by the a posteriori distributions; January - June \label{fig:UFF_BMA_A}}
\end{figure}

\begin{figure}
\center\subfloat[July]{\includegraphics[scale=0.2]{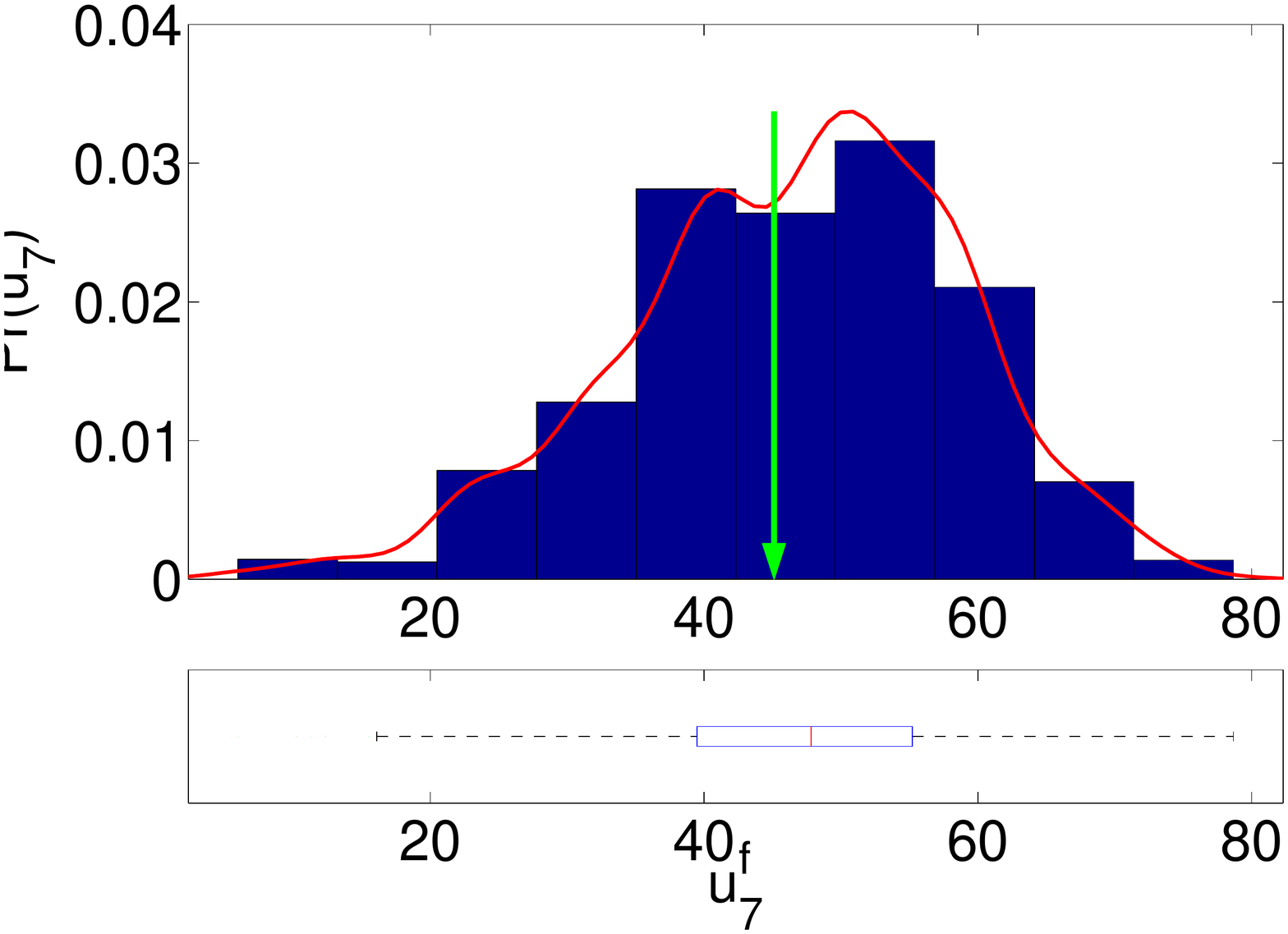}

}\subfloat[August]{\includegraphics[scale=0.2]{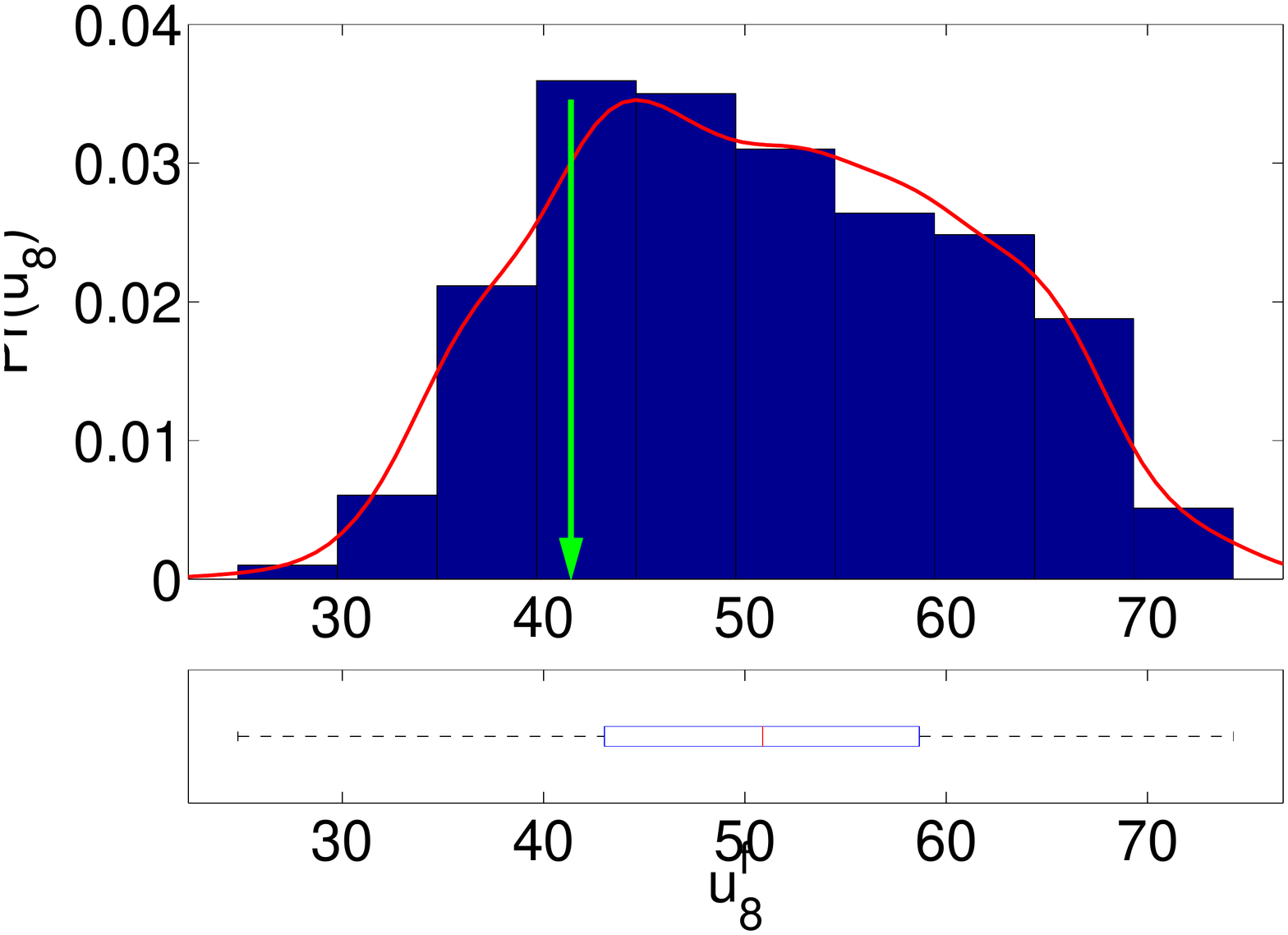}

}

\subfloat[September]{\includegraphics[scale=0.2]{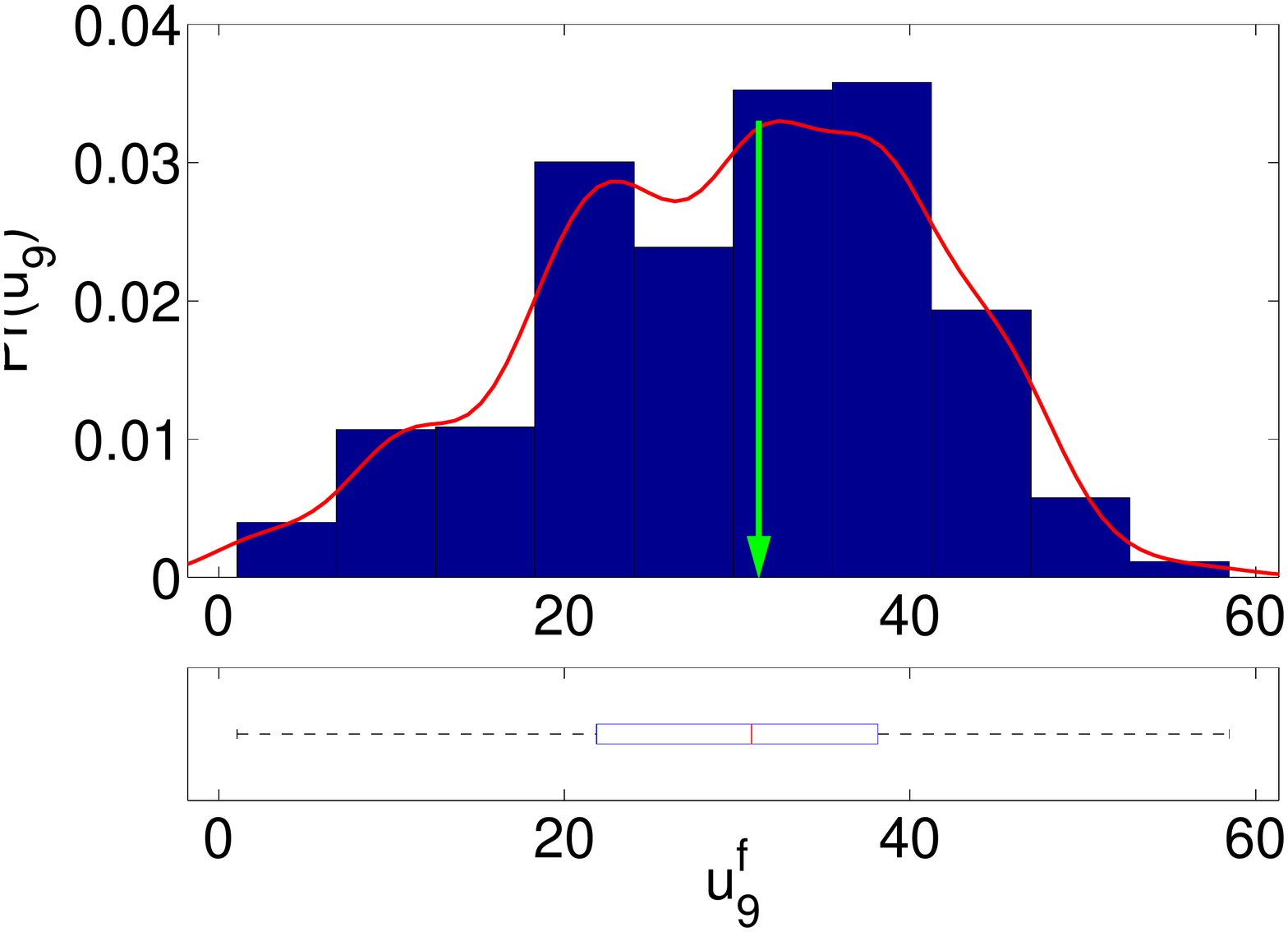}

}\subfloat[October]{\includegraphics[scale=0.2]{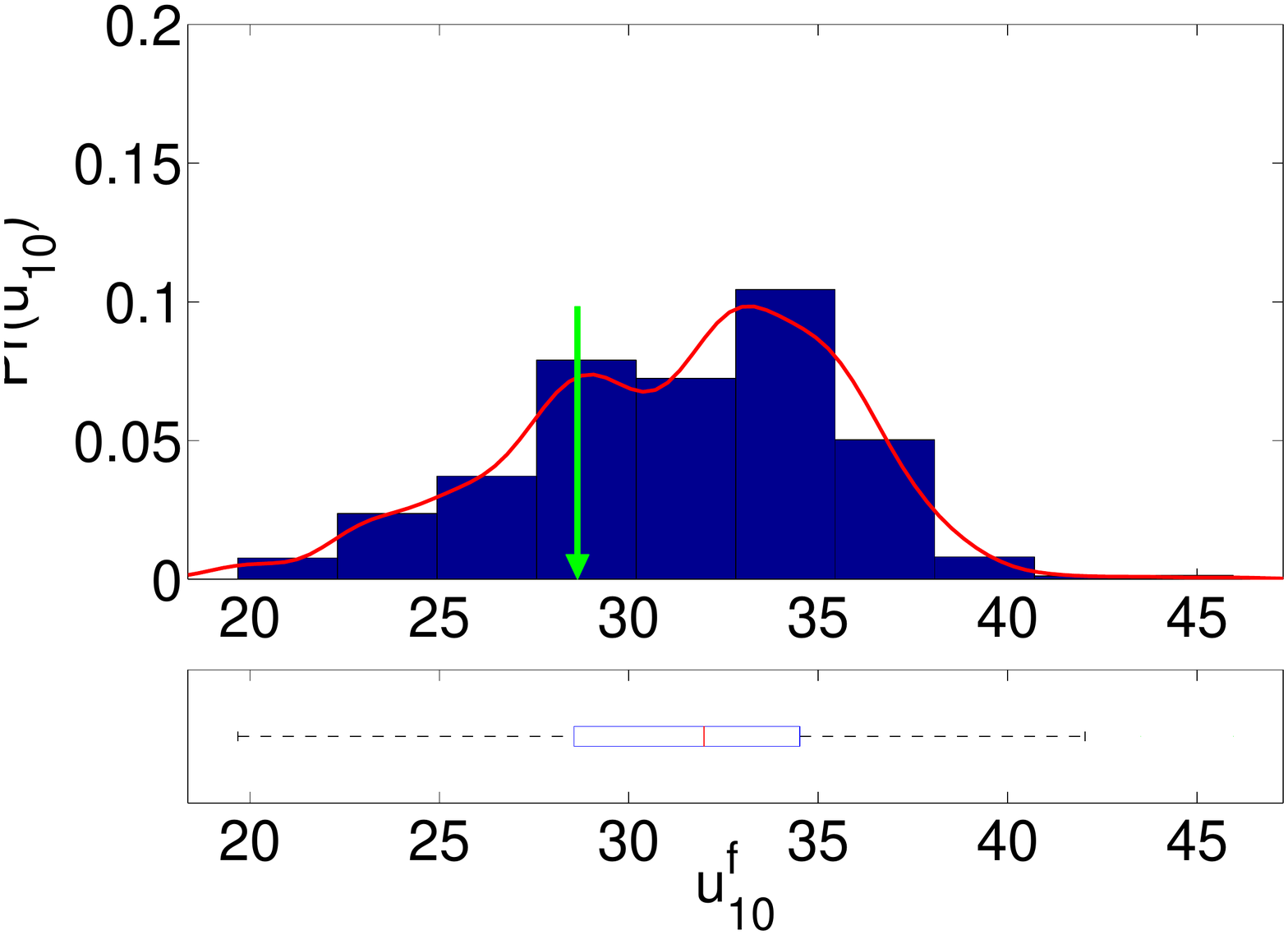}

}

\subfloat[November]{\includegraphics[scale=0.2]{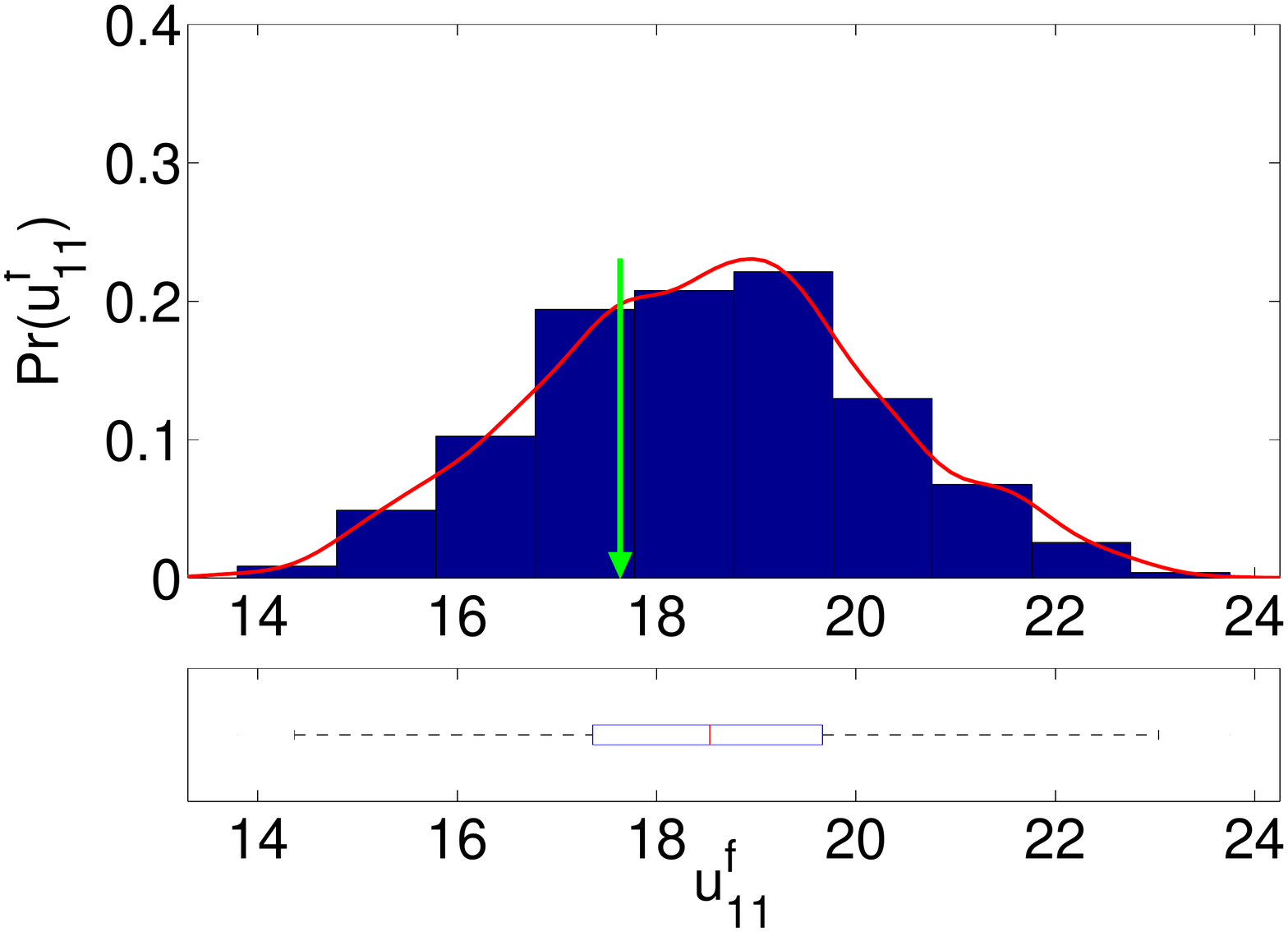}

}\subfloat[December]{\includegraphics[scale=0.2]{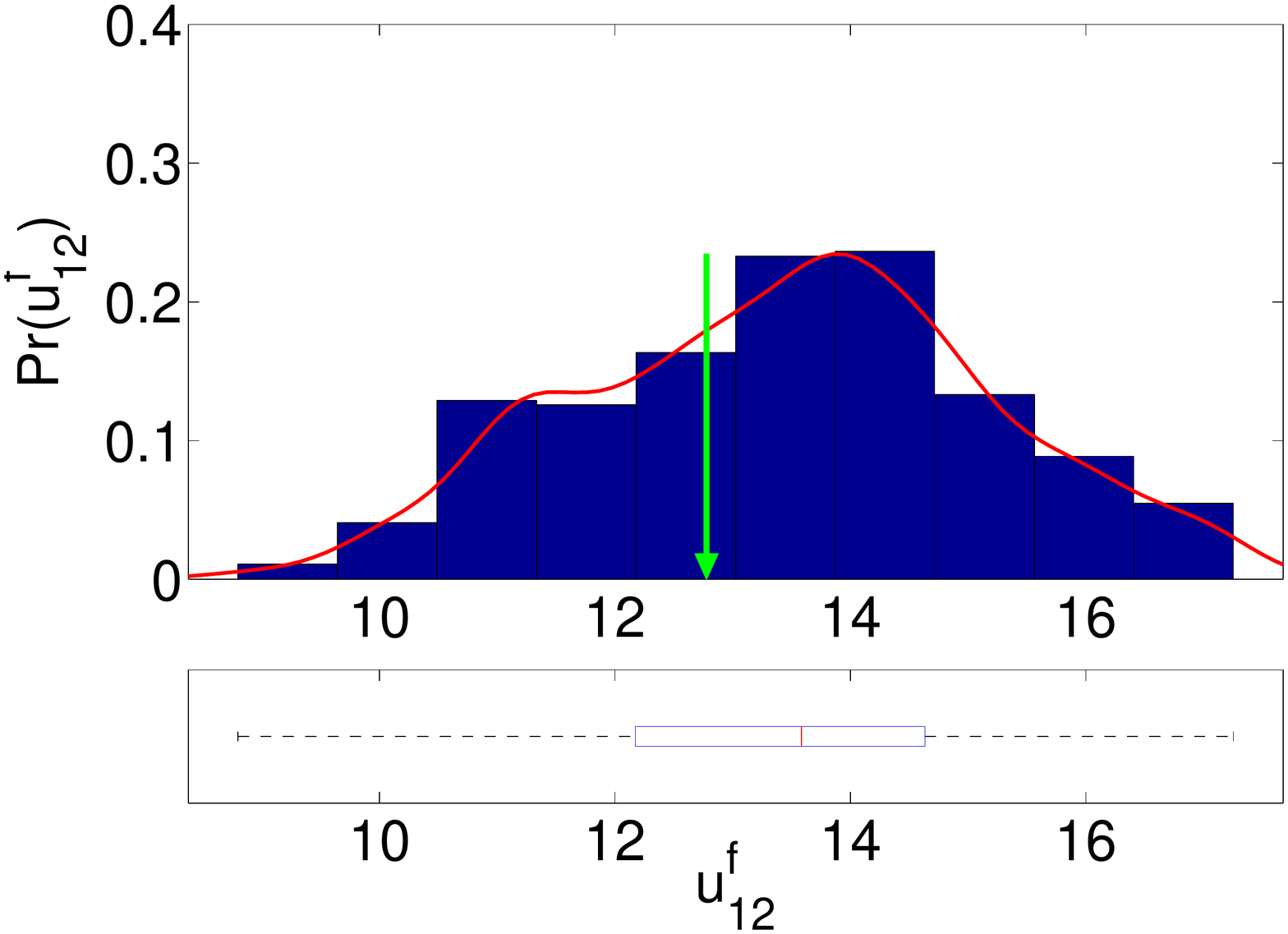}

}

\protect\caption{Distributions of the output LH associated to input parameters of CLM4
drawn by the a posteriori distributions; July - December \label{fig:UFF_BMA_B}}
\end{figure}

We validate the effectiveness of the Bayesian inversion procedure.
We evaluate the predictive distribution of the output parameter LH,
by using the derived gPC surrogate model, and the MCMC sample of the
input parameters generated by Algorithm \ref{alg:MCMCalg_inv}. In
Figures \ref{fig:UFF_BMA_A}-\ref{fig:UFF_BMA_B}, we present the
resulted predictive distributions of the output LH for the $12$ months.
The blue bars belong the histogram estimate, the red line is the kernel
density estimate of the predictive distribution, while the green arrow
represents the observed output value of LH. The plots show that for
each month the observed output value $u^{\text{f}}$ for the LH lie
below the modes of each of the marginal predictive distributions.
This implies that the proposed methodology is valid, and that the
surrogate model derived from the method is able to produce accurate
predictions.

\section{Conclusion \label{sec:Conclusions}}

In this study, we focused on evaluating uncertainties associated with
hydrologic parameters in CLM4, in the Bayesian framework. We presented
a Bayesian methodology for the uncertainty quantification (UQ) framework
that couples generalized Polynomial Chaos model with the Bayesian
variable selection methods. We presented a fully Bayesian methodology
that involves two steps: the construction of a surrogate model to
express the input-output mapping, and the evaluation of the posterior
distribution of the input parameters for a given value of the output
parameter LH.

For the construction of the surrogate model we propose a Bayesian
procedure, based on variable selection methods, that uses gPC expansions
and accounts for bases selection uncertainty. The advantage of this
approach is that it can quantify the significance of the gPC terms,
and hence the importance of the input parameters, in a probabilistic
manner. The input posterior distributions were evaluated according
to Bayesian inverse modeling. Our empirical results, showed that the
proposed method is suitable to perform inverse modeling of hydrologic
parameters in CLM4, and able to effectively describe the uncertainty
related to these parameters.

Our future work involves the comparison of the proposed method against
other methods based on neural networks, and generalized linear models
on which we are currently working.

\begin{singlespace}

\begin{acknowledgements} Guang Lin would like to thank the support
from the National Science Foundation (DMS-1555072, DMS-1736364, and
DMS-1821233) and U.S. Department of Energy, Office of Electricity
Delivery and Energy Reliability Advanced Grid Modeling Program. \end{acknowledgements}

\end{singlespace}

\begin{singlespace}\bibliographystyle{apalike}
\bibliography{citations}

\end{singlespace}
\end{document}